\documentclass[PAPER, cernpreprint, texlive=2023, UKenglish, texmf, orcidlogo]{atlasdoc}
 
\usepackage{atlaspackage}
\usepackage[block=none]{atlasbiblatex}
 
\usepackage[process=true]{atlasphysics}

\graphicspath{{logos/}{figures/}}
 
\usepackage{ANA-TOPQ-2018-40-PAPER-defs}

\AtlasTitle{
Measurement of the \ttbar\ production cross-section in $pp$ collisions at $\sqrt{s}=5.02$\,\TeV\ with the ATLAS detector
}
 
\AtlasAbstract{
The inclusive top-quark pair (\ttbar) production cross-section \xtt\ is
measured in proton--proton collisions at a centre-of-mass energy
$\sqrt{s}=5.02$\,\TeV, using
\intlumi\,\ipb\ of data collected in 2017 by the ATLAS experiment at the LHC.
The \ttbar\ cross-section is measured in both the dilepton and single-lepton
final states of the \ttbar\ system and then combined.
The combination of the two measurements yields
\[
\xtt = \ttxval \pm \ttxstat\,\mathrm{(stat.)} \pm \ttxsyst\,\mathrm{(syst.)} \pm \ttxlumi\,\mathrm{(lumi.)} \pm \ttxebeam\,\mathrm{(beam)}\,\mathrm{pb},
\]
where the four uncertainties reflect the limited size of the data sample,
experimental and theoretical systematic effects, and imperfect knowledge of both
the integrated luminosity and the LHC beam energy, giving a total uncertainty
of \ttxrel.
The result is in agreement with theoretical quantum chromodynamic calculations at next-to-next-to-leading
order in the strong coupling constant, including the resummation of next-to-next-to-leading logarithmic soft-gluon
terms, and constrains the parton distribution functions
of the proton at large Bjorken-$x$.
}
 
\author{The ATLAS Collaboration}

\AtlasNote{TOPQ-2018-40}
\PreprintIdNumber{CERN-EP-2022-119}
\AtlasJournalRef{JHEP 06 (2023) 138}
\AtlasDOI{10.1007/JHEP06(2023)138}

\hypersetup{pdftitle={ATLAS document},pdfauthor={The ATLAS Collaboration}}
 
\begin{document}
 
\maketitle
 
\tableofcontents
 
\section{Introduction}
\label{sec:intro}

The study of top quark--antiquark (\ttbar) production in proton--proton ($pp$)
collisions forms a central part of the physics programme of the
ATLAS experiment at the Large Hadron Collider (LHC).
Measurements of the inclusive \ttbar\ cross-section \xtt\ allow studies of
quantum  chromodynamics (QCD) at some of the highest accessible energy scales.
Predictions for \xtt\ in $pp$
collisions are available at next-to-next-to-leading-order (NNLO) accuracy
in the strong coupling constant \alphas, including the resummation of
next-to-next-to-leading logarithmic (NNLL) soft-gluon
terms~\cite{topxtheo1,topxtheo2,topxtheo3,topxtheo4,topxtheo5,topxtheo6}.
These predictions are in excellent agreement with measurements from the ATLAS,
CMS, and LHCb collaborations at $\sqrt{s}=7$, 8, and 13\,\TeV~\cite{
TOPQ-2013-04,
TOPQ-2016-08,
TOPQ-2018-17,
TOPQ-2020-02,
CMS-TOP-13-004,
CMS-TOP-12-006,
CMS-TOP-16-006,
CMS-TOP-17-001,
lhcbttbar1,
lhcbttbar3,
}.
 
During the data-taking period from 2015 to 2018, known as Run~2, the LHC provided samples of $pp$ collisions at \sxut, primarily to support the heavy-ion physics programme, but also to measure Standard Model (SM) physics processes
in this intermediate energy regime.
For \ttbar\ production, the lower $\sqrt{s}$ value increases the fraction
of \qqbar-initiated events from 11\% at \sxyt\ to about 25\%,
making this data complementary to the larger samples at higher centre-of-mass
energies and offering the potential for additional constraints on parton
distribution functions (PDFs).
Assuming a top-quark mass of $\mtop=172.5$\,\GeV,
the NNLO+NNLL prediction for \xtt\ at \sxut\ calculated with the \TOPpp\
program~\cite{toppp} is $68.2\pm 4.8^{+1.9}_{-2.3}$\,pb,
where the first uncertainty corresponds to PDF and \alphas\ uncertainties
and the second to QCD scale variations.
The PDF and \alphas\ uncertainties were
calculated using the PDF4LHC prescription~\cite{Butterworth:2015oua}\ with the
\MSTW~\cite{mstwnnlo1,mstwnnlo2}, \CT[10nnlo]~\cite{cttenpdf,cttennnlo},
and \NNPDF[2.3 5f FFN]~\cite{nnpdfffn} PDF sets.
The scale uncertainties were calculated from an envelope of predictions with
the QCD renormalisation and factorisation scales (\mur\ and \muf) increased
and decreased independently by a factor of two from their default values
of $\mur=\muf=\mtop$ \cite{qcdscale1,qcdscale2}, while never letting the scales
differ by more than a factor of two from each other.
The total uncertainty corresponds
to a relative precision of $^{+7.5}_{-7.7}$\%. The cross-section further
varies by $^{-3.2}_{+3.3}$\% for a $\pm 1$\,\GeV\ variation in \mtop.
The prediction agrees well with a measurement of $63.0\pm 5.1$\,pb from
the CMS Collaboration using data samples at $\sqrt{s}=5.02$~\TeV\ of 302\,\ipb\ recorded in 2017 and 27\,\ipb\ recorded in 2015~\cite{CMS-TOP-20-004}.
 
This paper reports a measurement of \xtt\ at \sxut\ using a \intlumi\,\ipb\
data sample recorded by the ATLAS Collaboration in 2017. The measurement was performed in both
the dilepton and single-lepton channels of the \ttbar\ decay.
The dilepton-channel final states are those
where the \Wboson\ bosons from both top quarks decay leptonically, i.e.
$\ttbar\rightarrow W^{+}bW^{-}\bar{b}\rightarrow\ell\ell^{\prime}\nu\bar{\nu}\bbbar$,
where $\ell$\ and $\ell^{\prime}$\ represent an electron or muon, including those produced
in leptonic decays of $\tau$-leptons ($W\rightarrow\tau\rightarrow \ell$).
Event samples with an opposite-charge pair of leptons, transverse-momentum imbalance arising from the
presence of two neutrinos,  and one or two jets
tagged as likely to contain $b$-hadrons were used to measure the rate of \ttbar\ production in
the dilepton channel, which is characterised by high \ttbar\ purity but
has a relatively small number of selected events.
This technique is similar to that  used in measurements of \xtt\ using $e\mu$
events~\cite{TOPQ-2013-04,TOPQ-2018-17} but also
includes the same-flavour $ee$ and $\mu\mu$ events.
The single-lepton final states are those where one \Wboson\ boson decays
leptonically and the other decays hadronically. This
results in a final state characterised by a charged lepton, transverse-momentum imbalance arising from a neutrino,
two jets from the bottom quarks, and two or more
jets arising from the hadronisation of the \Wboson-boson decay products.
The single-lepton event samples were defined by requiring a charged lepton
$\ell$, missing transverse momentum indicating
a neutrino, and two or more jets with at least one of the jets tagged as being likely to contain a $b$-hadron.
The single-lepton sample was separated into subsamples with different signal-to-background ratios,
increasing the precision of the measurement.
The measurement also used a multivariate technique to further separate the \ttbar\ signal from background events, using
techniques similar to those used in \xtt\ measurements at \sxyt\ \cite{TOPQ-2020-02}.
This approach combined the separation power of several variables and their correlations.
Finally, the dilepton and single-lepton measurements were combined, taking the correlated
systematic uncertainties into account.
 
This paper is structured as follows.
The ATLAS detector is described in Section~\ref{sec:detector}, followed by the data and simulation samples used in the analysis in Section~\ref{sec:samples}, and the event reconstruction
in Section~\ref{sec:selection}.
The dilepton measurement is described
in Section~\ref{sec:dilepton}\ while the single-lepton measurement is
described in Section~\ref{sec:leptonjets}.
Systematic uncertainties associated with the measurements are described in Section~\ref{sec:systematics} and the results of the individual channels are
reported in Section~\ref{sec:results}. The combined cross-section measurement
is presented in Section~\ref{sec:combination},
together with a comparison with predictions from various PDF sets, and an
illustration of how the gluon PDF is affected by
including this measurement in the ATLAS PDF fits.
The conclusions of this study are summarised in Section~\ref{sec:conclusions}.


\section{ATLAS detector}
\label{sec:detector}

The ATLAS detector~\cite{PERF-2007-01} at the
LHC is centred on the $pp$\ collision point and covers nearly the whole $4\pi$\
solid angle.\footnote{ATLAS uses a right-handed coordinate system with
its origin at the nominal interaction point (IP) in the centre of
the detector and the $z$-axis along the beam line.  The $x$-axis
points from the IP to the centre of the LHC ring, and the $y$-axis
points upwards.  Cylindrical coordinates $(r,\phi)$ are used in the
transverse plane, $\phi$ being the azimuthal angle around the
$z$-axis.  The pseudorapidity is defined in terms of the polar angle
$\theta$ as $\eta = -\ln \tan(\theta/2)$.  Angular distance is
measured in units of $\Delta R \equiv \sqrt{(\Delta\eta)^{2} +
(\Delta\phi)^{2}}$.}  It consists of an inner tracking detector
surrounded by a \SI{2}{\tesla} superconducting solenoid,
electromagnetic and hadronic calorimeters, and a muon spectrometer
incorporating three large superconducting toroid magnets.
 
The inner detector, including the insertable B-layer added as a new innermost layer in
2014~\cite{ATLAS-TDR-19,atlasibl}, provides charged-particle
tracking information from silicon pixel and microstrip detectors in
the pseudorapidity range $|\eta| < 2.5$ and a transition radiation
tracker covering $|\eta| < 2.0$.
 
The calorimeter system covers the pseudorapidity range $|\eta| < 4.9$
and measures the positions and energies of electrons, photons, and
charged and neutral hadrons.  Within the region $|\eta|< 3.2$,
electromagnetic calorimetry is provided by barrel and endcap
high-granularity lead and liquid-argon sampling calorimeters.  The
hadronic sampling calorimeter uses either scintillator tiles or liquid
argon as active material and steel, copper or tungsten as absorber.
 
The muon spectrometer comprises separate trigger and high-precision
tracking chambers measuring the tracks of muons in a magnetic field
generated by superconducting air-core toroid magnets.  The precision
chamber system covers the region $|\eta| < 2.7$, while the muon
trigger system covers the range $|\eta| < 2.4$.
 
A two-level trigger system is used to select which events to save for
offline analysis~\cite{TRIG-2016-01}.  The first level is
implemented in hardware/firmware and uses a subset of the detector
information to reduce the event rate from the \SI{40}{\MHz} proton bunch crossings to less
than \SI{100}{\kHz}.  This is followed by a software-based high-level
trigger that reduces the event rate to
approximately \SI{1}{\kHz}.
An extensive software suite~\cite{ATL-SOFT-PUB-2021-001} is used in the
reconstruction and analysis of real and simulated data, in detector operations,
and in the trigger and data acquisition systems of the experiment.


\section{Event samples}
\label{sec:samples}

The analysis was performed on $pp$\ collision data
collected at \sxut\ in November 2017 by the ATLAS detector, with an integrated
luminosity of \intlumi\,\ipb\ after data quality requirements~\cite{DAPR-2018-01}.
Events were required to pass either a single-electron
or a single-muon trigger, both of which reach the efficiency plateau region
in data for leptons
with an offline-reconstructed transverse momentum $\pt>15$\,\GeV~\cite{TRIG-2018-05,TRIG-2018-01}.
The trigger efficiencies were measured using a tag-and-probe method
applied to $Z\rightarrow\ell\ell$ decays (see Section~\ref{ssec:det}).
Most triggered events also included signals from additional
inelastic $pp$ collisions in the same bunch crossing, referred to as \pileup.
The mean number of inelastic $pp$ collisions per bunch crossing \mmu\
varied between about 0.5 and 4, with an average of $\mmu\approx 2$~\cite{ATLAS-CONF-2020-023}.
 
Dedicated \sxut\ Monte Carlo (MC) simulated event samples were used to
develop the analysis procedures, evaluate signal and background contributions,
and compare the predicted distributions with data.
All samples were processed using the full ATLAS
detector simulation~\cite{SOFT-2010-01} based on the \GEANT framework~\cite{geant4}.
The effects of \pileup were simulated by generating additional inelastic $pp$ collisions with
the \PYTHIA[8.186] event generator~\cite{pythia8} using the A3 set of parameter values~\cite{ATL-PHYS-PUB-2016-017}, a dedicated \PYTHIA\ tune that provides
a good description of inclusive $pp$ collisions. Several of these events were
overlaid on each primary simulated event, and the
combined events were then processed using the same reconstruction and analysis chain as the data. As discussed in Section~\ref{ssec:det},
small corrections derived from comparisons of data and simulation at both
\sxut\ and \sxyt\ were applied as scale factors to the
simulated lepton-trigger and reconstruction efficiencies,
to improve agreement with the response observed in data.
 
The \sxut\ simulation samples were created using event-generator configurations
similar to those developed for the analysis of \sxyt\ data~\cite{TOPQ-2018-17,TOPQ-2020-02}.
The nominal simulated \ttbar\ sample was produced using the NLO
event generator \POWHEGBOX[v2]~\cite{Frixione:2007nw,Nason:2004rx,Frixione:2007vw,Alioli:2010xd}\
with the \NNPDF[3.0nlo]\ PDF set~\cite{Ball:2014uwa} and employed
the \PYTHIA[8.210]\ model with the \NNPDF[2.3lo]\ PDF set and the A14 tune~\cite{ATL-PHYS-PUB-2014-021}
for the parton-shower, hadronisation, and underlying-event modelling.
In the \POWHEG{}
configuration, the cut-off scale for the first gluon emission (represented by the \hdamp\ parameter) was set to $\frac{3}{2} \mtop$ and the factorisation (\muf) and
renormalisation (\mur) scales were set to
$\muf=\mur=\sqrt{\mtop^2+p_{\mathrm{T},t}^2}$, where the top-quark transverse
momentum ($p_{\mathrm{T},t}$) is evaluated before radiation~\cite{ATL-PHYS-PUB-2016-020}.
 
Alternative \ttbar\ simulation samples were generated in order to assess
systematic uncertainties.
One sample used the \POWHEG{} MC generator with the \HERWIG[7.1.6]{}
parton-shower and hadronisation model~\cite{Bahr:2008pv,Bellm:2015jjp}\ employing the H7UE tune~\cite{Bellm:2015jjp}.
Another sample was generated using
the \MGNLO[2.3.3.p1]{} generator (referred to hereafter as
\AMCatNLO)~\cite{Alwall:2014hca}\ with the \NNPDF[3.0nlo]\ PDF set while using the \PYTHIA[8]\
parton-shower and hadronisation model.
Uncertainties in the amount of parton-shower radiation were evaluated by
reweighting the nominal \ttbar\ sample so as to effectively change QCD
scales and shower parameters, and by generating
an additional \POWPY{8} sample with $\hdamp=3\mtop$.
The top-quark mass was set to 172.5\,\GeV\ in all top-quark samples,
and the \EVTGEN{}\ program~\cite{evtgen} was used to handle the decays
of $b$- and $c$-flavoured hadrons.
All samples were normalised using
the NNLO+NNLL \ttbar\ cross-section prediction discussed in Section~\ref{sec:intro}.
 
The backgrounds in these analyses arise from single-top-quark production,
$W$ and $Z$ bosons produced in association with hadronic jets, and diboson
production.
The $t$-channel, $s$-channel, and $Wt$ associated production processes for
single top quarks were
simulated using the \POWHEG[v2]~\cite{Frederix:2012dh,wtinter1}\ generator
with the \NNPDF[3.0nlo]\ PDF set and employing \PYTHIA[8]\ with the A14 tune as
the parton-shower and hadronisation model.
The diagram removal scheme~\cite{wtdr}\ was used to handle the interference
between the \ttbar\ and $Wt$ final states. The $Wt$ sample
was normalised to a cross-section of $6.05\pm 0.57$\,pb, obtained by
extrapolating \cite{lhctopSingle} the
approximate NNLO calculation \cite{wttheo,wttheo2} using the
\MSTW{} NNLO PDF set \cite{mstwnnlo1,mstwnnlo2} to \sxut,
and taking into account PDF and QCD scale uncertainties.
The $t$- and $s$-channel samples were normalised to the cross-section
predictions from the \HATHOR~\cite{hathor} NLO MC generator, and the uncertainties
were conservatively taken to be 9.5\% for both processes, the same as for
$Wt$.
Two alternative $Wt$ samples were generated: one using the \POWPY{8} MC generator but with
the diagram subtraction scheme~\cite{wtinter1,wtinter2} and the other using the \POWHEG{} MC generator with
the \HERWIG[7.1.6]{} parton-shower
and hadronisation model employing the H7UE tune. A $t$-channel sample using the \POWHER{7.1.6} scheme similar to the
alternative $Wt$ sample was also generated.
 
The \Zjets\ and \Wjets\ events were simulated with
the \SHERPA[2.2.5] generator~\cite{sherpa22} using NLO matrix elements for up to
two partons, and LO matrix elements for up to four partons,
as discussed in Ref.~\cite{TOPQ-2020-02}.
The samples were generated using the \NNPDF[3.0nlo]{} PDF set and
normalised using an NNLO cross-section prediction~\cite{zxsecnnlo}.
Alternative \Zjets\ samples were generated using the
\POWPY{8} event generator.
Smaller backgrounds from diboson production ($WW$, $WZ$, and $ZZ$) with additional
jets were simulated using the \SHERPA[2.1.1]~\cite{sherpa21} generator with
the \CT[10] PDF set~\cite{cttenpdf}, as discussed in Ref.~\cite{TOPQ-2016-11}.


\section{Event reconstruction}
\label{sec:selection}

The dilepton and single-lepton measurements made use of reconstructed
electrons, muons and jets, as well as the momentum imbalance in the transverse
plane to infer the presence of neutrinos.
 
Electron candidates were reconstructed from a localised cluster
of energy deposits in the electromagnetic calorimeter matched to a track
in the inner detector and passing the `Medium' likelihood-based requirements
of Ref.~\cite{EGAM-2018-01}, giving a combined efficiency of about 80\%
for electrons from $Z\rightarrow ee$ decays.
They were further required to have transverse momentum
$\pt>18$\,\GeV\ and pseudorapidity $|\eta|<2.47$. Only the dilepton measurement included electrons
reconstructed in the transition region $1.37<|\eta|<1.52$ between the
barrel and endcap electromagnetic calorimeters,
to increase the reconstruction efficiencies for the final states with electrons.
Electrons were also required to satisfy requirements on the transverse impact-parameter significance
calculated relative to the beam line of $|d_0|/\sigma_{d_0}<5$ and
on the longitudinal impact parameter $\Delta z_0$\ calculated relative to the event primary
vertex of $|\Delta z_0\sin\theta|<0.5$\,mm, where $\theta$ is the polar
angle of the track. A vertex was defined as having at  least two associated tracks with $\pt>0.5$\,\GeV\ and the primary vertex of an event was the vertex with the
highest sum of $\pt^2$ of the associated tracks. To reduce
background from misidentified and non-prompt electrons, the electron candidates were required to be isolated using requirements on the summed calorimeter energy within a cone of
size $\Delta R=0.2$ around the electron cluster and on the sum of track
$\pt$ within a cone of variable size
$\Delta R=\mathrm{min}(0.2,10\,\GeV/\pt(e))$
around the electron track direction,\footnote{{\em i.e.}\ a cone whose
size $\Delta R$ shrinks at high electron \pt, {\em e.g.} to $\Delta R=0.1$
for $\pt=100$\,\GeV, but whose size is limited to $\Delta R=0.2$ for
$\pt<50$\,\GeV.}
both divided by the electron \pt.
 
Muon candidates were reconstructed by combining matching tracks reconstructed
in the inner detector and the muon spectrometer, and were required
to satisfy the `Medium' requirements of Ref.~\cite{MUON-2018-03},
giving an efficiency of about 97\% for muons from $Z\rightarrow\mu\mu$ decays.
The muon
candidates were also required to have $\pt>18$\,\GeV\ and $|\eta|<2.5$,
and to be isolated using the same calorimeter isolation variable as the
electrons and a track-based isolation requirement based on a cone of size
$\Delta R=\mathrm{min}(0.3,10\,\GeV/\pt(\mu))$.
Requirements on the muon-candidate track's impact parameters similar to those
of the electron candidates were made for the single-lepton measurement only.
 
Jets were reconstructed using the \antikt\ algorithm~\cite{antikt,Fastjet}
with radius parameter $R=0.4$. Particle-flow objects that combine
information from topological clusters of calorimeter energy deposits and
inner-detector tracks~\cite{PERF-2015-09} were used during reconstruction,
and were calibrated according to the standard calibration used
for \sxyt\ high-pileup data \cite{JETM-2018-05}.
An additional calibration specific to the \sxut\ $pp$ data sample was performed
to account for the modified calorimeter thresholds used for this
low-pileup data sample.
This calibration used a sample of $\Zboson$\,+\,jet events to determine a
correction to the \pT\ scale for the jets as a function of jet \pT\ and $\eta$.
The \pT\ of the well-measured $\Zboson\rightarrow \ell^+\ell^-$\
system was compared with the \pT\ of the jet recoiling opposite the \Zboson~boson in azimuth.
The correction to the jet energy scale was typically 2\%--8\%.
Jets were required to have
$|\eta|<2.5$, and $\pt>25$\,\GeV\ for the dilepton selection and $\pt>20$\,\GeV\ for the single-lepton selection,
corresponding to the optimal requirements for each analysis.
Jets with $\pt<60$\,\GeV\ and $|\eta|<2.4$ were subject to additional \pileup rejection criteria using a multivariate jet-vertex tagger~\cite{PERF-2014-03}.
This algorithm makes use of the tracks associated to the jet and their
matching to each of the reconstructed $pp$ collision vertices.
 
To prevent the double-counting
of electron energy deposits as jets, the closest jet to an electron candidate
was removed if it was within $\Delta R_y=0.2$
of the electron, where $\Delta R_y=\sqrt{(\Delta y)^2+(\Delta\phi)^2}$
and $\Delta y$ is the difference in rapidity between the jet and the electron.
Furthermore, to reduce the contribution of leptons
from heavy-flavour hadron decays inside jets, leptons within $\Delta R_y=0.4$
of selected jets were discarded, unless the lepton was a muon and the jet
had fewer than three associated tracks, in which case the jet was discarded
(thus avoiding an efficiency loss for high-energy muons undergoing significant
energy loss in the calorimeters that gives rise to a jet candidate).
 
Jets likely to contain $b$-hadrons were tagged using the DL1r
algorithm~\cite{FTAG-2018-01}, a multivariate discriminant based on deep-learning
techniques making use of track impact parameters and reconstructed secondary
vertices.  A tagger working point with 85\% efficiency for tagging $b$-quark
jets from top-quark decays in simulated \ttbar\ events was used for the dilepton-event selection,
corresponding to rejection factors of about 3 against
charm jets and 40 against light-quark and gluon jets.
A tighter working point with 70\% efficiency was used
in the single-lepton channel with rejection factors of 6 against charm jets
and 200 against light-quark and gluon jets.
 
The missing transverse momentum was
reconstructed as the negative vector sum of the transverse momenta
of all identified physics objects (electrons, muons, and jets),
together with a `soft term' built from all tracks associated with the
reconstructed primary vertex but not with any of the identified physics
objects~\cite{PERF-2016-07}.
The magnitude of the missing transverse momentum is denoted by \etmiss.


\section{Dilepton cross-section measurement}
\label{sec:dilepton}

The measurement in the dilepton channel is based on counting the numbers of
events with an opposite-charge dilepton pair and one or two $b$-tagged
jets, and fitting these event counts to predictions based on the assumed
\ttbar\ cross-section, leptonic selection efficiencies and estimated
non-\ttbar\ backgrounds. In the same-flavour channels ($ee$ and $\mu\mu$),
the dilepton invariant mass and \etmiss\ are also exploited to separate
signal events from the dominant \Zjets\ background.
This method allows the efficiency for reconstructing and $b$-tagging
a $b$-jet from the top quark decay to be determined from the data, and
also minimises uncertainties due to the modelling of additional jets
from QCD radiation in the \ttbar\ signal events.
 
Selected events in the dilepton channel were required to have exactly two
opposite-charge lepton candidates
with at least one lepton matched to a corresponding electron or muon trigger.
Events with no reconstructed primary vertex, with an electron and muon
separated in angle by
$|\Delta\theta|<0.15$ and $|\Delta\phi|<0.15$, or where at least one jet
with $\pt>20$\,\GeV\ failed quality requirements~\cite{ATLAS-CONF-2015-029},
were rejected. In the $e\mu$ channel, 
\mll\ was required to satisfy $\mll>15$\,\GeV. In the same-flavour channels
a tighter requirement of $\mll>40$\,\GeV\ was imposed
in order to match the phase space of the \Zjets\ simulation sample, and
the \Zjets\ background was reduced by requiring $\etmiss>30$\,\GeV.
In all channels, selected events were further required to have exactly one or
exactly two $b$-tagged jets, irrespective of the number of untagged jets present.
 
In the $e\mu$ channel, the inclusive \ttbar\ cross-section was determined
from the number of opposite-charge events with one ($N_1$) or two ($N_2$)
$b$-tagged jets using the formalism of Refs.~\cite{TOPQ-2013-04,TOPQ-2018-17}.
The two event counts satisfy the tagging equations
\begin{equation}
\begin{array}{lllll}
N_1^{e\mu} & = & L \xtt\ \epsem 2\epsbem (1-\cbem\epsbem) & + & \displaystyle\sum\limits_{k=\mathrm{bkg}}
s_1^k\, N_1^{e\mu,k}\ \textrm{and}  \\
N_2^{e\mu} & = & L \xtt\ \epsem \cbem(\epsbem)^2 & + & \displaystyle\sum\limits_{k=\mathrm{bkg}} s_2^k\, N_2^{e\mu,k},
\end{array}\label{e:tags}
\end{equation}
where $L$ is the integrated luminosity of the sample, \epsem\ is the efficiency
for a \ttbar\ event to pass the opposite-charge $e\mu$ selection (including
the $W\rightarrow\ell\nu$ branching ratios), and \cbem\ is a tagging correlation
coefficient close to unity. The combined probability for a jet from the quark
$q$ in the $t\rightarrow Wq$
decay to fall within the acceptance of the detector, be reconstructed as a jet
with transverse momentum above the selection threshold, and be tagged as a
$b$-jet is denoted by \epsbem. The correlation coefficient \cbem\ is defined by
$\epsbb/(\epsbem)^2$, where \epsbb\ is the probability to reconstruct
and tag both $b$-jets from the top-quark decays. In practice, it was evaluated
from simulation as $\cbem=4\nttem N^{\ttbar}_2/(N^{\ttbar}_1+2N^{\ttbar}_2)^2$,
where \nttem\ is the number of selected $e\mu$ \ttbar\ events and $N^{\ttbar}_1$
and $N^{\ttbar}_2$ are the numbers of such events with one and two $b$-tagged
jets. Evaluated in this way, it also
accounts for extra $b$-tagged jets from mistagged light jets, and extra
heavy-flavour jets in the \ttbar\ event.
In the baseline \ttbar\ simulation, $\epsem\approx 1.35$\%, compared to
the $\ttbar\rightarrow e\mu X$ branching ratio of 3.2\% including the
$W\rightarrow\tau\rightarrow e/\mu$ contributions,
and $\cbem\approx 1.011$, indicating a small positive correlation between the
reconstruction and
$b$-tagging of the two jets produced in the top-quark decays. Background
from sources other than \ttbar\ events with two prompt leptons also
contributes to the event counts $N_1$ and $N_2$ in Eqs.~(\ref{e:tags}). It
was divided into four sources indexed by $k$: $Wt$ events, \Zjets events,
diboson events, and events with at least one misidentified lepton.
The estimate of each background was scaled by a factor $s_1^k$ or $s_2^k$
for events with one or two $b$-tagged jets.
 
This formalism was extended to the same-flavour channels by also exploiting
the invariant mass \mll\ of the lepton pair to provide discrimination against
the dominant \Zjets\ background. The events were divided into six bins of
\mll\ indexed by subscript $m$, with lower bin boundaries at 40, 71, 81, 101,
111 and 151\,\GeV, the last bin including all events with $\mll>151\,\GeV$. The
numbers of opposite-charge $\ell\ell$ events in each bin $m$ with one
and two $b$-tagged jets, $N_{1,m}^{\ell\ell}$ and $N_{2,m}^{\ell\ell}$
can then be expressed as
\begin{equation}
\begin{array}{lllll}
N_{1,m}^{\ell\ell} & = & L \xtt\,\epsll\, 2\epsbll (1-\cbll\epsbll)\,f_{1,m}^{\ell\ell,t\bar{t}} & + &
\displaystyle\sum\limits_{k=\mathrm{bkg}} s_1^k\, f_{1,m}^{\ell\ell,k} N_{1}^{\ell\ell,k}\ \textrm{and} \\
N_{2,m}^{\ell\ell} & = & L \xtt\,\epsll\,\cbll(\epsbll)^2\,f_{2,m}^{\ell\ell,t\bar{t}} & + &
\displaystyle\sum\limits_{k=\mathrm{bkg}} s_2^k\, f_{2,m}^{\ell\ell,k} N_{2}^{\ell\ell,k},
\end{array}\label{e:tagsf}
\end{equation}
with separate selection efficiencies \epsll\ and correlation coefficients
\cbll\ for each same-flavour channel ($\ell\ell=ee$ or $\mu\mu$).
The coefficients $f_{1,m}^{\ell\ell,k}$ and $f_{2,m}^{\ell\ell,k}$ represent the
integrals of the \mll\ distributions, giving the fractions of events for each dilepton flavour,
background source and $b$-tagged jet multiplicity that appear in each mass bin.
The estimated total numbers of background events for each source $k$ were
scaled by $s_1^k$ and $s_2^k$, whose values were taken to be common to all
three dilepton channels.
In the baseline
\ttbar\ simulation, $\epsee\approx 0.48$\% and $\epsmm\approx 0.55$\%, compared
to the $\ttbar\rightarrow \ell\ell$ branching ratio of 1.6\% for a single
lepton flavour $\ell$ including $W\rightarrow\tau\rightarrow\ell$,
and the \cbll\ values  are compatible with that of the $e\mu$ channel.
 
The \ttbar\ cross-section was determined by comparing the observed event
counts $N_1$ and $N_2$ for the $e\mu$ channel, and the observed counts in
each dilepton mass bin $N_{1,m}^{\ell\ell}$ and $N_{2,m}^{\ell\ell}$ for each
of the same-flavour channels, with the predictions from Eqs.~(\ref{e:tags})
and~(\ref{e:tagsf}), maximising a likelihood consisting of the product of
Poisson probabilities for each event count. The values of \xtt, \epsbllp\ for
each dilepton flavour (with $\ell\ell^{\prime}=ee$, $e\mu$, or $\mu\mu$),
and $s_1^k$ and $s_2^k$ for the \Zjets background
(referred to as \rza\ and \rzb\ below) were left free in the fit. This choice
allows the level of $Z$\,+\,$b$-tagged-jet background, which has large
uncertainties when predicted from simulation, to be determined from data.
The values
of $s_1^k$ and $s_2^k$ for all other background sources were fixed to unity,
as these backgrounds are either small, or in the case of $Wt$, well predicted
by simulation.
All other parameters, i.e.\ \epsllp\ and \cbllp, and $N_i^{e\mu,k}$
and $N_i^{\ell\ell,k}$, $f_{i,m}^{\ell\ell,\ttbar}$ and $f_{i,m}^{\ell\ell,k}$
(with $\ell\ell=ee$ or $\mu\mu$), were determined from
simulation, with systematic uncertainties taken into account as discussed in
Section~\ref{sec:systematics}.
 
This maximum-likelihood fit determines the product of $b$-tagging efficiencies
and jet acceptance (\epsbllp) for all three dilepton channels largely from data,
minimising the dependence on the modelling of jets in the same way as
in the pure $e\mu$-based
analyses of Refs.~\cite{TOPQ-2013-04,TOPQ-2018-17}. The normalisation of the
\Zjets\ background in all channels was determined from the fits to the \mll\
distribution in the same-flavour channels, minimising the dependence on
the modelling of heavy-flavour jets produced in association with a $Z$ boson.
The backgrounds from $Wt$ and diboson events were determined from simulation.
Simulation was also used to predict the small background from events with
misidentified leptons, which comprises up to 2\% of the one $b$-tagged jet
samples, but is  significantly smaller in the two $b$-tagged jet samples.
This background is dominated by events where one lepton candidate is a real
prompt lepton from a $W$ or $Z$ decay, and the other is an electron from a
photon conversion, a real lepton produced in the decay of a
heavy-flavour hadron, or a misidentified hadron. The
analysis procedure was validated using simulation-based pseudo-experiments with
various input \ttbar\ cross-section values. These tests verified that the
fit is unbiased and gives correct uncertainty estimates.
 
\begin{figure}[tp]
\vspace{-7mm}
\parbox{83mm}{\includegraphics[width=76mm]{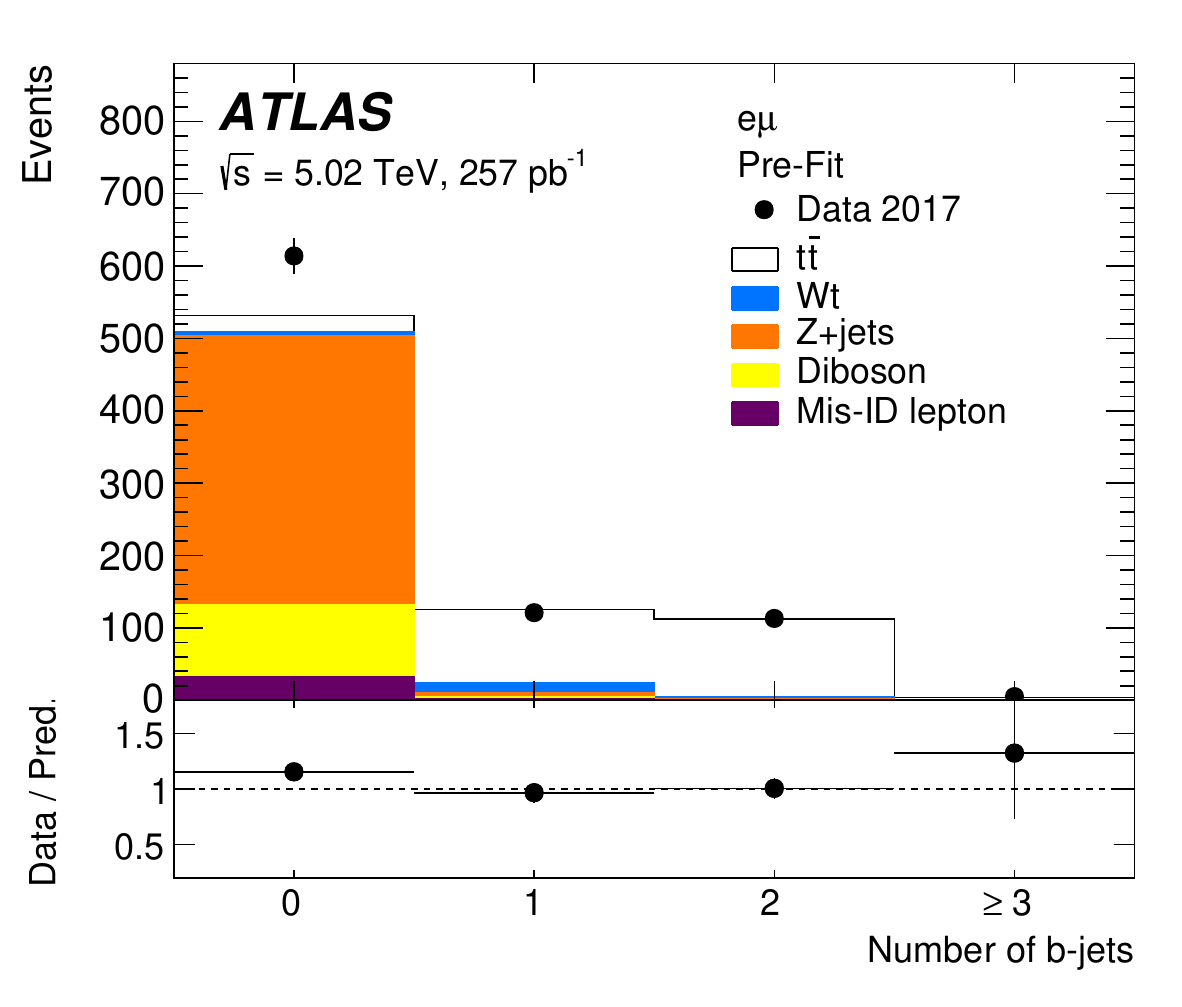}\vspace{-6mm}\center{(a)}}
\parbox{83mm}{\includegraphics[width=76mm]{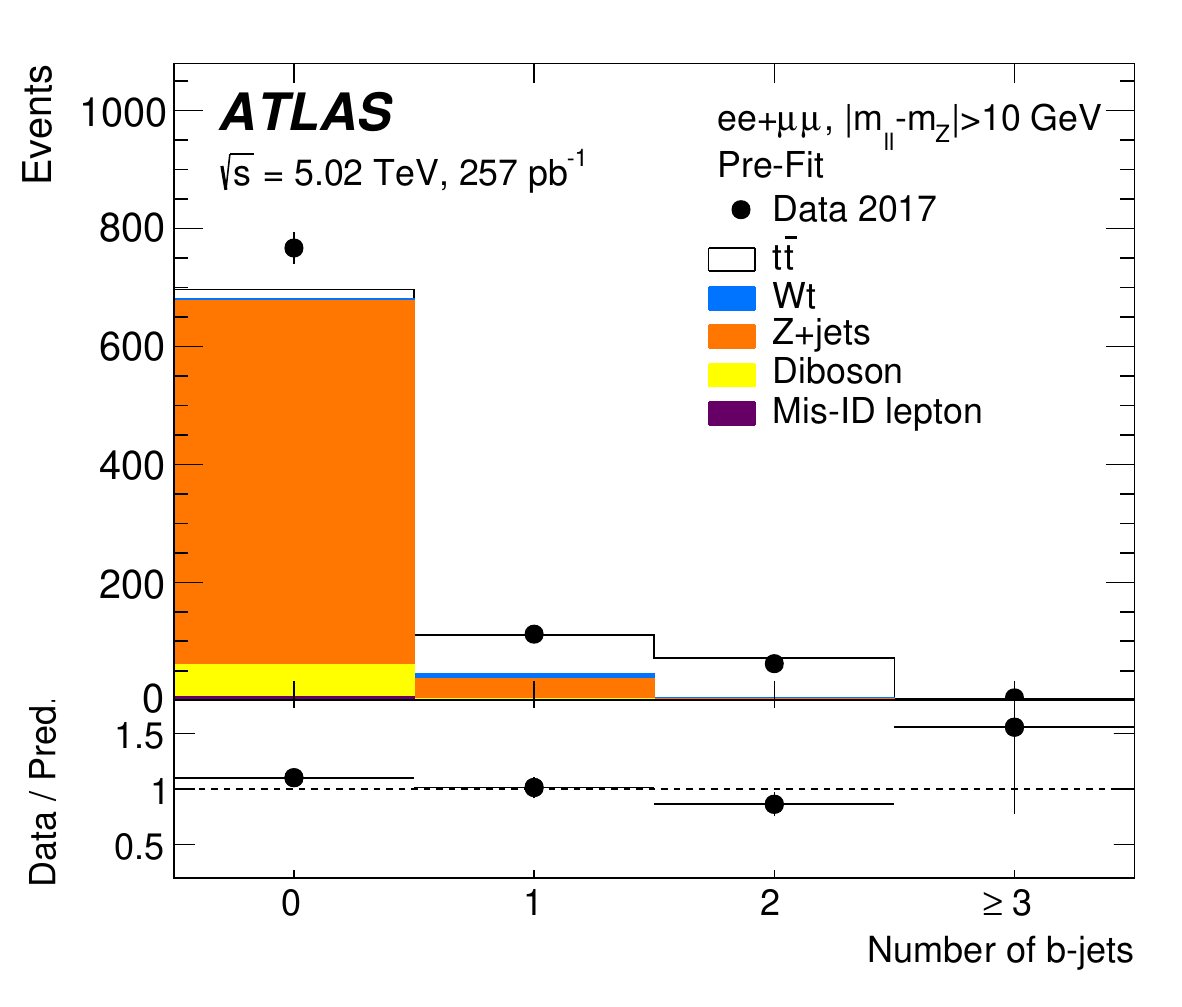}\vspace{-6mm}\center{(b)}}
\caption{\label{f:dmcjet}Distributions of the number of $b$-tagged jets in
selected opposite-charge events in (a) the $e\mu$ channel and (b) the
combined $ee$ and $\mu\mu$ same-flavour channels, additionally requiring
$|\mll-m_Z|>10$\,\GeV. The data are compared with the prediction from simulation
normalised to the same integrated luminosity as the data. The expected
contributions of \ttbar, $Wt$, \Zjets, dibosons and events with misidentified
electrons or muons are shown separately. The lower panels of the figure show
the ratios of data to simulation, with the error bars indicating the
statistical uncertainty.}
\end{figure}

\begin{figure}[!tp]
\vspace{-7mm}
\parbox{84mm}{\includegraphics[width=80mm]{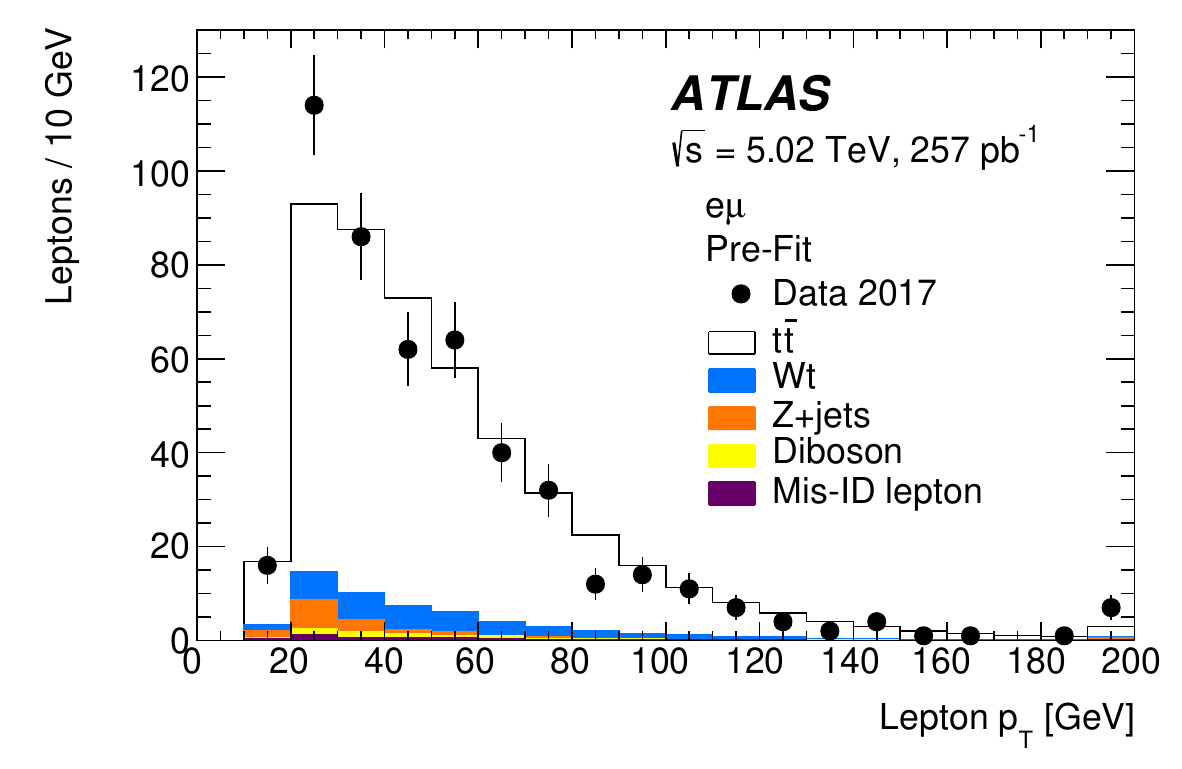}\vspace{-6mm}\center{(a)}}
\parbox{84mm}{\includegraphics[width=80mm]{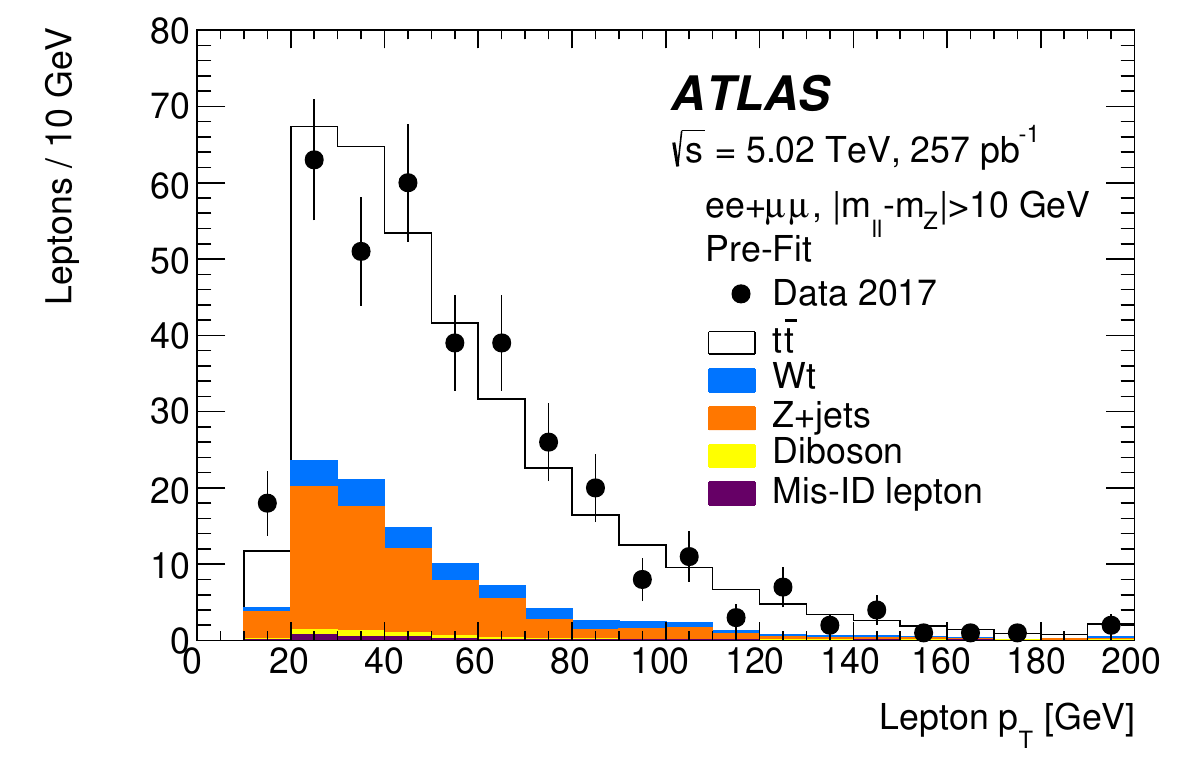}\vspace{-6mm}\center{(b)}}
\parbox{84mm}{\includegraphics[width=80mm]{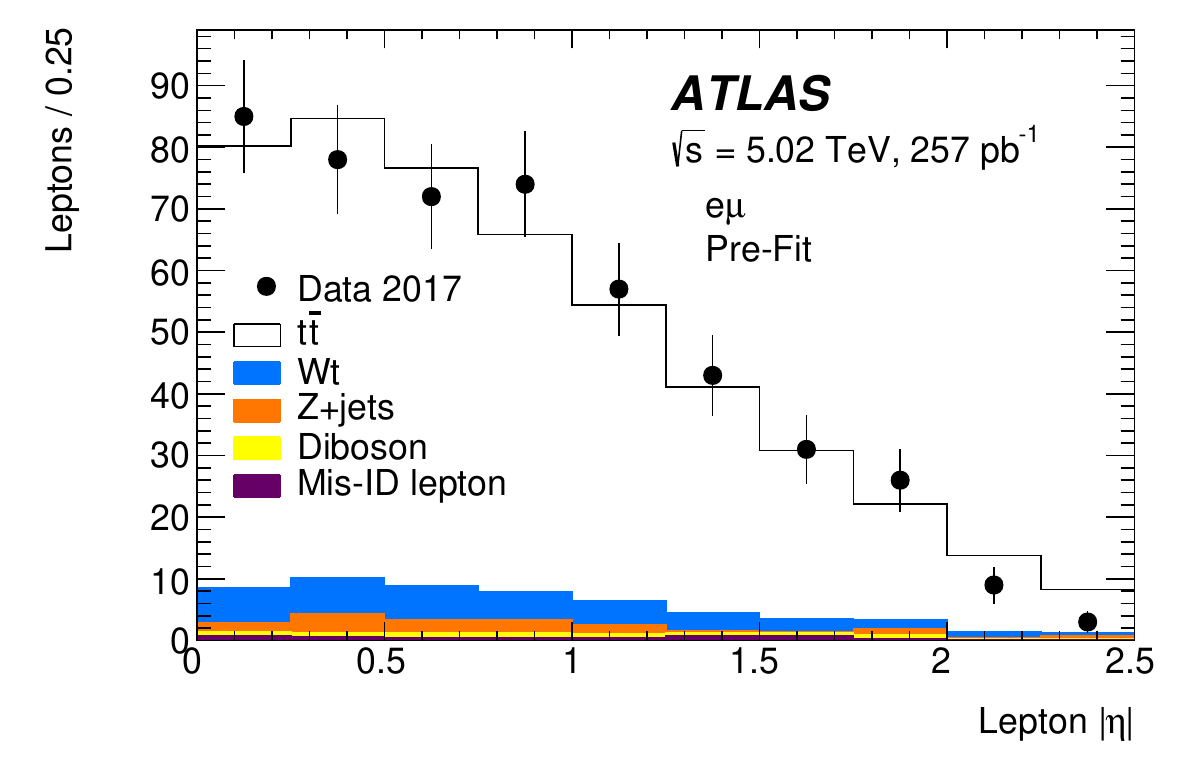}\vspace{-6mm}\center{(c)}}
\parbox{84mm}{\includegraphics[width=80mm]{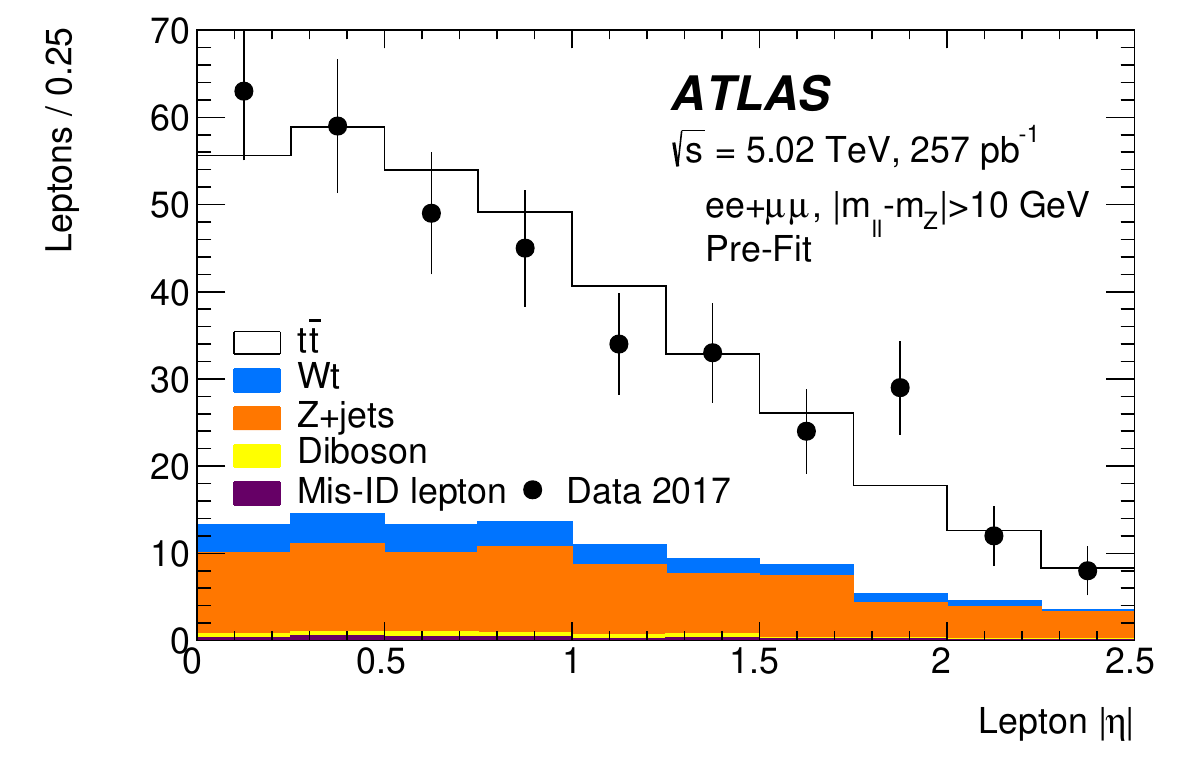}\vspace{-6mm}\center{(d)}}
\parbox{84mm}{\includegraphics[width=80mm]{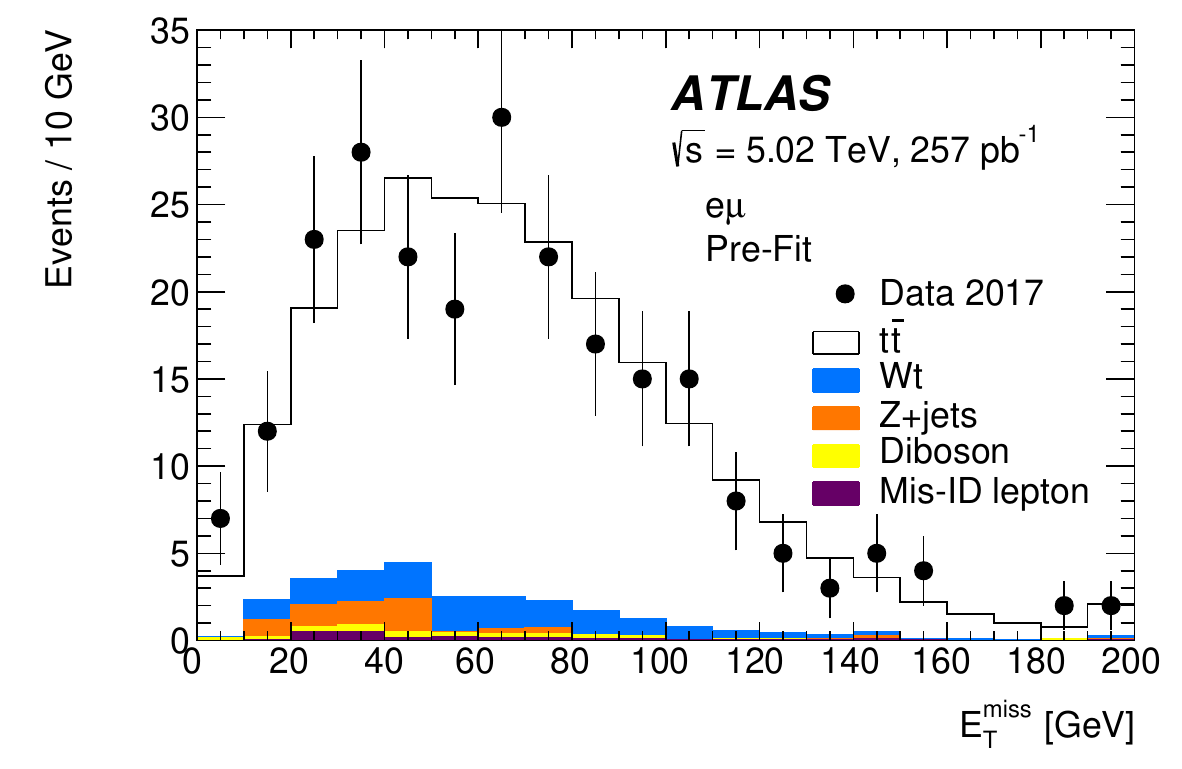}\vspace{-6mm}\center{(e)}}
\parbox{84mm}{\includegraphics[width=80mm]{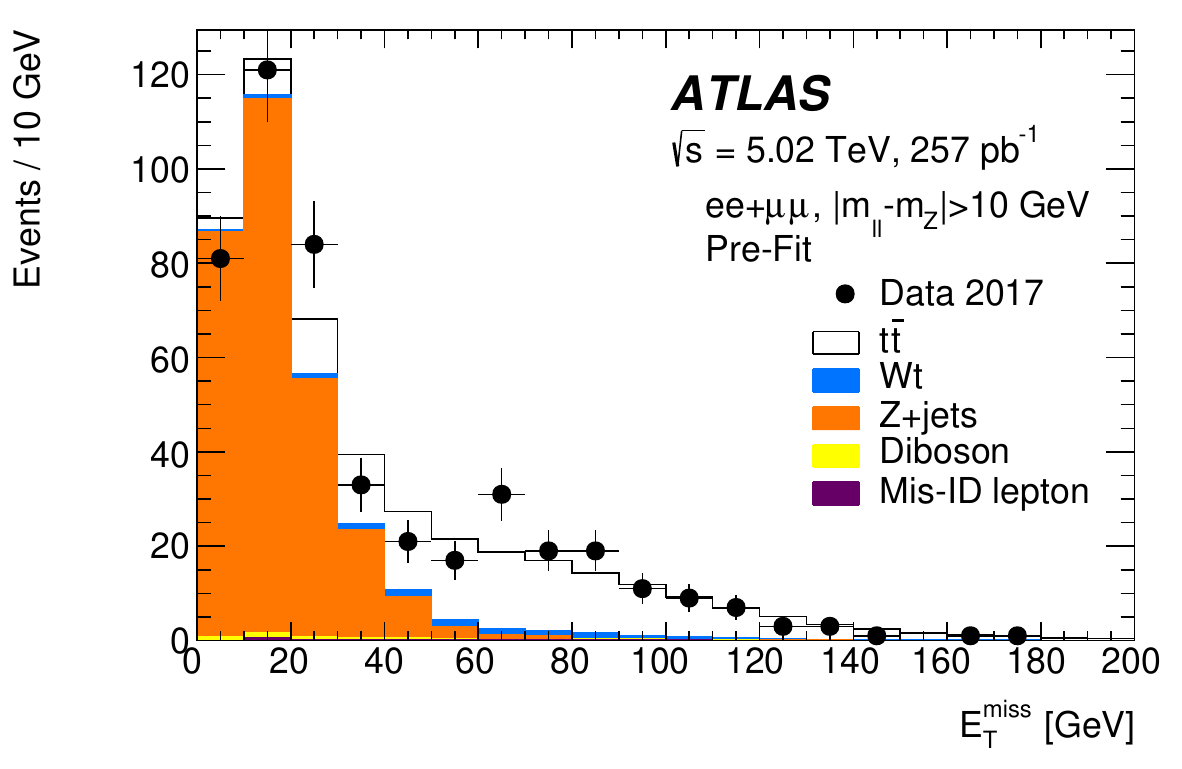}\vspace{-6mm}\center{(f)}}
\caption{\label{f:dmclept}Distributions of
(a, b) the lepton transverse momentum, (c, d) the absolute value of the lepton
pseudorapidity, and (e, f) the missing transverse momentum
for opposite-charge dilepton events with at least one $b$-tagged jet.
The left column (a, c, e) shows selected $e\mu$ events
and the right column (b, d, f) shows selected same-flavour events with
an additional requirement of $|\mll-m_Z|>10$\,\GeV.
The same-flavour requirement of $\etmiss>30$\,\GeV\ is not applied in
(f). The data are compared with the prediction from simulation normalised
to the same number of selected events as the data, and the expected contributions
from \ttbar, $Wt$, \Zjets, dibosons, and events with misidentified electrons
or muons are shown separately. The last bin includes the overflows in the
\pt\ and \etmiss\ distributions.}
\end{figure}

Figure~\ref{f:dmcjet} shows the $b$-tagged-jet multiplicity in the $e\mu$
channel, and in the same-flavour channels away from the peak of the $Z$
resonance ($|\mll-m_Z|>10$\,\GeV, referred to as off-$Z$),
comparing the data with the simulation prediction.
In the $e\mu$ channel, the one and two $b$-tagged-jet samples are expected
from simulation to be about 80\% and 96\% pure in \ttbar\ events,
assuming a \ttbar\ cross-section of 68.2\,pb, and the largest background
comes from $Wt$ events.
The backgrounds from \Zjets\ are much larger in the same-flavour samples even
off the $Z$ resonance, where the average \ttbar\
purities are about 60\% and 94\% for one and two $b$-tagged jet events.
Near the peak of the $Z$ resonance, the \Zjets\ contributions are even larger.
The \ttbar\ simulation describes the data well, except for a data excess in the
zero $b$-tagged-jet bins of both the $e\mu$ and same-flavour channels.
Qualitatively similar discrepancies have also been seen in the $e\mu$ channel at
other
centre-of-mass energies \cite{TOPQ-2013-04,TOPQ-2018-17}, and may indicate
mismodelling of the contributions of \Zjets\ or diboson events without
accompanying heavy-flavour jets.
However, zero $b$-tagged-jet events are not used in the fit, and this
discrepancy has no effect on the analysis. Figure~\ref{f:dmclept} shows
the lepton transverse momentum, pseudorapidity, and
\etmiss\ distributions for $e\mu$ and off-$Z$ same-flavour events. The total
prediction is normalised to the same number of selected events
as in the data, to focus on shape comparisons.
In general, the description of the data by the simulation is good, given the
limited size of the data sample.
 
\section{Single-lepton cross-section measurement}
\label{sec:leptonjets}

The single-lepton final state arising from \ttbar\ decay is characterised by
a charged lepton, a neutrino, and high jet multiplicity with several $b$-tagged
jets. The primary backgrounds in this final state arise from
\Wjets production, \Zjets production, $Wt$ associated single-top-quark production, misidentified and non-prompt leptons, and diboson
production.
 
\begin{figure}[tp]
\begin{center}
\subfloat[]{\includegraphics[width=0.45\textwidth]{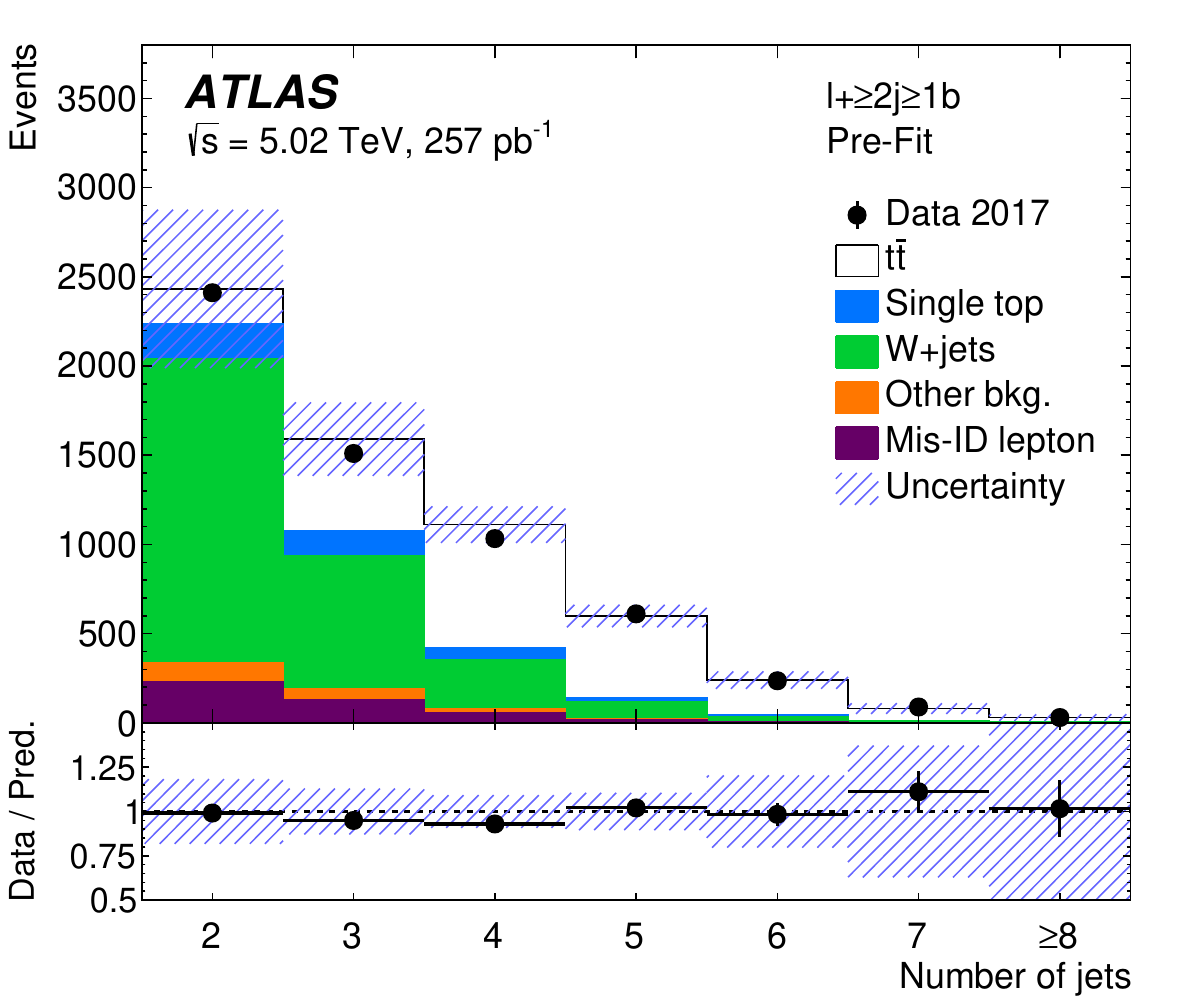} }\quad
\subfloat[]{\includegraphics[width=0.45\textwidth]{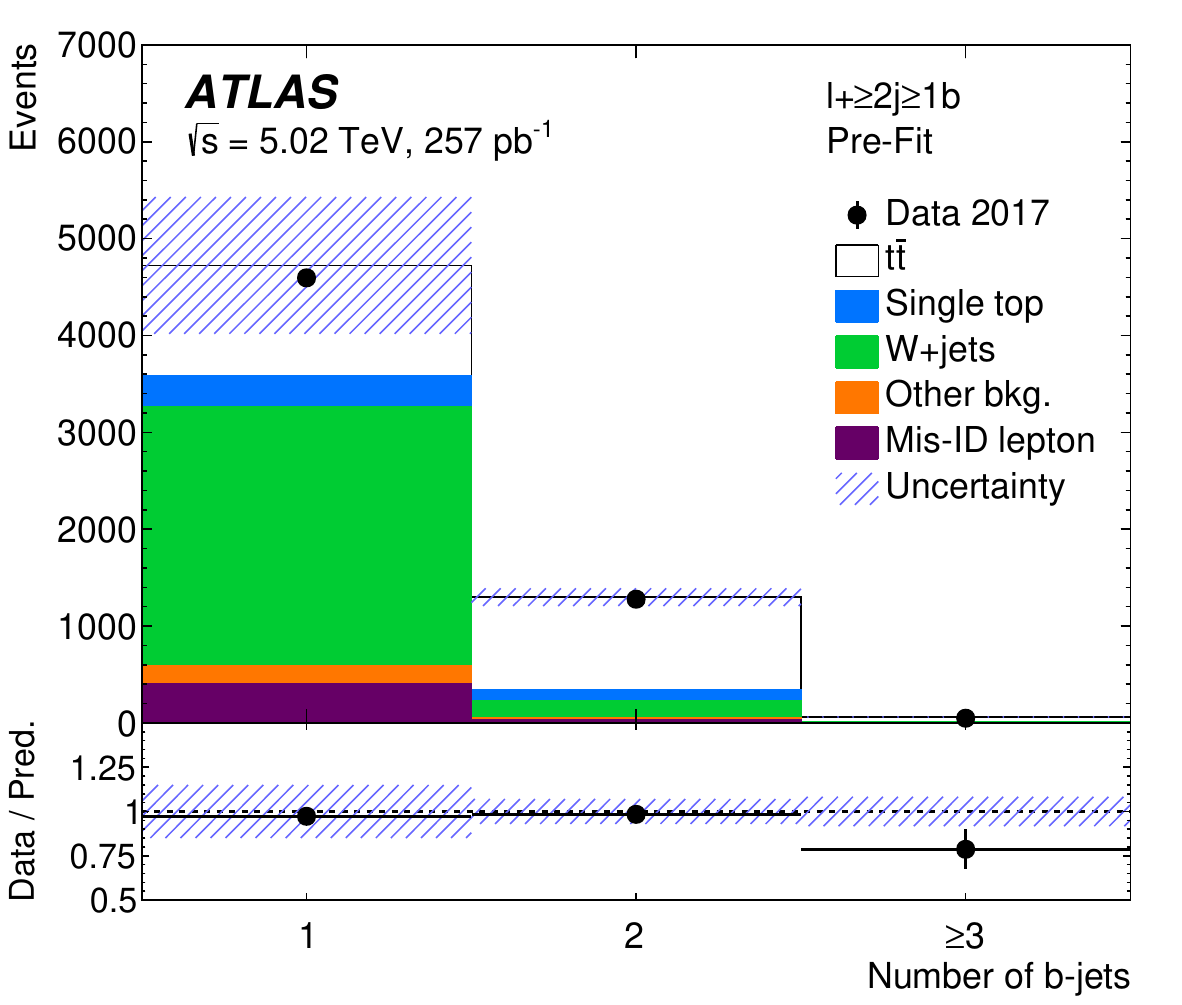} }
\caption{(a) Jet multiplicity and (b) $b$-tagged jet multiplicity distributions
in the single-lepton final state.
The uncertainties are prior to the fit (`pre-fit') and the last bin contains the overflow.
The `Other bkg.' includes \Zjets and diboson backgrounds.
The blue-hashed lines correspond to the total uncertainty of the prediction in a given bin.
The lower panels show the ratio of the data to the prediction, along with the uncertainty in the ratio.}
\label{fig:njets_prefit}
\end{center}
\end{figure}

Selected events in the single-lepton channel of the \ttbar\ decay were required to have exactly one electron or muon candidate with $\pt>25$\,\GeV, $\met>30$\,\GeV, and two or more jets with at least one of the jets being $b$-tagged.
The minimum lepton \pt\ requirement is higher than in the dilepton analysis because
of the larger background from misidentified leptons.
Cuts on the transverse mass\footnote{$\mtw=\sqrt{2p^{\ell}_{\mathrm T}\met\,(1-\cos\phi)}$, where $p^{\ell}_\mathrm{T}$ is the transverse momentum of the charged lepton and $\phi$
is the opening azimuthal angle between the charged lepton and the missing
transverse momentum.} of the \Wboson boson, \mtw,
and $\met$ were applied depending on the number of jets in an
event in order to reduce the non-prompt lepton background.
Events with two, three, or four jets were required to have  $\mtw > 30$\,\GeV, as were events with five or more jets where one
of the jets was $b$-tagged.
Events with five or more jets and at least two of these being $b$-tagged were required to pass the looser requirement
$\met + \mtw > 60$\,\GeV.
Figure~\ref{fig:njets_prefit}\ shows the jet multiplicity and $b$-tagged jet
multiplicity distributions of the events passing this selection.
The predicted backgrounds and the expected \ttbar\ contribution assuming a cross-section of 68.2\,pb are
also shown. The total prediction is found to be in excellent agreement with data.
 
The \Wjets, \Zjets, single-top-quark, and diboson
backgrounds were modelled using the MC samples described in Section~\ref{sec:samples}.
The backgrounds arising from misidentified leptons were determined using a `matrix method' technique~\cite{ATLAS-CONF-2014-058}.
Events were selected using
looser isolation or identification requirements for the lepton and were then
weighted according to the efficiencies for both the prompt and background
(misidentified and non-prompt) leptons to pass the tighter baseline selection.
In order to validate the method, the predictions were compared with data in two dedicated validation regions with a larger
fraction of misidentified-lepton candidates than expected in the analysis regions.
Good agreement between data and the prediction in these validation regions was found.
 
Events passing the selection requirements were further split into six orthogonal regions
based on the total number of jets (whether $b$-tagged or not) and the
number of $b$-tagged jets:
$\ell+2$~jets and ${\ge}1$~$b$-tagged jet,
$\ell+3$~jets and 1~$b$-tagged jet,
$\ell+3$~jets and 2~$b$-tagged jets,
$\ell+ {\ge}4$~jets and 1~$b$-tagged jet,
$\ell+4$~jets and 2~$b$-tagged jets, and
$\ell+{\ge}5$~jets and 2~$b$-tagged jets.
This separation created subsamples with different levels of signal and background, which provided
additional constraints on the estimated backgrounds.
The observed and predicted event yields in the six regions are given in Table~\ref{tab:PrefitYield}.
The inclusion of the $\ell+3$-jet regions increased the statistical power of the measurement, while the $\ell+2$-jet
region was used to improve the background modelling.
In simulation, the total \ttbar\ event selection efficiency integrated across
all six regions is ${\sim}12$\%.

\begin{table}[tp]
\caption{Estimated event yields in the six regions after passing the selection requirements.
The `Other bkg.' category contains the \Zjets and diboson contributions.
The uncertainties in the signal and background yields combine the
statistical and systematic uncertainties.
}
\label{tab:PrefitYield}
\centering
 
\begin{tabular}{l
r@{$\,\pm\,$}l
r@{$\,\pm\,$}l
r@{$\,\pm\,$}l
r@{$\,\pm\,$}l
r@{$\,\pm\,$}l
r@{$\,\pm\,$}l
}
\toprule
& \multicolumn{2}{c}{\CRzero} & \multicolumn{2}{c}{\CRone}  & \multicolumn{2}{c}{\CRtwo}  & \multicolumn{2}{c}{\SRone}    & \multicolumn{2}{c}{\SRtwo} & \multicolumn{2}{c}{\SRthree}    \\
\midrule
$\ttbar{}$     & 194 & 27    & 310 & 33  & 199 & 24 & 690 & 60 & 318 & 32 & 380 & 60 \\
Single top     & 195 & 22    & 98 & 12   & 38 & 5 & 67 & 9 & 22 & 4 & 15.9 & 2.7 \\
\Wjets         & 1700 & 400  & 690 & 210 & 58 & 23 & 350 & 120 & 30 & 14 & 19 & 10 \\
Other bkg.     & 110 & 40    & 55 & 23   & 7.2 & 3.0 & 29 & 12 & 3.5 & 1.5 & 3.7 & 1.7 \\
Misidentified leptons & 250 & 130   & 110 & 60  & 10 & 5 & 60 & 30 & 6 & 3 & 8 & 5 \\
\midrule
Total          & 2500 & 400 & 1260 & 210 & 312 & 34 & 1200 & 160 & 380 & 40 & 430 & 70 \\
\midrule
Data           & \multicolumn{2}{c}{2411}  & \multicolumn{2}{c}{1214} & \multicolumn{2}{c}{293} & \multicolumn{2}{c}{1135} & \multicolumn{2}{c}{375} & \multicolumn{2}{c}{444} \\
 
\bottomrule
\end{tabular}
\end{table}
 
\begin{figure}[tp]
\begin{center}
\subfloat[]{\includegraphics[width=0.45\textwidth]{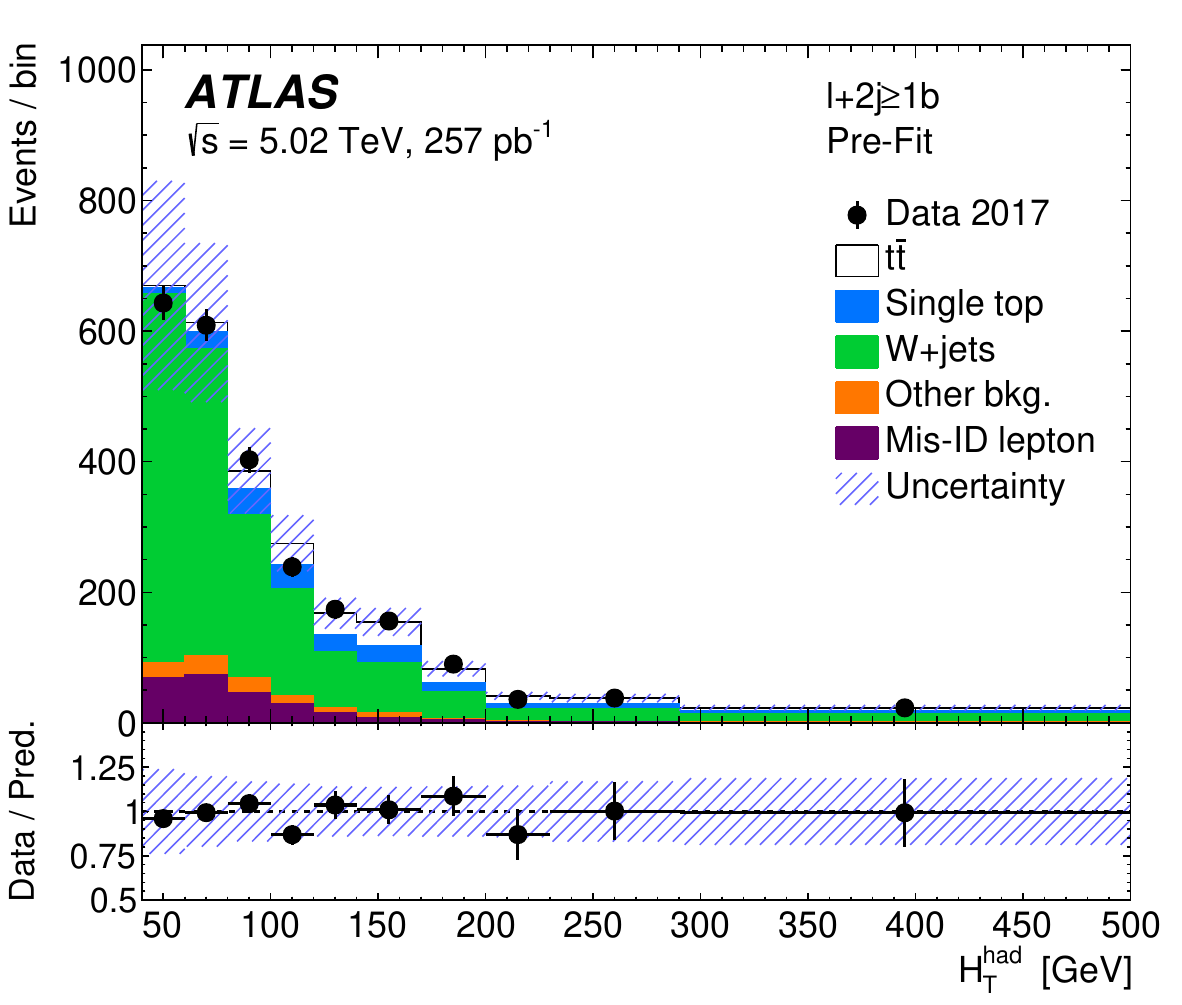}}
\subfloat[]{\includegraphics[width=0.45\textwidth]{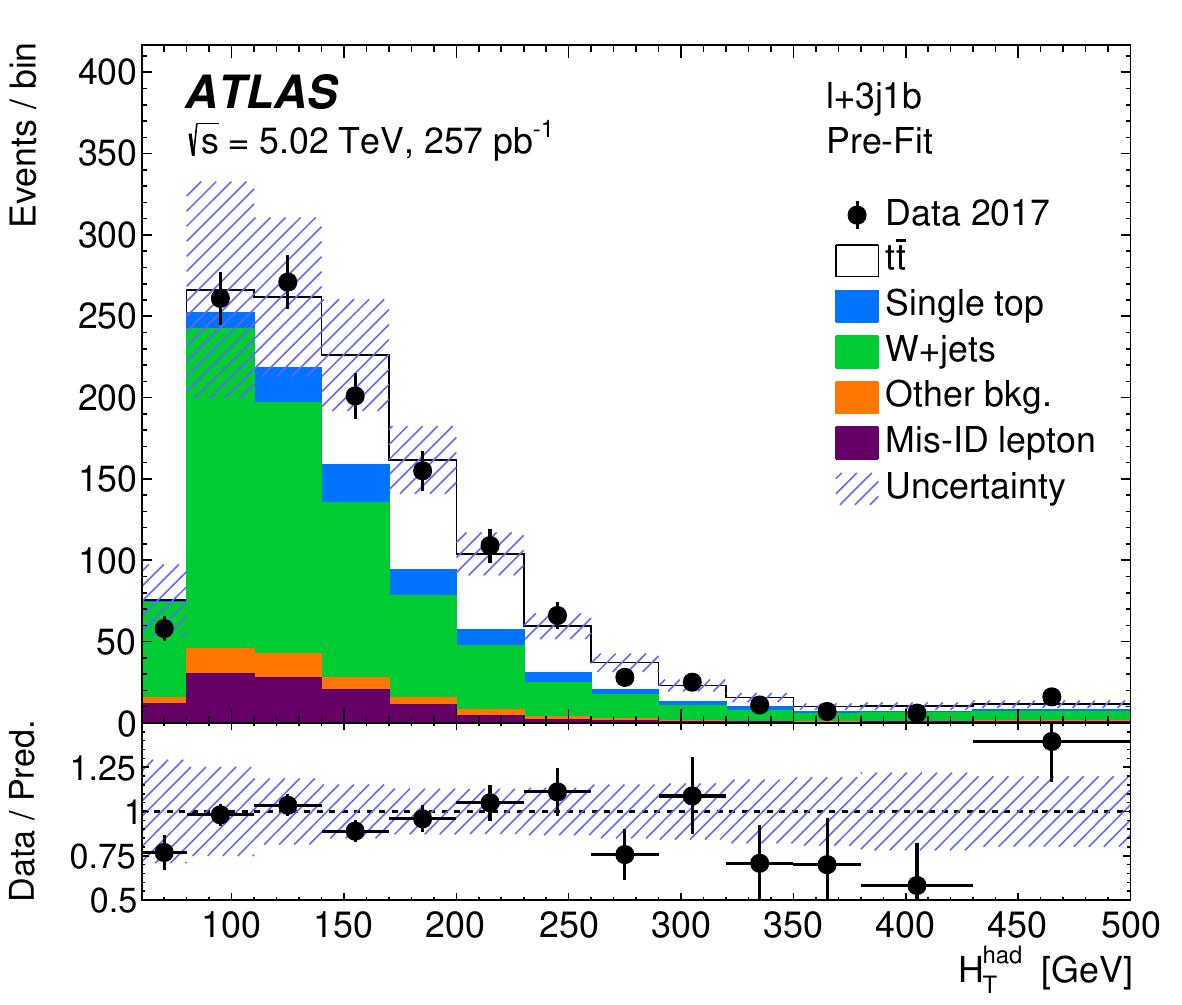}}\\
\subfloat[]{\includegraphics[width=0.45\textwidth]{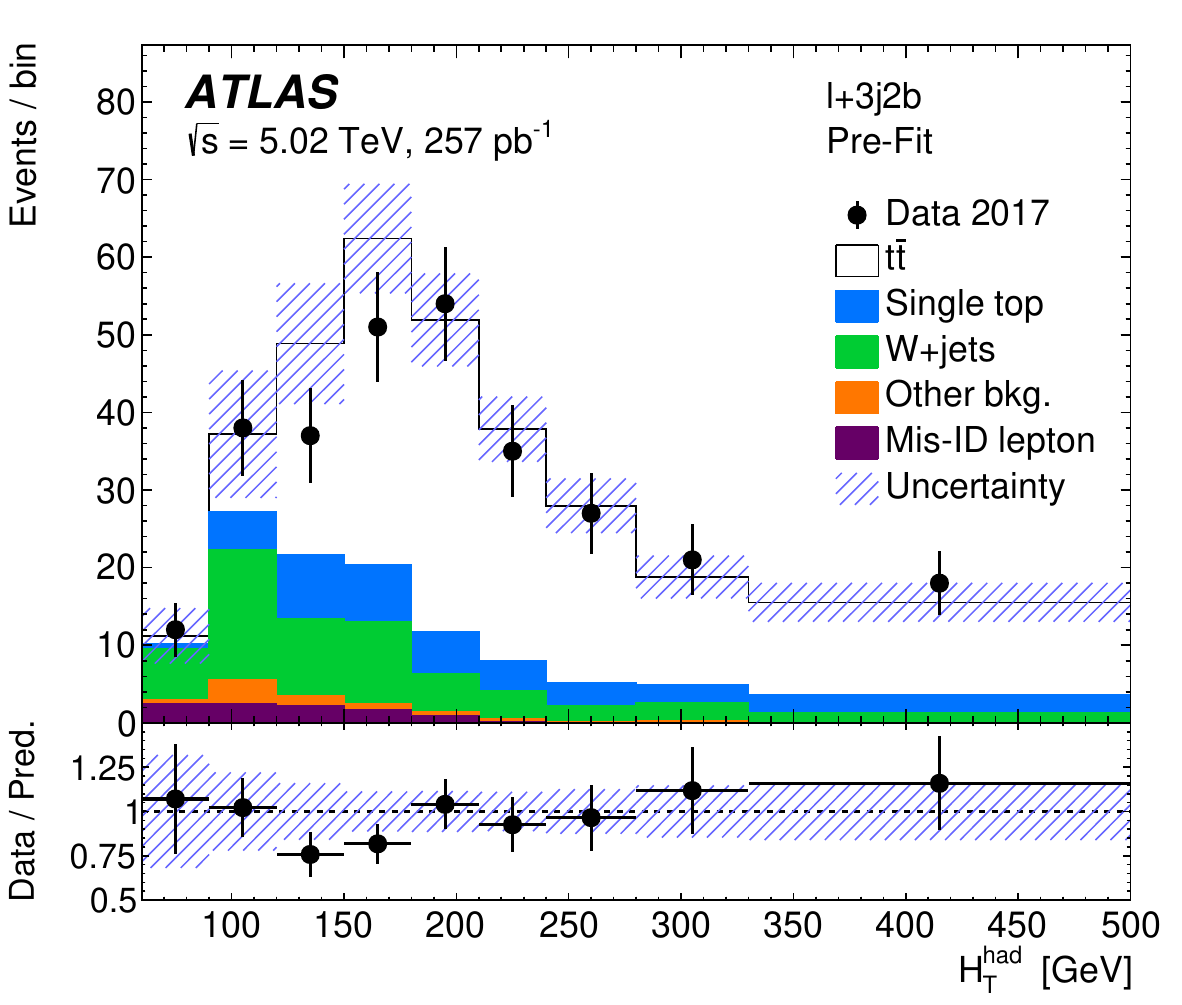}}
\subfloat[]{\includegraphics[width=0.45\textwidth]{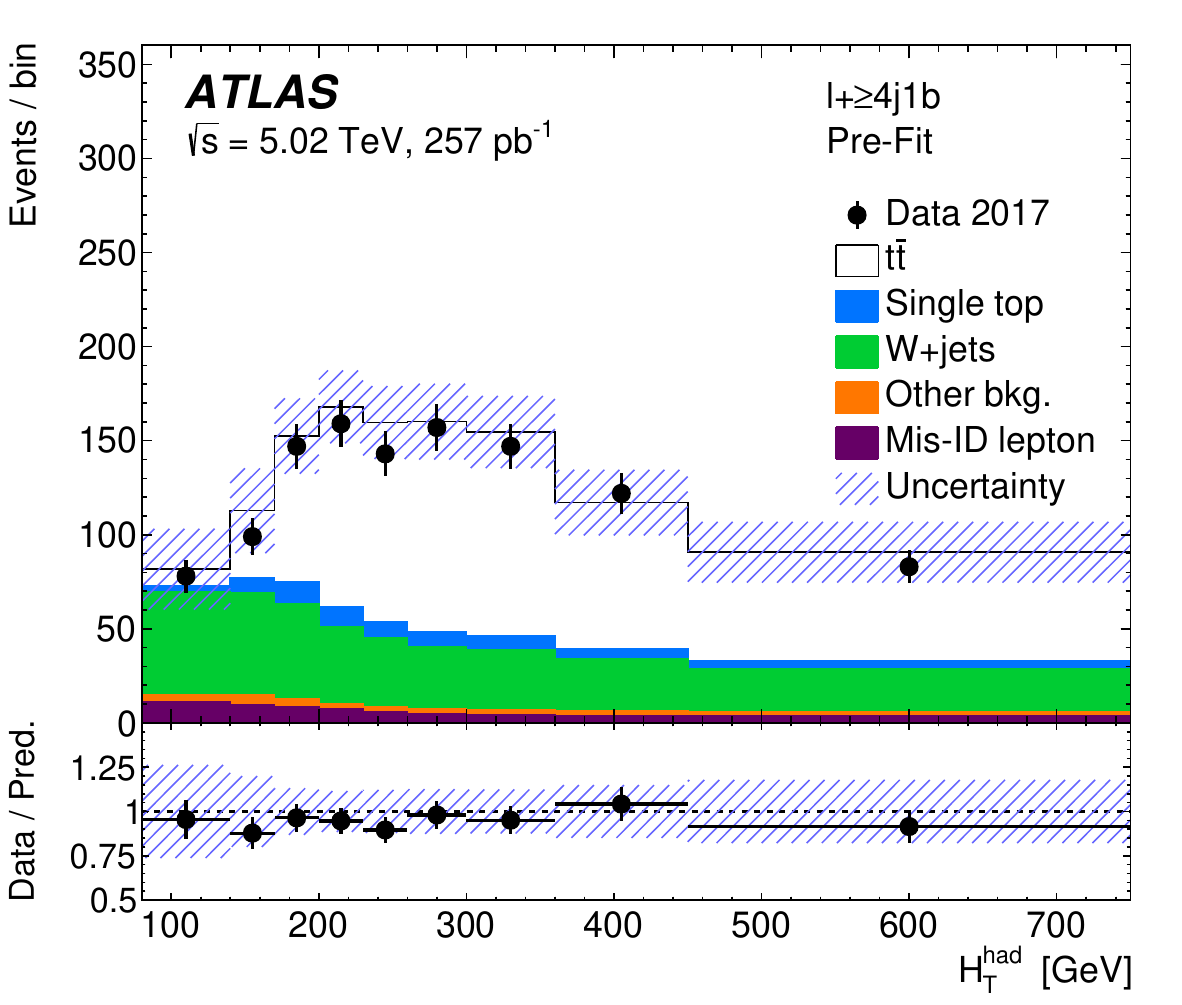}}\\
\subfloat[]{\includegraphics[width=0.45\textwidth]{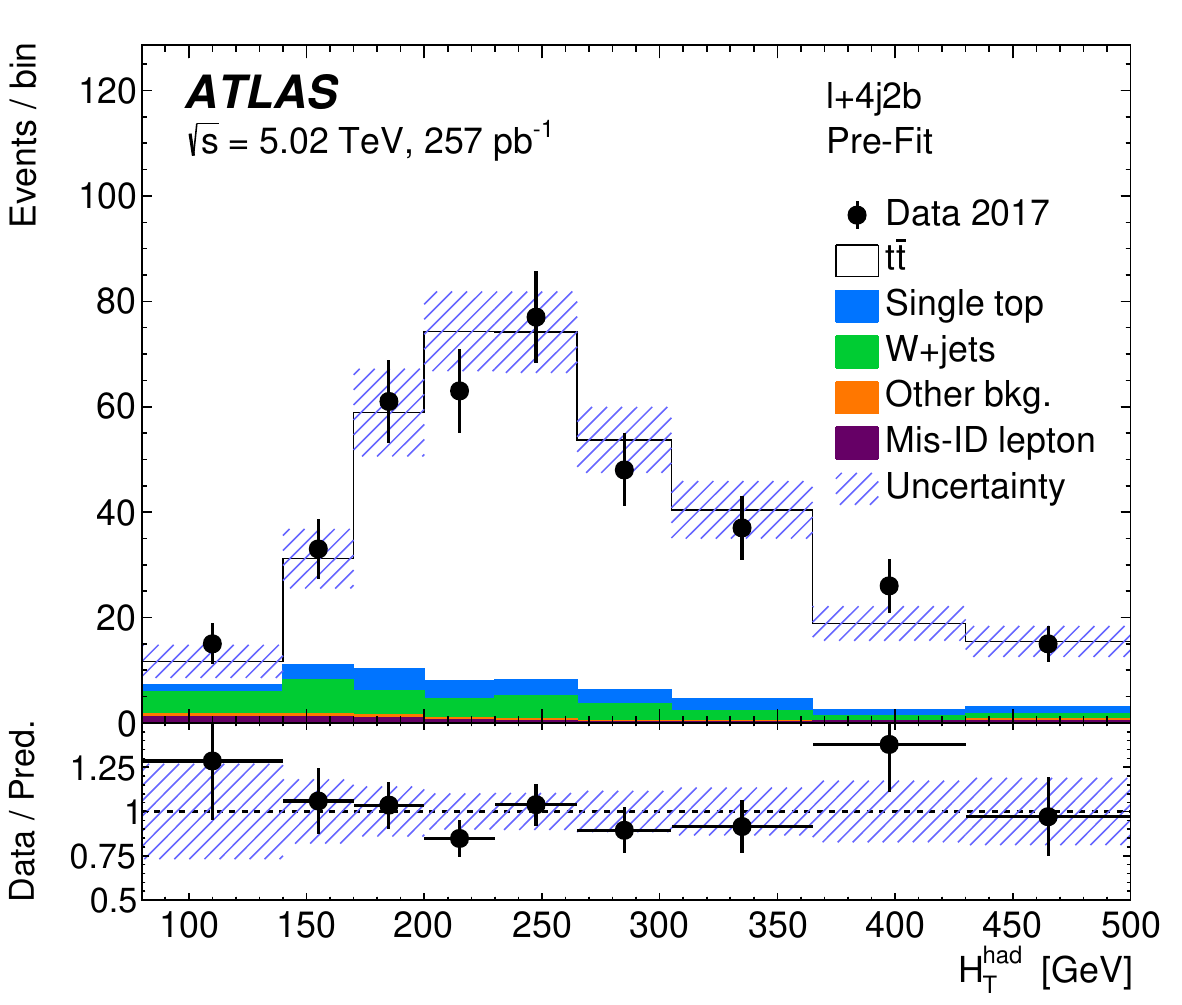}}
\subfloat[]{\includegraphics[width=0.45\textwidth]{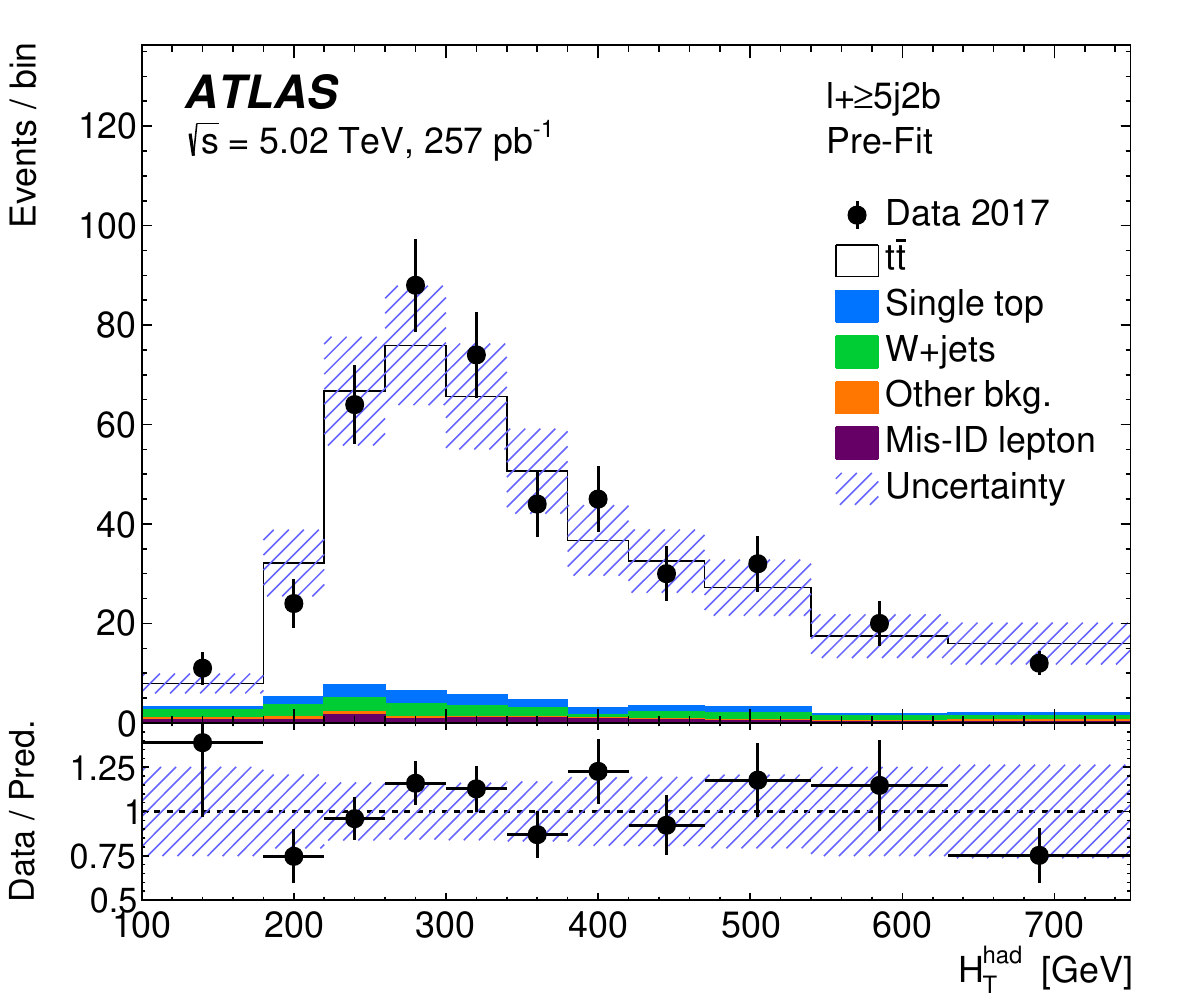}}
\caption{The observed and predicted distributions of $\hthad$ in the
(a) {\CRzero} region,
(b) {\CRone} region,
(c) {\CRtwo} region,
(d) {\SRone} region,
(e) {\SRtwo} region, and
(f) {\SRthree} region.
The pre-fit uncertainties are shown and the last bin contains the overflow.
The `Other bkg.' includes \Zjets and diboson backgrounds.
The blue hashed lines correspond to the total uncertainty of the prediction in a given bin.
The lower panels show the ratio of the data to the prediction, along with the uncertainty in the ratio.
}
\label{fig:HT_v1}
\end{center}
\end{figure}

The \hthad\ variable, defined as the scalar sum of the transverse momenta of the jets,  was found to provide the single largest
signal-to-background separation amongst the many kinematic variables tested.
The \hthad\ distribution is shown in Figure~\ref{fig:HT_v1} for the six single-lepton regions.
The predicted rates and shapes are found to be in good agreement with the observed distributions.
To further increase the separation of signal from background, two
boosted decision trees (BDT) using six input variables each were
created using the Toolkit for Multivariate Analysis (TMVA)~\Cite{hoecker2009tmva}.
One BDT was trained using MC signal and background events exclusively in the two-jet and three-jet
regions (\CRzero, \CRone, \CRtwo) whereas the second BDT
was trained in the four-jet and five-jet regions (\SRone, \SRtwo, \SRthree).
In both cases, the variables chosen were ones that were found to individually
provide good separation of signal from background and in combination
provided greater separation than other choices that were considered.
In the first BDT, the six input variables chosen are
the $\hthad$ variable,
the second Fox--Wolfram moment~\cite{FoxWolfram:1978}\ computed using all jets and the lepton (FW2 (l+j)),
lepton $\eta$,
the median $\Delta R$ between $b$-tagged jets and the lepton ($\Delta R_{b\ell}$),
the median $\Delta R$ between two jets ($\Delta R_{jj}$), and
the invariant mass of two jets with the smallest $\Delta R$ ($m_{jj}^{\mathrm{min.} \Delta R}$).
The second BDT uses as input
the $\hthad$ variable,
FW2 (l+j),
lepton $\eta$,
$\Delta R_{b\ell}$,
the median $\Delta R$ between any two untagged jets ($\Delta R_{\mathrm{uu}}$), and
the invariant mass of two untagged jets with the smallest $\Delta R$ ($m_{\mathrm{uu}}^{\mathrm{min} \Delta R}$).
The use of the median $\Delta R$\ provides more rejection power than other
measures of the average separation of $b$-tagged jets and the lepton.
 
\begin{figure}[tp]
\begin{center}
\subfloat[]{
\includegraphics[width=0.45\textwidth]{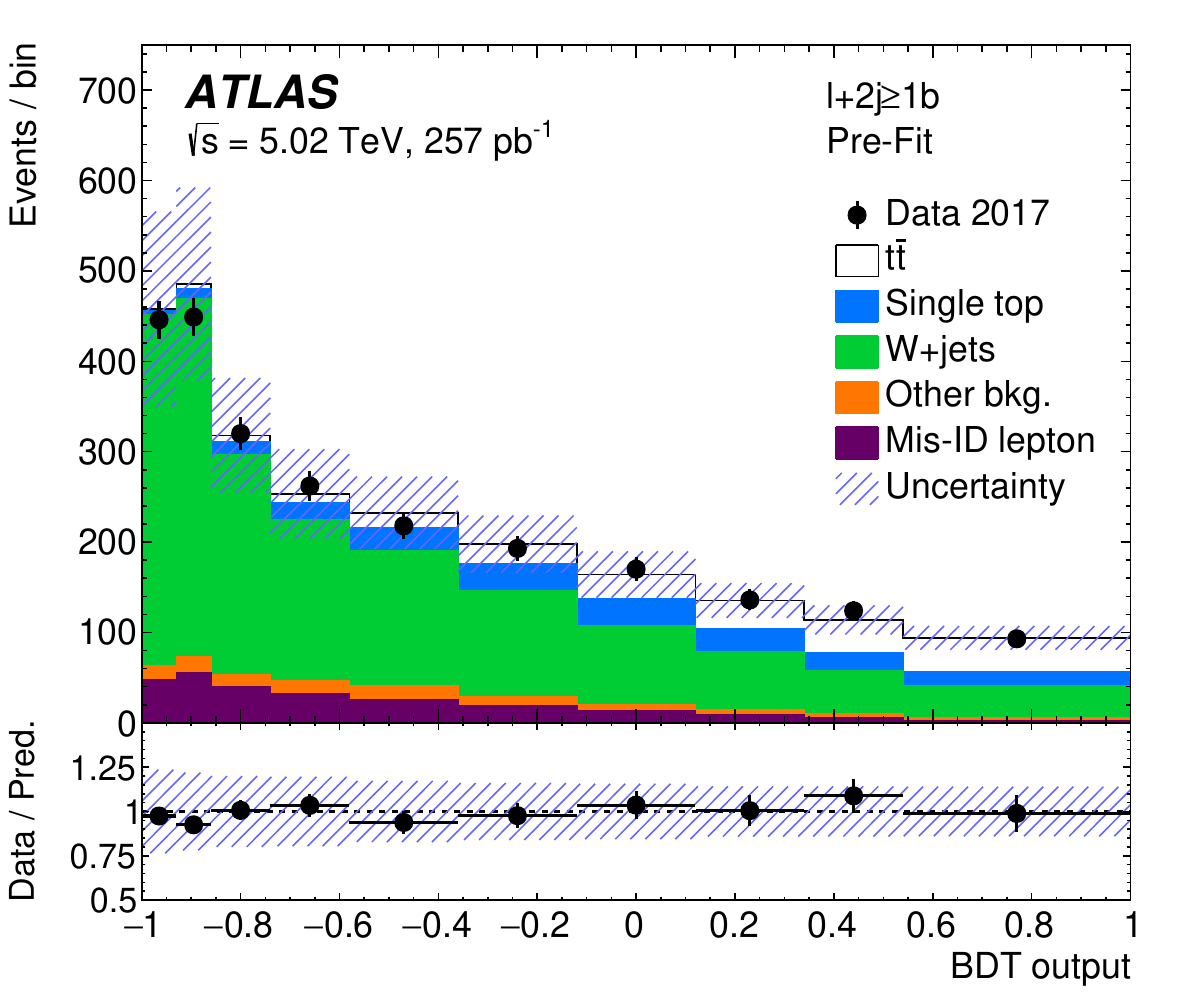}}
\subfloat[]{\includegraphics[width=0.45\textwidth]{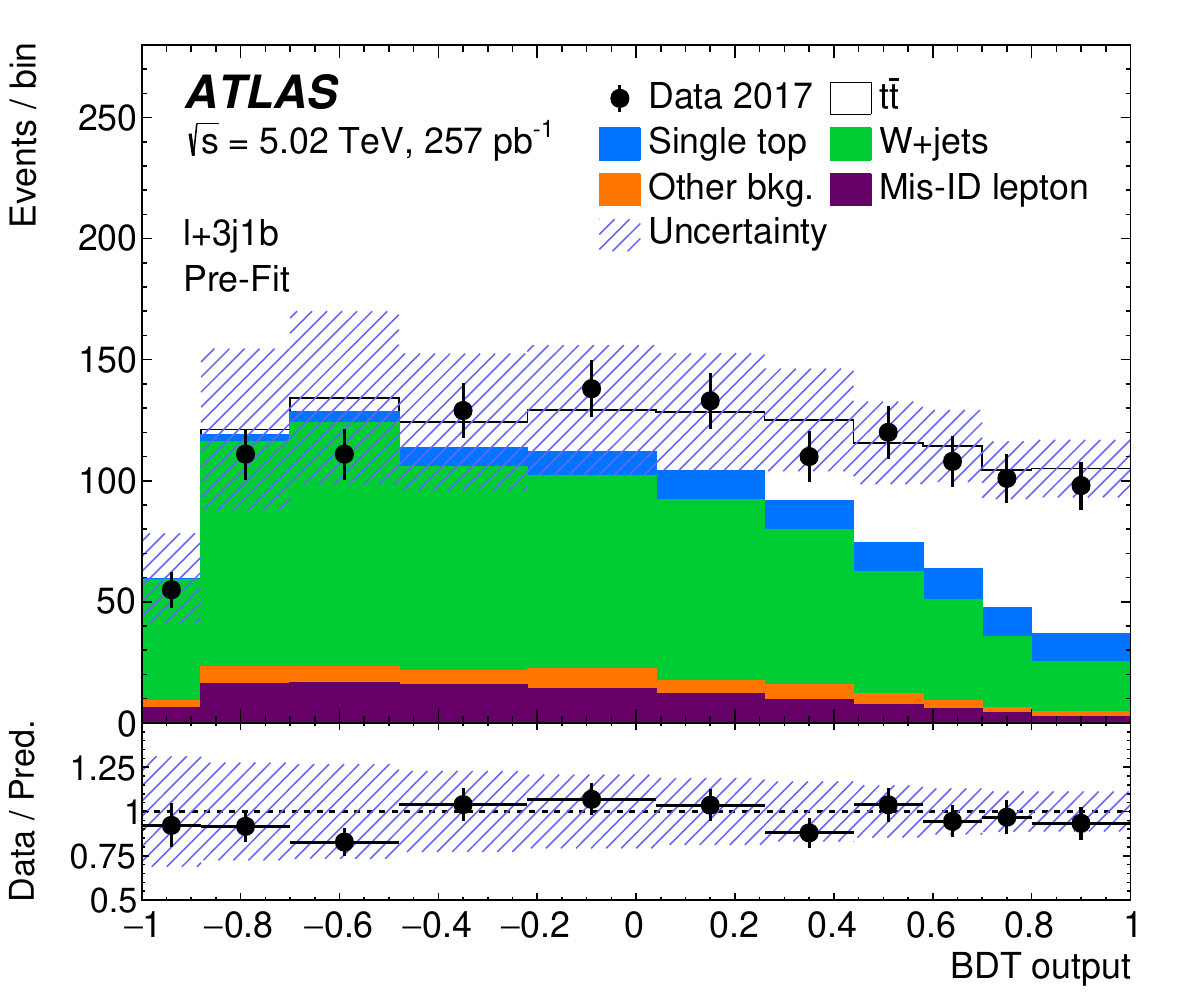}} \\
\subfloat[]{\includegraphics[width=0.45\textwidth]{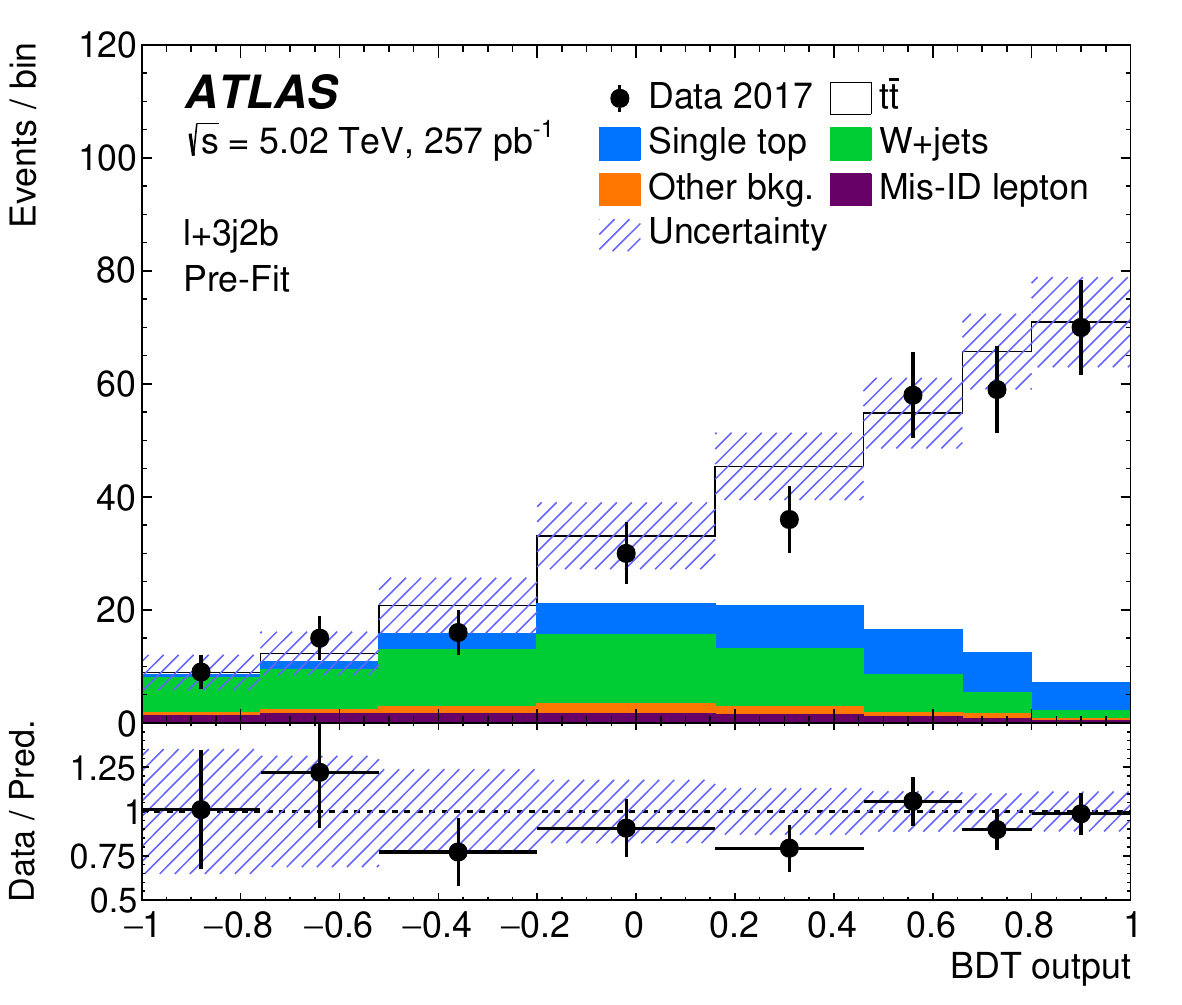}}
\subfloat[]{\includegraphics[width=0.45\textwidth]{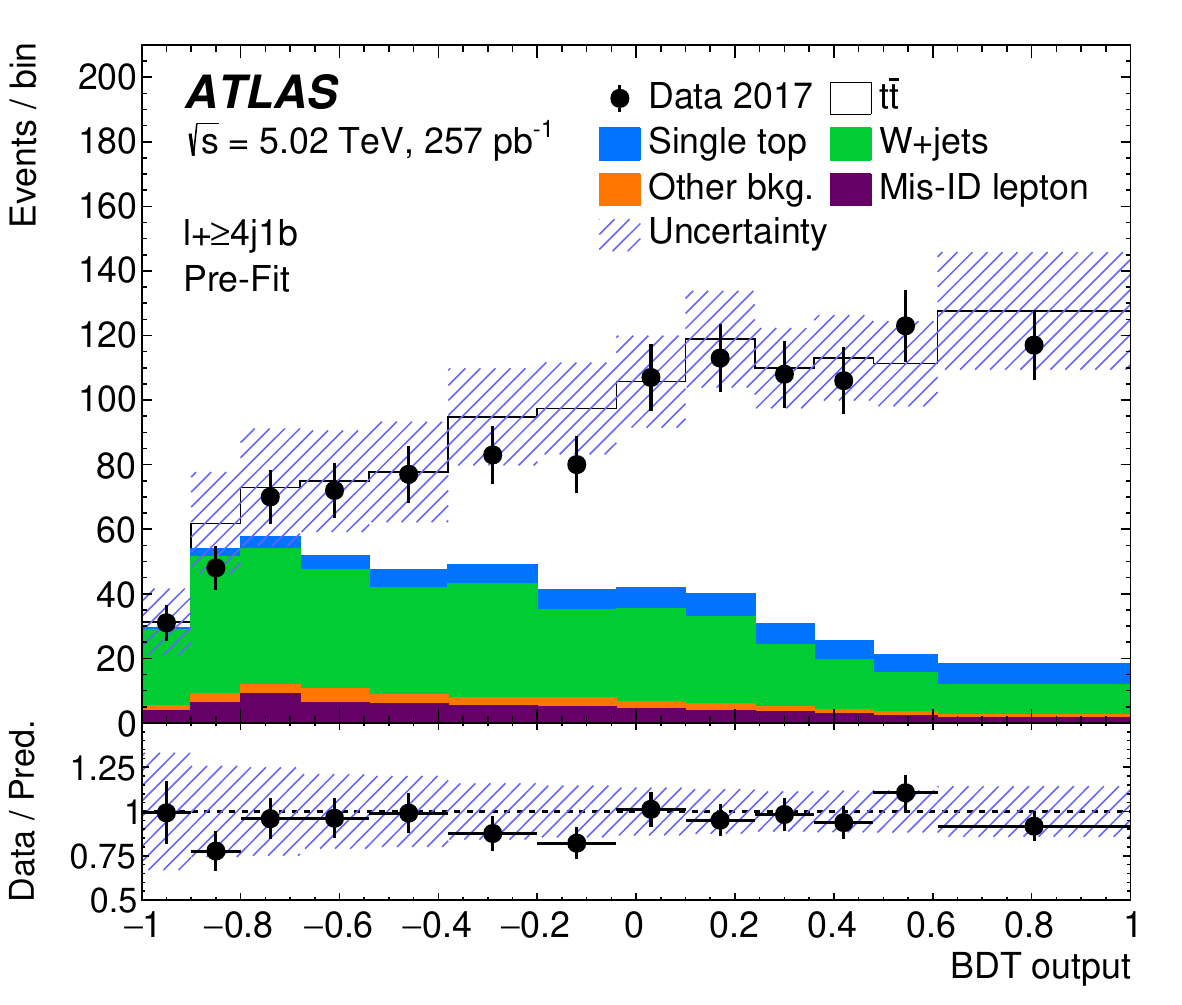}} \\
\subfloat[]{\includegraphics[width=0.45\textwidth]{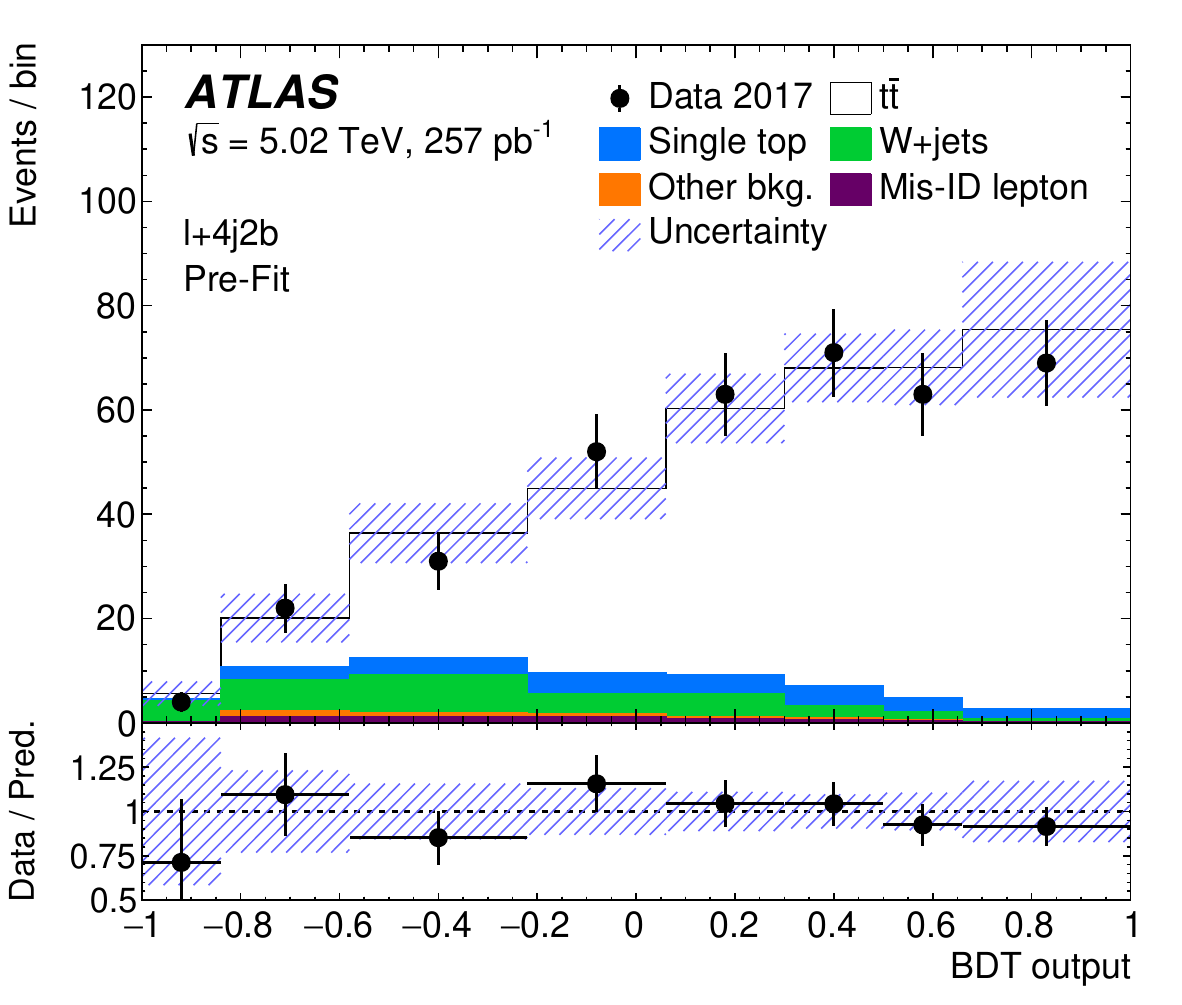}}
\subfloat[]{\includegraphics[width=0.45\textwidth]{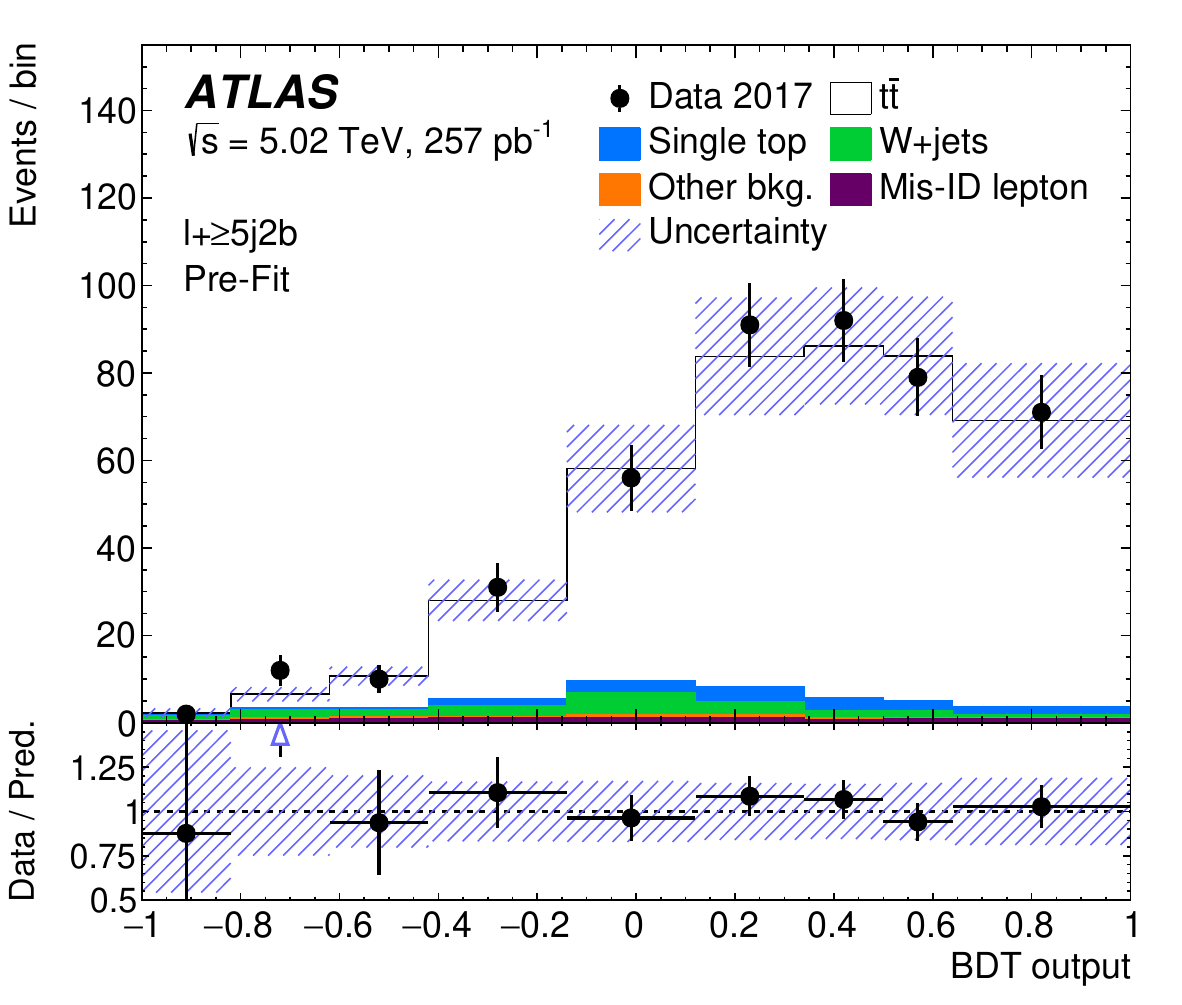}}
\caption{Comparison of the data and predicted BDT output distributions for the
(a) {\CRzero} region,
(b) {\CRone} region,
(c) {\CRtwo} region,
(d) {\SRone} region,
(e) {\SRtwo} region, and
(f) {\SRthree} region.
The pre-fit uncertainties are shown.
The `Other bkg.' includes \Zjets and diboson backgrounds.
The blue hashed lines correspond to the total uncertainty of the prediction in a given bin.
The lower panels show the ratio of the data to the prediction, along with the
uncertainty in the ratio.
}
\label{fig:prefit_BDT}
\end{center}
\end{figure}

The BDTs were applied to data and simulation  events in the regions where they were trained and the
resulting distributions of the outputs of the two BDTs used in different jet-multiplicity regions are shown in Figure~\ref{fig:prefit_BDT}.
There is good agreement between the shapes of the BDT output distributions for
data and simulation in each region. Since the \ttbar\ purity is already
quite high in the regions with large jet and/or $b$-tagged jet multiplicity
requirements, some of the selected \ttbar\ events
are kinematically similar to the bulk of the remaining background,
and are thus assigned low BDT scores.
The BDT output distributions were interpreted by a statistical model that employs the
expected distributions for both the background and signal contributions in
the six regions.
This model was fitted to the observed BDT output distributions in each region
to determine \xtt\ as described in Section~\ref{sec:results}.
 
\section{Systematic uncertainties}
\label{sec:systematics}

The systematic uncertainties in the \xtt\ measurement are broken down into
two categories:
those arising from the modelling of signal- and background-related processes,
and the detector-related uncertainties.
Table~\ref{tab:sys_impact_data} in Section~\ref{sec:combination}
gives a summary of all uncertainties for the two separate measurements
and their combination.

\subsection{Modelling uncertainties}
 
Modelling uncertainties arise from the MC predictions
of the \ttbar\ signal and the background processes.
The \ttbar\ signal has uncertainties arising from the NLO generator, the parton-shower and hadronisation model, initial/final-state radiation (I/FSR), the renormalisation and factorisation scales, the value of the \hdamp\ parameter, and the proton PDFs.
 
The uncertainty due to the choice of NLO generator was assessed by comparing the nominal \POWPY{8} sample with the alternative \MGNLOPY{8} sample.
Similarly, uncertainties in the parton-shower and hadronisation model were evaluated by comparing the nominal \POWPY{8} sample with the alternative \POWHER{7} sample.
The uncertainty due to ISR was estimated by using the Var3c A14 tunes, which
correspond to variations of \alphas\ for ISR in the A14 tune~\cite{ATL-PHYS-PUB-2014-021}.
Further effects on the ISR were seen when varying the renormalisation and factorisation scales \muR\ and \muF, and the \hdamp parameter. The renormalisation and factorisation scales were varied independently by factors of 0.5 and 2.0.
The uncertainty from the value of \hdamp\ was evaluated by comparing the nominal
\ttbar\ sample with an alternative sample that has $\hdamp= 3m_{\text{top}}$.
The effect of the FSR uncertainties was evaluated by
increasing and decreasing the renormalisation scale for emissions from the
parton shower by factors of 0.5 and 2.0.
The PDF uncertainties affecting the \ttbar\ signal were evaluated by using the PDF4LHC15 Hessian
uncertainties~\cite{Butterworth:2015oua}.
A similar procedure was used to evaluate the PDF uncertainties affecting the $Wt$ background in the dilepton measurement only.
 
The predictions for \cbllp\ in the dilepton measurement are
sensitive to the fraction of \ttbar\ events with extra \bbbar\ or \ccbar\
pairs, which was found to be underestimated by the nominal
\POWPY{8} samples at \sxyt\ \cite{TOPQ-2018-17,TOPQ-2017-12}.
The corresponding uncertainty at
\sxut\ was evaluated by reweighting the nominal \POWPY{8} \ttbar\ sample so as
to increase
the rate of three-$b$-jet events at the generator level by 40\%, the value best
fitting the \sxyt\ data \cite{TOPQ-2018-17}. This variation  gives an
uncertainty of 0.2\% in \xtt, and is included
as part of the \ttbar\ parton-shower/hadronisation uncertainty in
Table~\ref{tab:sys_impact_data}.
 
The uncertainties in the cross-sections for the $Wt$ associated, $t$-channel, and $s$-channel single-top-quark background processes were taken to be 9.5\%
as discussed in Section~\ref{sec:samples}.
The uncertainty arising from interference between the \ttbar\ and $Wt$ final states was evaluated by comparing the nominal MC sample generated using the diagram removal scheme with an alternative sample that used the diagram subtraction approach \cite{wtdr,wtinter1,wtinter2}.
The uncertainties in the parton-shower and hadronisation model for $Wt$ associated single-top-quark production were evaluated by comparing the nominal \POWPY{8} samples with an alternative \POWHER{7} sample. A similar comparison was performed for $t$-channel single-top-quark production in the single-lepton measurement
(the effect is negligible in the dilepton measurement).
In both measurements, the uncertainties in relevant single-top backgrounds from modelling of ISR, renormalisation and factorisation scales, and FSR were considered in a manner similar to that for \ttbar\ signal.
 
The \Wjets\ modelling uncertainties in the single-lepton measurement were
evaluated by first splitting the \Wjets\ background into three categories
$W+{\ge}1c$, $W+{\ge}1b$, and $W+{\ge}1q/g$ (light quarks or gluons)
based on the flavour of additional jets in the event.
In each category the events were further separated into categories with 2, 3, 4, or ${\geq}5$ jets, and a
normalisation uncertainty based on Berends scaling ($4\% \oplus 24\%$ per jet) \cite{Berger2011} was applied for each
of the 12 categories.
The renormalisation and factorisation scales were varied to estimate additional
shape uncertainties for the three categories of \Wjets\ background.
 
In the dilepton measurement, the normalisation of the \Zjets\ background
was fitted in the same-flavour channels with a requirement of $\etmiss>30$\,\GeV. This normalisation factor was then propagated to the $e\mu$ channel which has
no \etmiss\ requirement, implicitly assuming that the \etmiss\ distributions
are well modelled in simulation.
Uncertainties in this  procedure were assessed by comparing the fractions of
$Z\rightarrow ee/\mu\mu$ events with one and two $b$-tagged jets
which have $\etmiss>30$\,\GeV\
in data and simulation, probing the modelling of the \etmiss\ distributions.
There were 24\% more data events than predicted from the simulation sample with one $b$-tagged jet and
19\% more events with two $b$-tagged jets, after averaging over same-flavour
events. These values were taken as additional uncertainties of the \Zjets\
backgrounds in the $e\mu$ channel. An additional modelling uncertainty  from the
derivation of the \mll\ distributions  (i.e. the
$f_{1,m}^{\ell\ell,Z+\mathrm{jets}}$ and $f_{2,m}^{\ell\ell,Z+\mathrm{jets}}$ fractions)
using the \POWPY{8} MC samples rather than the \Sherpa\ \Zjets\ samples was also applied in the dilepton measurement. In the single-lepton measurement, a 50\% normalisation uncertainty in the \Zjets\ background was used to take into account
the uncertainties in the cross-section, acceptance and modelling evaluated
by varying the normalisation and matching scales in the \Sherpa\ MC prediction.
 
The uncertainty in the diboson background
was conservatively taken to be 20\%. This variation accounts for
uncertainties in the acceptances, and in
the cross-sections calculated using the \MCFM\ MC generator~\cite{mcfm}.
 
In the dilepton \xtt\ measurements at $\sqrt{s}=7$, 8, and~13\,\TeV,
the misidentified-lepton background was found to be well modelled by simulated events.
Therefore, simulation was also used to predict the misidentified-lepton background in the \sxut\ dilepton
measurement.
Variations of 50\%\ and 100\%, approximately twice the uncertainties derived at \sxyt~\cite{TOPQ-2018-17},
were applied to the predicted misidentified-lepton backgrounds for the one
$b$-jet and two $b$-jet samples, respectively. In the single-lepton measurement,
the uncertainty in the misidentified-lepton background was first categorised
according to the flavour of the misidentified lepton ($e$ or $\mu$) and then
further split into shape and normalisation components for each of the six
regions, giving 24 individual components. The shape uncertainty was estimated
by comparing the nominal misidentified-lepton
background sample with a sample where the electron- and
muon-misidentification rates were determined using different kinematic
variables and different numbers of $b$-tagged jets for the parameterisation.
A 50\% relative
uncertainty was used to account for all uncertainties affecting the rate of the
misidentified-lepton background~\cite{ATLAS-CONF-2014-058}.
 
\subsection{Detector-related uncertainties}\label{ssec:det}
 
The electron identification
efficiency, energy scale, and energy resolution were measured using $Z\rightarrow ee$ events in the \sxut\ data sample
using techniques described in Ref.~\cite{EGAM-2018-01}.
Uncertainties in the identification were implemented as variations of the corresponding simulation-to-data correction factors.
The $Z\rightarrow ee$ sample was also used to assess uncertainties
in the modelling of electron-charge misidentification in the dilepton measurement.
 
The muon identification efficiencies
and momentum scale calibration were taken from high-\pileup \sxyt\ data
\cite{MUON-2018-03}, with additional uncertainties to account for the extrapolation to the low-\pileup\ \sxut\ data.
 
The efficiencies of the lepton-isolation requirements were determined using
tag-and-probe techniques applied to \sxut\ $Z\rightarrow ee$ and
$Z\rightarrow\mu\mu$ events. The differences between the efficiencies
measured in data and simulation samples were studied as functions of
lepton \pt\ and $\eta$, and the activity around the lepton.
Applying these shifts
to leptons in simulated \ttbar\ events gave per-lepton efficiency changes
of up to $+0.4$\% for electrons and $-0.5$\% for muons. Given the small sizes
of these shifts, no corrections were applied to the nominal simulation,
and the largest shifts were taken as uncertainties in the efficiencies
and background estimates.
 
The electron-trigger efficiency was measured using $Z\rightarrow ee$ events
in low-\pileup datasets recorded at both \sxut\ and \sxyt, using the method
described in \cite{TRIG-2018-05}.
The muon-trigger efficiency was measured using $Z\rightarrow\mu\mu$ events in
high-\pileup \sxyt\ data recorded immediately before the low-\pileup datasets
\cite{TRIG-2018-01}.
 
The jet calibration is based on the jet-energy
scale determined for \sxyt\ high-\pileup data~\cite{JETM-2018-05}, with
an additional correction to the jet-energy scale derived from the $Z$\,+\,jet
\pt-balancing technique applied to \sxut\ data and simulation, as discussed in
Section~\ref{sec:selection}. The uncertainty of 1\%--2\%\ in this correction was
incorporated into the systematic uncertainties of the jet-energy scale,
including the modelling uncertainty of the $Z$\,+\,jet final state.
An uncertainty from the jet-vertex tagger arises from the efficiency correction factors for this algorithm~\cite{PERF-2014-03}.
 
Predictions for the $b$-tagging efficiency (for the single-lepton
measurement), \cbllp\ (for the dilepton measurement) and background rejection
were taken from \sxut\ simulation samples, corrected by the standard
simulation-to-data scale factors obtained from
studies of \ttbar\ and dijet events at \sxyt\ \cite{FTAG-2018-01,ATLAS-CONF-2018-001,ATLAS-CONF-2018-006}.
This choice is supported by the values of \epsbllp\ obtained from the dilepton
measurement at \sxut\ as discussed in Section~\ref{sec:results}, and
from simulation studies which show that the $b$-tagging performance
is similar in the two environments.
 
The uncertainties in the calibrations of leptons and jets discussed above were propagated into the \etmiss\
calculation along with uncertainties in the modelling of the soft term~\cite{PERF-2016-07}.
 
The uncertainty in the integrated luminosity is 1.6\%, evaluated as discussed
in Ref.~\cite{ATLAS-CONF-2019-021} and using the LUCID-2
detector~\cite{LUCID2} for the primary luminosity measurement.
The corresponding uncertainty in \xtt\ is larger than 1.6\%
in the dilepton measurement because the luminosity
uncertainty affects the evaluation of the $Wt$ and diboson backgrounds
from simulation.
 
The LHC beam energy is known to be within 0.1\% of the nominal value based on the LHC magnetic model and comparisons of the
revolution frequencies of proton and lead-ion beams~\cite{ebeam2}.
A 0.1\% variation in $\sqrt{s}$ corresponds to a 0.3\% variation in the predicted \xtt\ according to the NNLO+NNLL
predictions of  \TOPpp.
This uncertainty is included as a separate uncertainty in \xtt\ to facilitate comparisons with theoretical predictions at exactly \sxut.


\section{Results}
\label{sec:results}

\begin{figure}[tp]
\parbox{83mm}{\includegraphics[width=76mm]{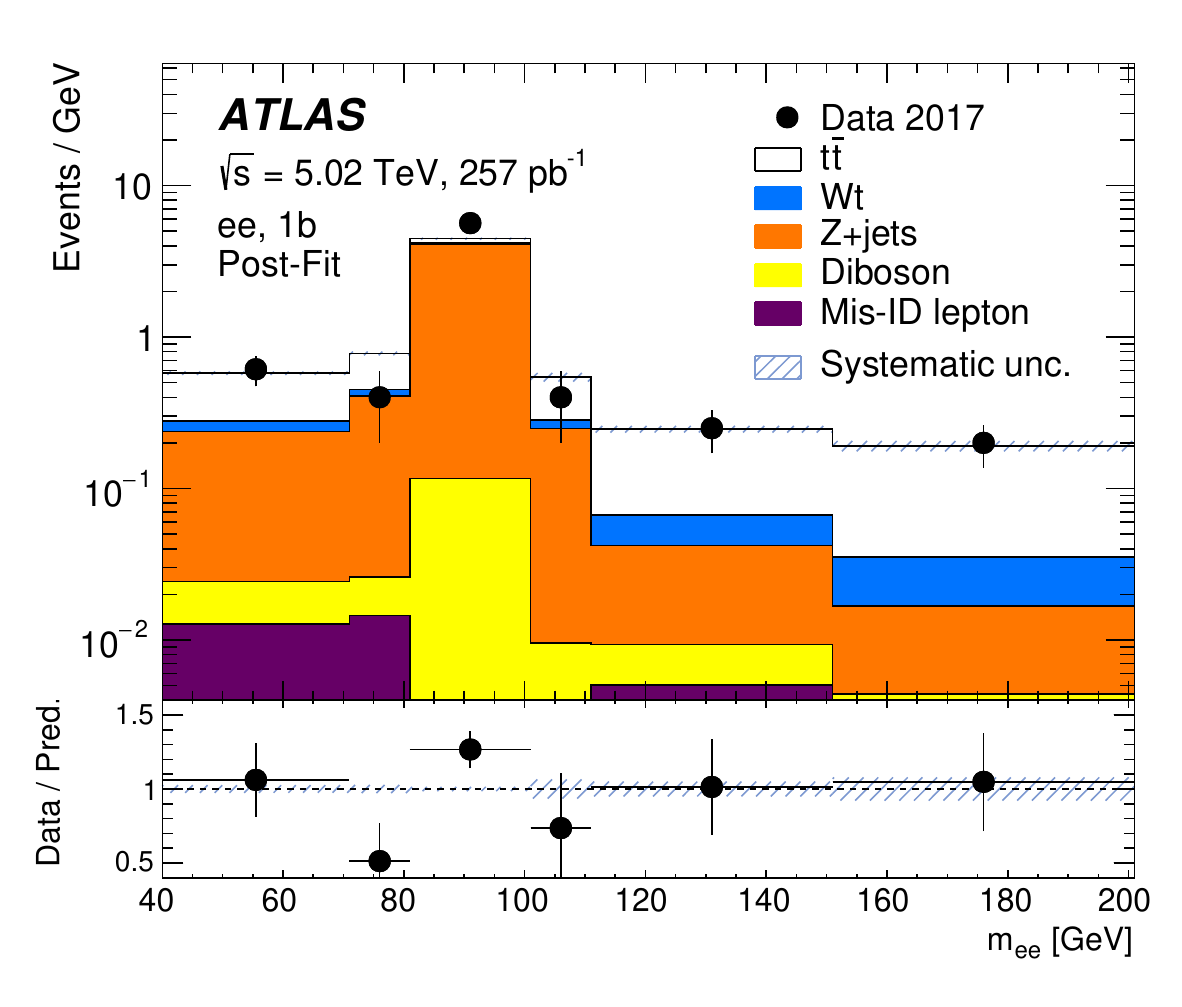}\vspace{-6mm}\center{(a)}}
\parbox{83mm}{\includegraphics[width=76mm]{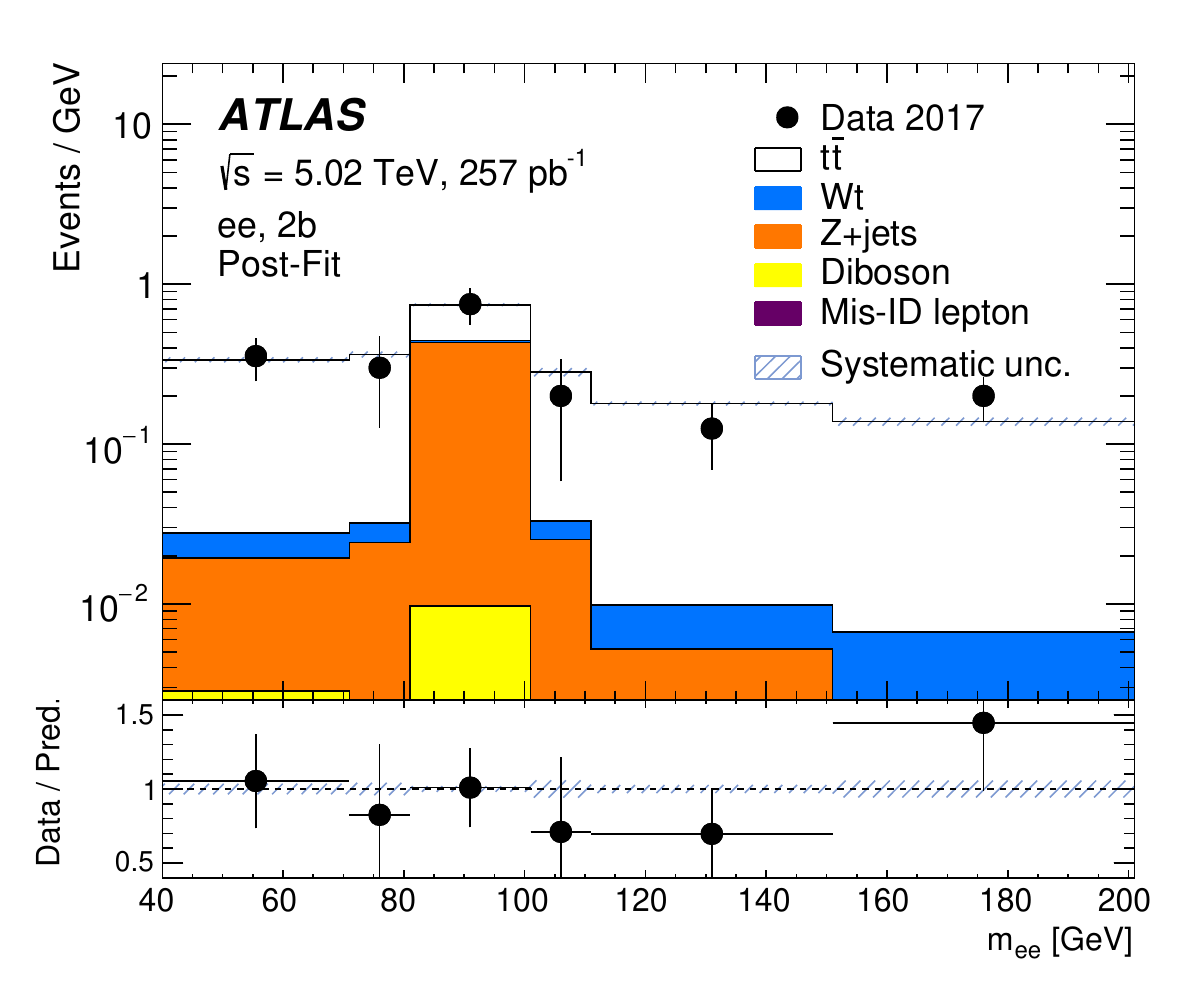}\vspace{-6mm}\center{(b)}}
 
\hspace{40mm}\parbox{83mm}{\includegraphics[width=76mm]{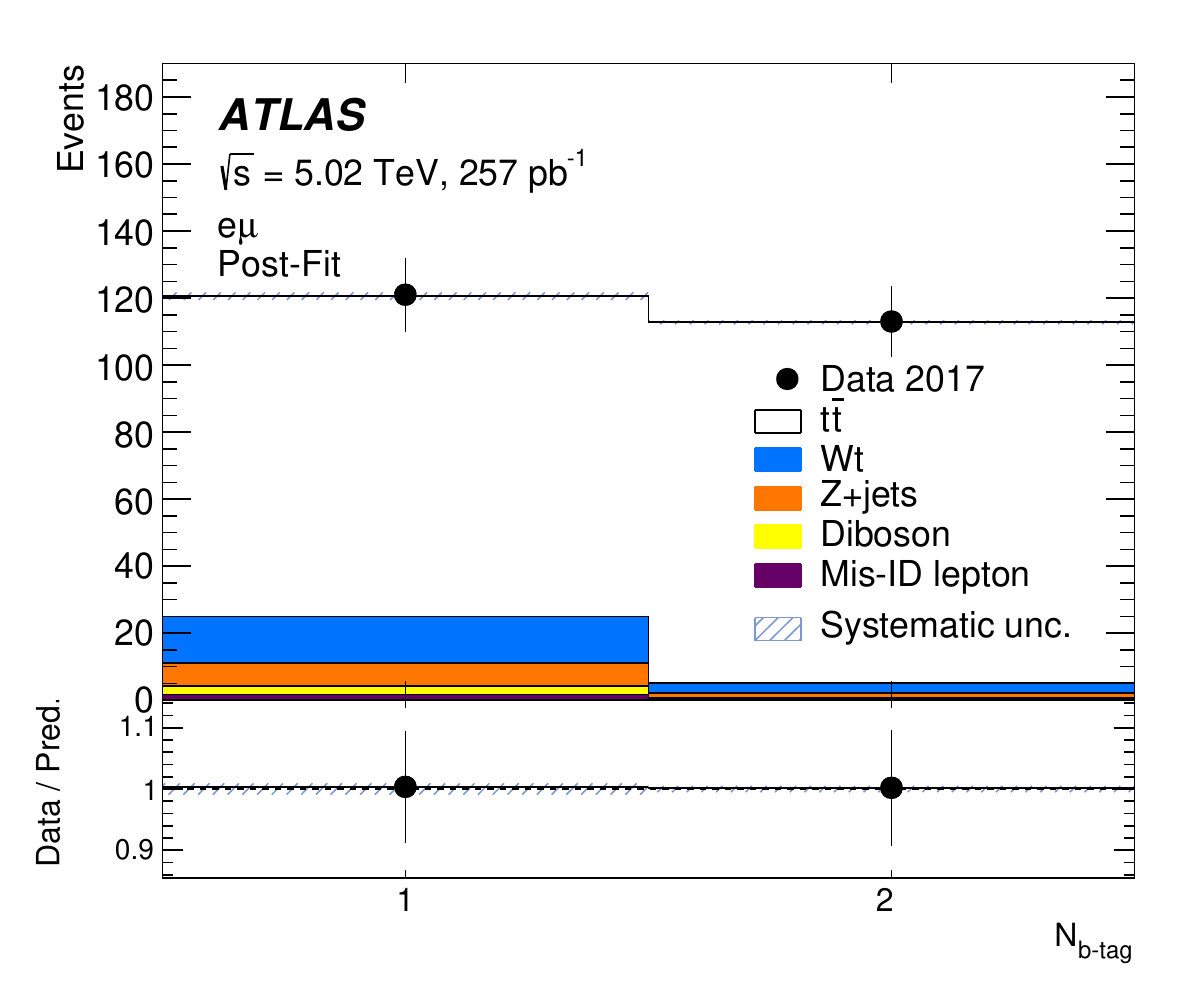}\vspace{-6mm}\center{(c)}}
\\
 
\parbox{83mm}{\includegraphics[width=76mm]{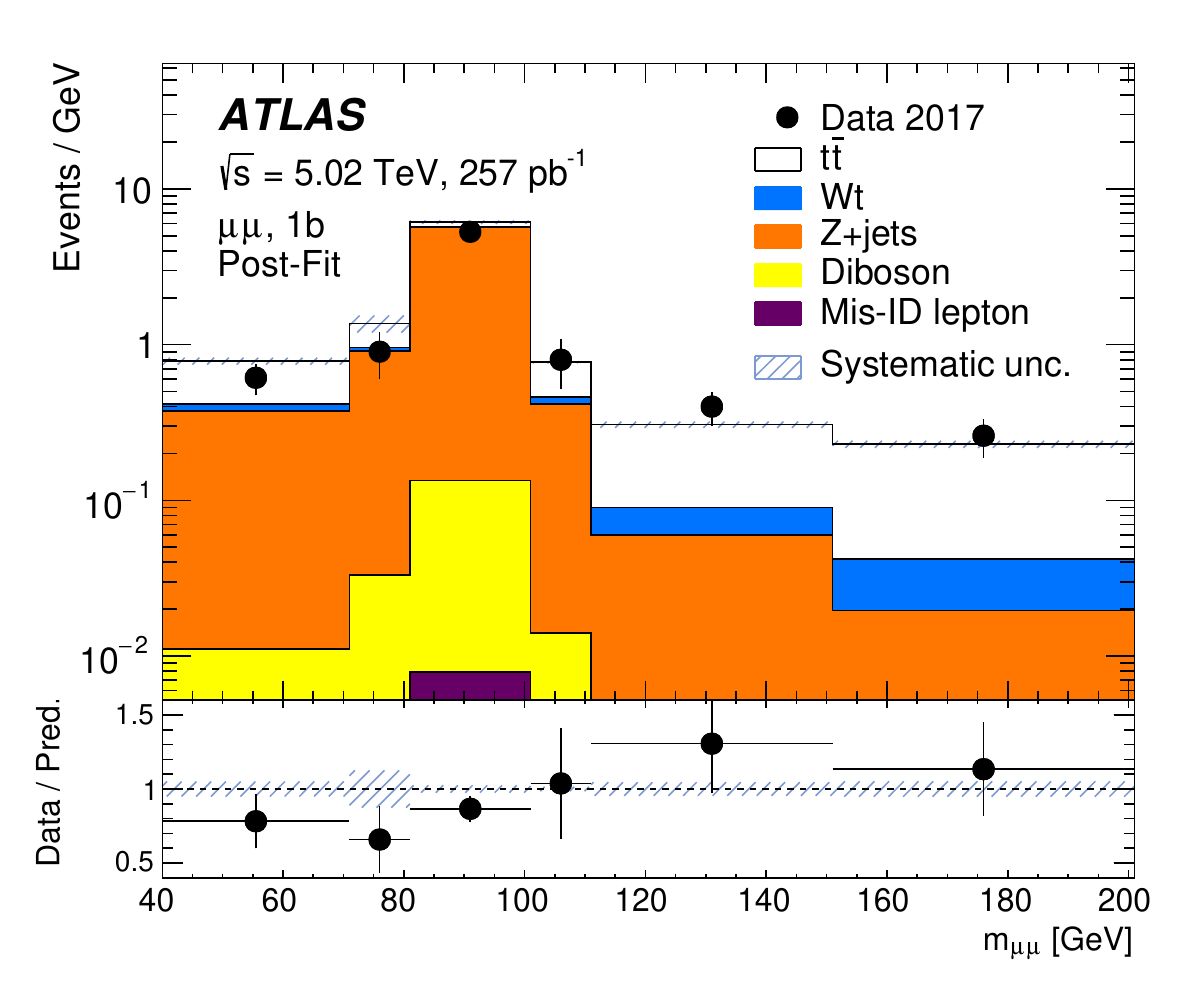}\vspace{-6mm}\center{(d)}}
\parbox{83mm}{\includegraphics[width=76mm]{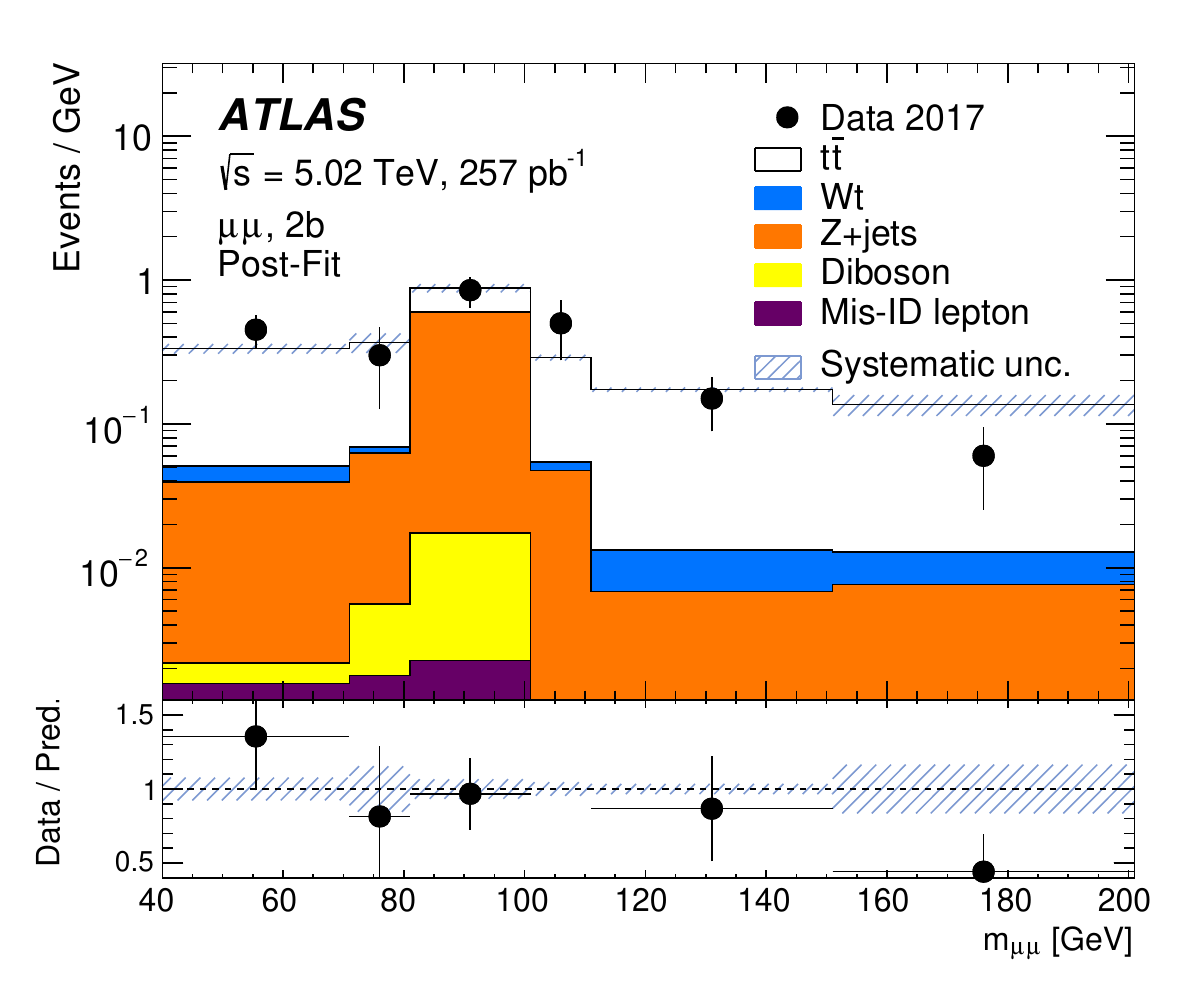}\vspace{-6mm}\center{(e)}}
\caption{\label{f:dilfitres}Results of the fit to data in the dilepton channel,
showing the invariant mass distributions for the one $b$-tagged jet and
two $b$-tagged jet samples in the $ee$ (top)
and $\mu\mu$ (bottom) channels, and the $b$-tagged jet multiplicity in the
$e\mu$ channel (centre). The data are shown by the points with uncertainties,
and are compared with the results of the fit, showing the scaled contributions
from \ttbar, $Wt$, \Zjets, dibosons, and events with misidentified leptons.
The total systematic uncertainty of the fit prediction in each bin is shown by
the hatched band. The lower panels show the ratios of data to the fit
predictions. In the invariant mass distributions, the last bin includes
the overflow with $\mll>200$\,\GeV\ but is normalised to the displayed bin
width.}
\end{figure}

 
Table~\ref{t:evtcount} and Figure~\ref{f:dilfitres} show the results
of applying the fit described in Section~\ref{sec:dilepton} to
the dilepton data sample, comparing the observed event counts with the fit
predictions including the \ttbar\ signal component and all backgrounds.
In Table~\ref{t:evtcount}, the same-flavour channel event counts are divided
into those near the peak of the $Z\rightarrow\ell\ell$ resonance
($|\mll-m_Z|<10$\,GeV and referred to as on-$Z$) and those off the $Z$-boson
resonance (off-$Z$), to illustrate the changing background level and
composition. The fit describes the data well,
with the exception of the $Z$-boson region in the same-flavour channels with one $b$-tagged jet, as noted earlier.
The fitted \ttbar\ production cross-section in the dilepton channel is
\begin{equation}\nonumber
\xtt = \ttxdlval \pm \ttxdlstat\,\mathrm{(stat.)}\ \pm \ttxdlsyst\,\mathrm{(syst.)}\ \pm \ttxdllumi\,\mathrm{(lumi.)}\ \pm \ttxdlebeam\,\mathrm{(beam)}\,\mathrm{pb},
\end{equation}
where the four uncertainties are due to the size of the data sample, experimental and theoretical systematic effects, and imperfect knowledge of both the
integrated luminosity and the LHC beam energy.
The total uncertainty of \ttxdlerrtot\,pb corresponds to a relative uncertainty of \ttxdlrel, of which \ttxdlrstat\ is the data statistical uncertainty.
The `dilepton' column of Table~\ref{tab:sys_impact_data} in
Section~\ref{sec:combination}
shows a breakdown of the uncertainties. The result is quoted for a top-quark mass of 172.5\,\GeV, and changes by $\mp\dmtopvaldl$\% for a $\pm 1$\,\GeV\ change in \mtop.
 
\begin{table}[tp]
\caption{\label{t:evtcount}Observed numbers of opposite-charge dilepton
events with one (upper block) and two (lower block) $b$-tagged jets in data,
together with the \ttbar\ and background event counts from the fit prediction,
including the associated statistical and systematic uncertainties.
The five columns show the $ee$ channel
with $|\mll-m_Z|>10$\,\GeV\ (off-$Z$) and $|\mll-m_Z|<10$\,\GeV\ (on-$Z$),
the $e\mu$ channel, and the $\mu\mu$ channel including off-$Z$ and on-$Z$
contributions. The uncertainty in the total prediction is in some cases smaller
than the individual uncertainties due to correlations induced by the fit.}
\centering
 
\begin{tabular}{l|
r@{$\,\pm\,$}l
r@{$\,\pm\,$}l|
r@{$\,\pm\,$}l|
r@{$\,\pm\,$}l
r@{$\,\pm\,$}l
}
\toprule
Event counts $\rule[-3mm]{0mm}{7mm}$ & \multicolumn{2}{l}{$N_{1,\mathrm{off-Z}}^{ee}$} & \multicolumn{2}{l|}{$N_{1,\mathrm{on-Z}}^{ee}$} & \multicolumn{2}{l|}{$N_{1}^{e\mu}$} & \multicolumn{2}{l}{$N_{1,\mathrm{off-Z}}^{\mu\mu}$} & \multicolumn{2}{l}{$N_{1,\mathrm{on-Z}}^{\mu\mu}$} \\
\hline
Data& \multicolumn{2}{c}{47}& \multicolumn{2}{c|}{113}& \multicolumn{2}{c|}{121}& \multicolumn{2}{c}{65}& \multicolumn{2}{c}{106} \\
\hline
\ttbar & 30.1 & 3.8 & 6.0 & 0.7 & 95.6 & 10.0 & 36.6 & 3.6 & 7.5 & 0.8 \\
$Wt$ single top & 4.0 & 0.6 & 0.75 & 0.12 & 13.9 & 1.5 & 4.5 & 0.5 & 0.90 & 0.12 \\
$Z$\,+\,jets & 14.7 & 1.5 & 80.1 & 6.0 & 6.8 & 1.7 & 27.1 & 3.3 & 111.3 & 8.8 \\
Diboson & 0.83 & 0.17 & 2.3 & 0.5 & 2.6 & 0.5 & 0.90 & 0.18 & 2.5 & 0.5 \\
Misidentified leptons & 0.9 & 0.4 & 0.05 & 0.03 & 1.7 & 0.8 & 0.23 & 0.17 & 0.16 & 0.15 \\
\hline
Total prediction & 50.5 & 3.7 & 89.2 & 5.9 & 120.6 & 9.7 & 69.4 & 4.7 & 122.4 & 8.6 \\
\hline
\hline
Event counts $\rule[-3mm]{0mm}{7mm}$ & \multicolumn{2}{l}{$N_{2,\mathrm{off-Z}}^{ee}$} & \multicolumn{2}{l|}{$N_{2,\mathrm{on-Z}}^{ee}$} & \multicolumn{2}{l|}{$N_{2}^{e\mu}$} & \multicolumn{2}{l}{$N_{2,\mathrm{off-Z}}^{\mu\mu}$} & \multicolumn{2}{l}{$N_{2,\mathrm{on-Z}}^{\mu\mu}$} \\
\hline
Data& \multicolumn{2}{c}{31}& \multicolumn{2}{c|}{15}& \multicolumn{2}{c|}{113}& \multicolumn{2}{c}{31}& \multicolumn{2}{c}{17} \\
\hline
\ttbar & 28.7 & 5.3 & 6.0 & 1.1 & 107.6 & 10.3 & 26.6 & 5.9 & 5.5 & 1.2 \\
$Wt$ single top & 0.88 & 0.15 & 0.21 & 0.07 & 3.0 & 0.6 & 1.00 & 0.27 & 0.17 & 0.05 \\
$Z$\,+\,jets & 1.2 & 1.0 & 8.5 & 2.6 & 1.3 & 0.5 & 2.8 & 1.0 & 11.5 & 3.8 \\
Diboson & 0.06 & 0.01 & 0.16 & 0.03 & 0.20 & 0.04 & 0.06 & 0.01 & 0.30 & 0.06 \\
Misidentified leptons & 0.15 & 0.15 & 0.04 & 0.04 & 0.6 & 0.6 & 0.10 & 0.11 & 0.05 & 0.06 \\
\hline
Total prediction & 31.0 & 5.1 & 14.9 & 2.6 & 112.8 & 10.3 & 30.6 & 5.6 & 17.6 & 3.5 \\
\bottomrule
\end{tabular}
\end{table}
 
The products of jet acceptance and $b$-tagging efficiencies in the three
dilepton channels are $\epsbee=0.65\pm 0.06$, $\epsbem=0.68\pm 0.03$, and
$\epsbmm=0.59\pm 0.06$. All three values are consistent with each other and the predictions from simulation, validating the modelling of $b$-tagging in
the simulation. The scale factors for the $Z$\,+\,jets background are
$\rza=1.23\pm 0.09$ and $\rzb=1.68\pm 0.52$, suggesting that the \Sherpa\
$Z$\,+\,jets simulation slightly underestimates this background.
Consistent results were obtained from fits to the $e\mu$ and $ee+\mu\mu$
channels alone, which have total uncertainties of \ttxdlemrel\ and \ttxdlsfrel, respectively. The result is stable against variations of the \etmiss\ requirement.
 
The potential effect of the difference  between the $Z$-boson $ee$ and
$\mu\mu$ yields with one $b$-tagged jet visible in Table~\ref{t:evtcount} and
Figures~\ref{f:dilfitres}(a) and~\ref{f:dilfitres}(d) was
addressed by fitting the two same-flavour channels separately. This gave
compatible values for \xtt\ and \rzb, but a three-standard-deviation
discrepancy between the values of \rza\ fitted in the $ee$ and $\mu\mu$
channels. However, this discrepancy disappeared when lowering the \etmiss\
requirement to $\etmiss>20$\,\GeV, which increased the \Zjets background
by a factor of three. The inclusive $Z\rightarrow ee$ and $Z\rightarrow\mu\mu$
yields (without requirements on additional jets) were also compared and
found to be consistent within 1\%, validating the modelling of the
electron and muon efficiencies. The difference between the observed and
predicted yields seen in Figure~\ref{f:dilfitres}
was therefore attributed to a statistical fluctuation.
 
\begin{figure}[tp]
\begin{center}
\subfloat[]{\includegraphics[width=0.45\textwidth]{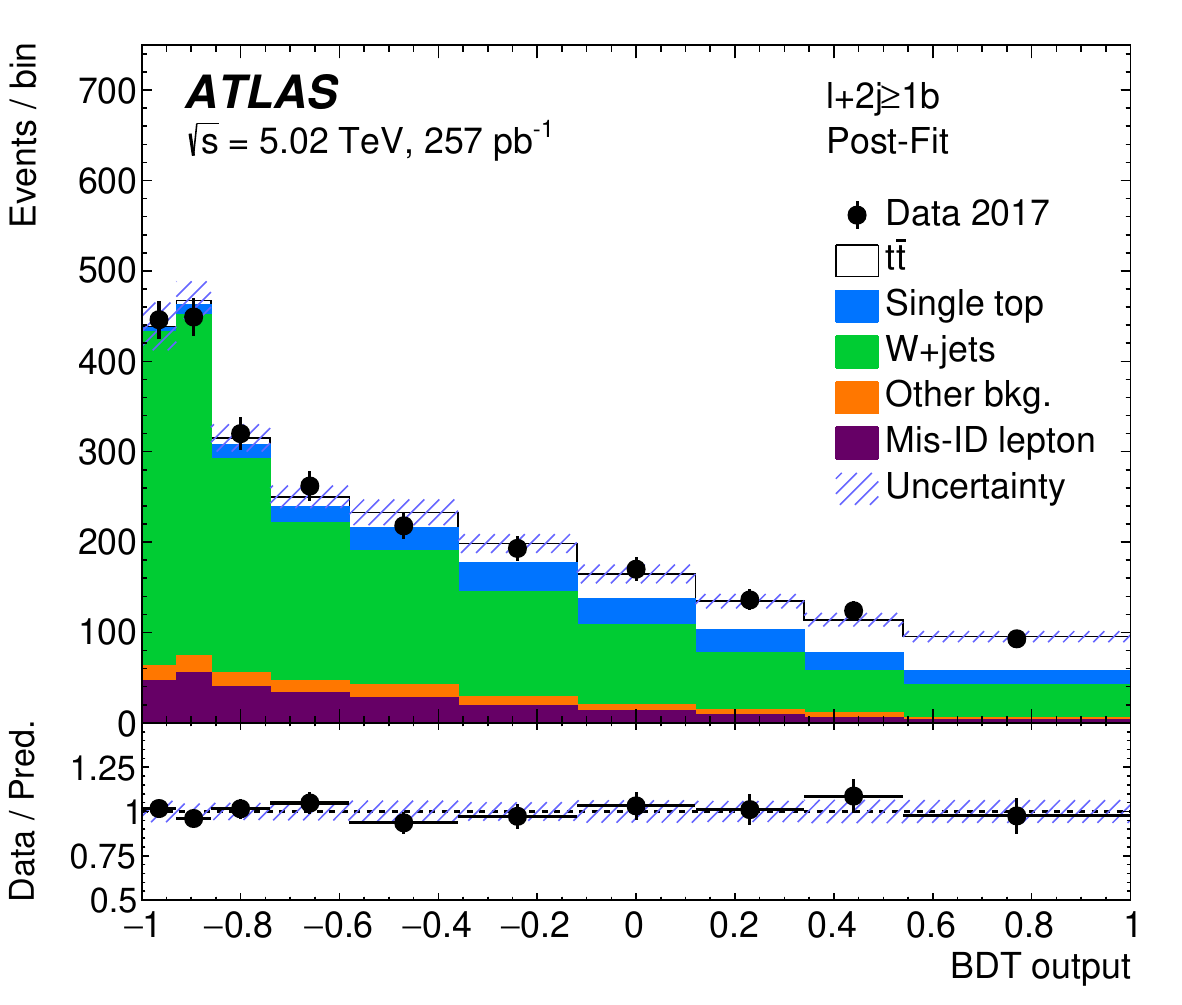}}
\subfloat[]{\includegraphics[width=0.45\textwidth]{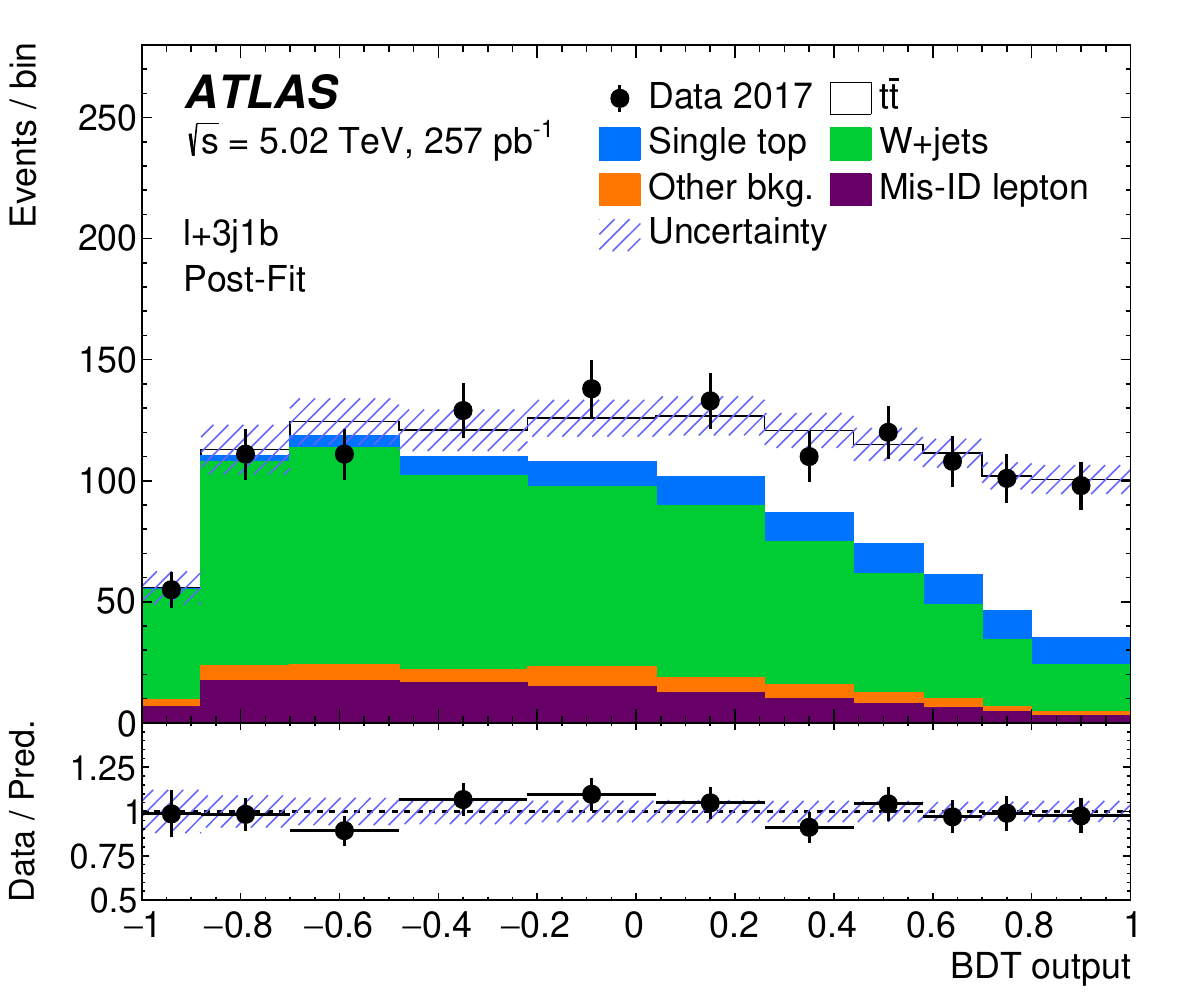}} \\
\subfloat[]{\includegraphics[width=0.45\textwidth]{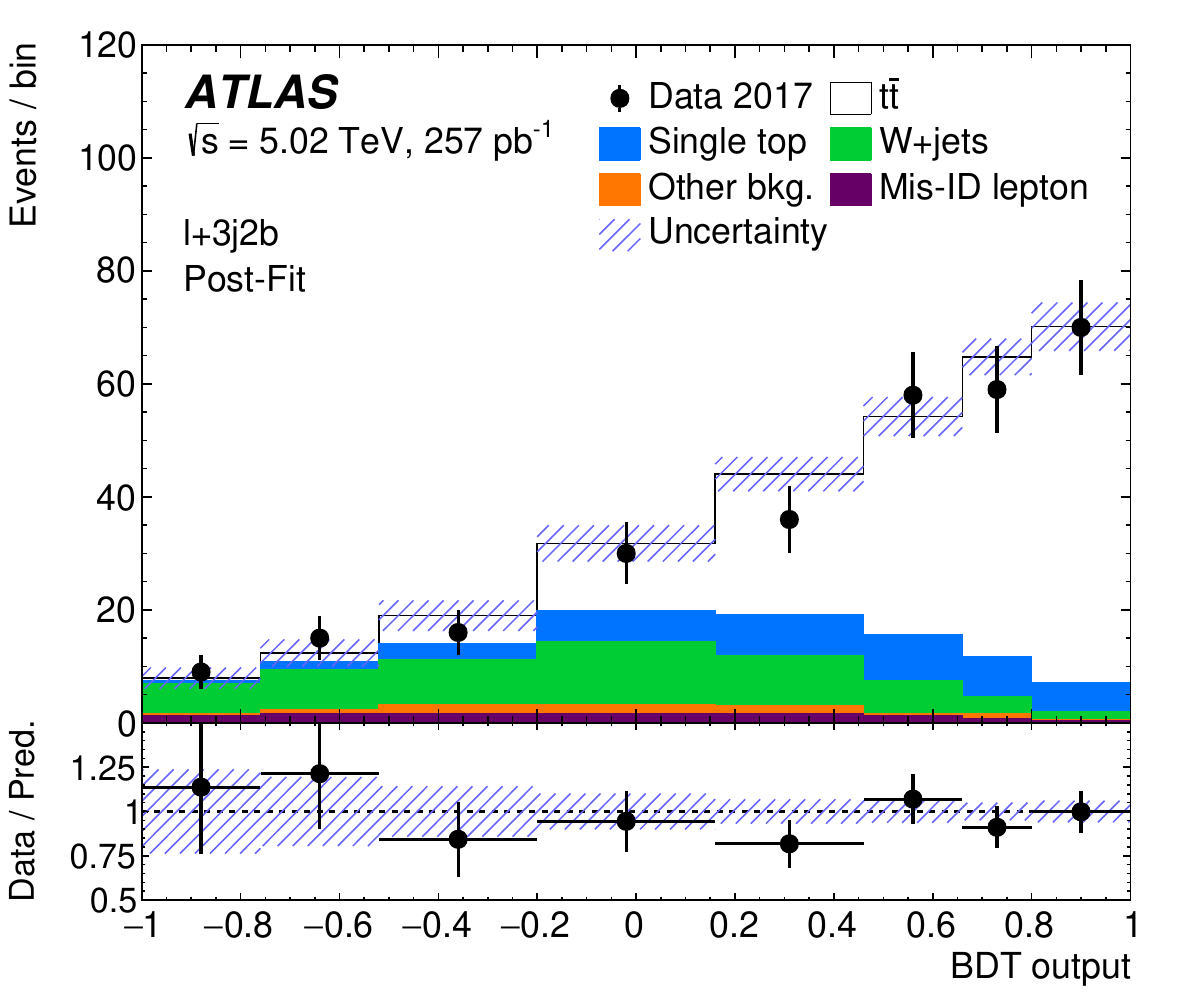}}
\subfloat[]{\includegraphics[width=0.45\textwidth]{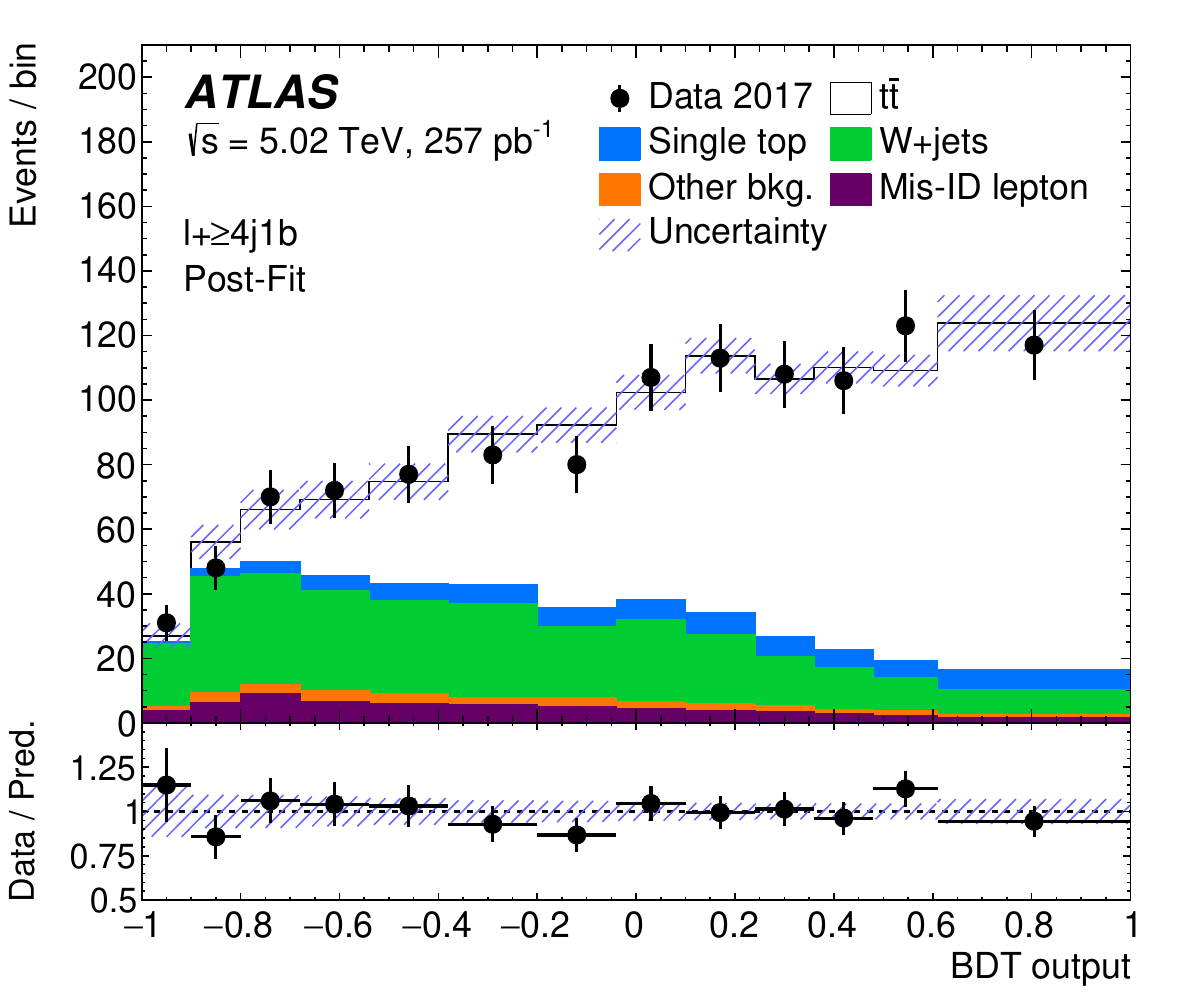}} \\
\subfloat[]{\includegraphics[width=0.45\textwidth]{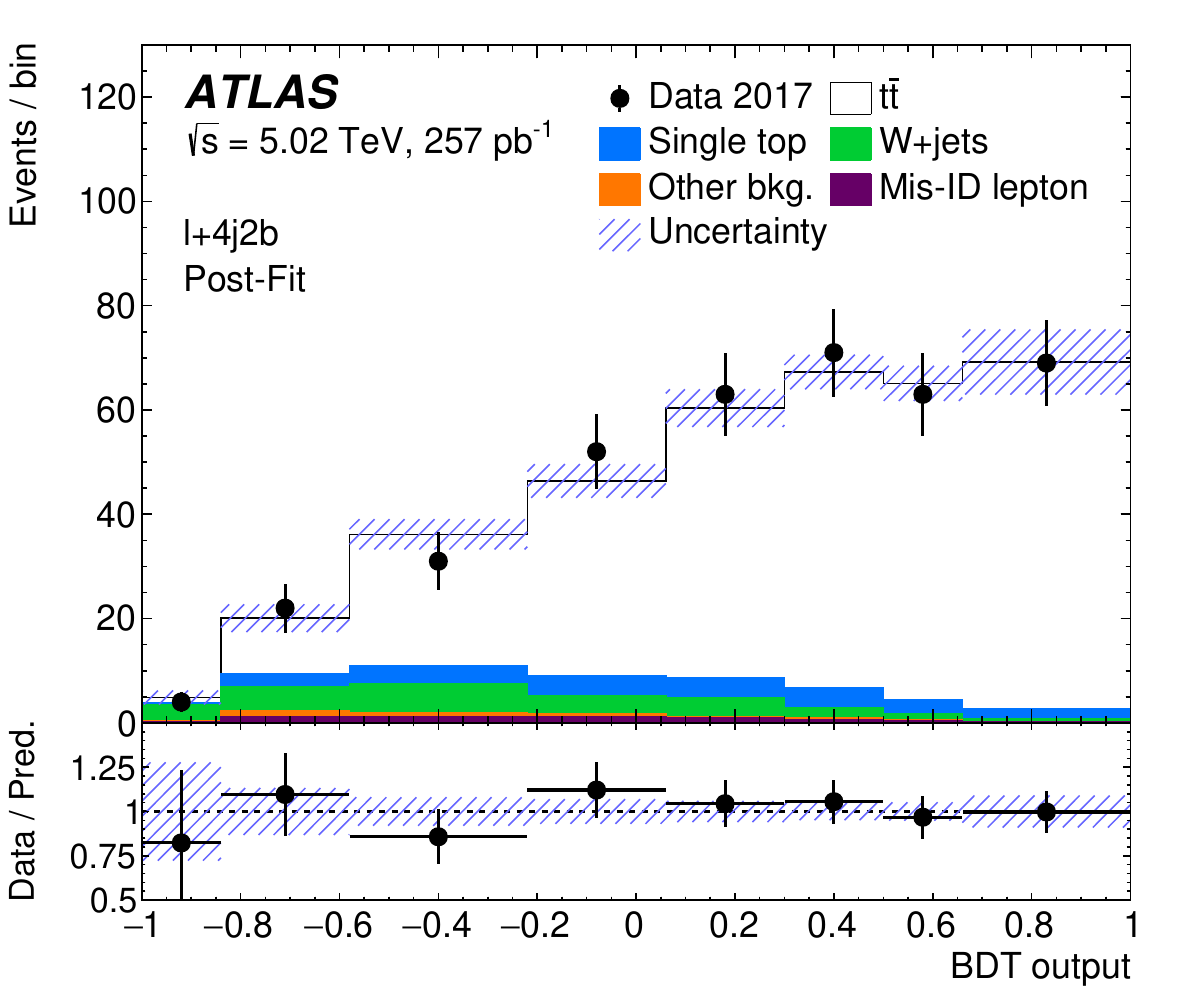}}
\subfloat[]{\includegraphics[width=0.45\textwidth]{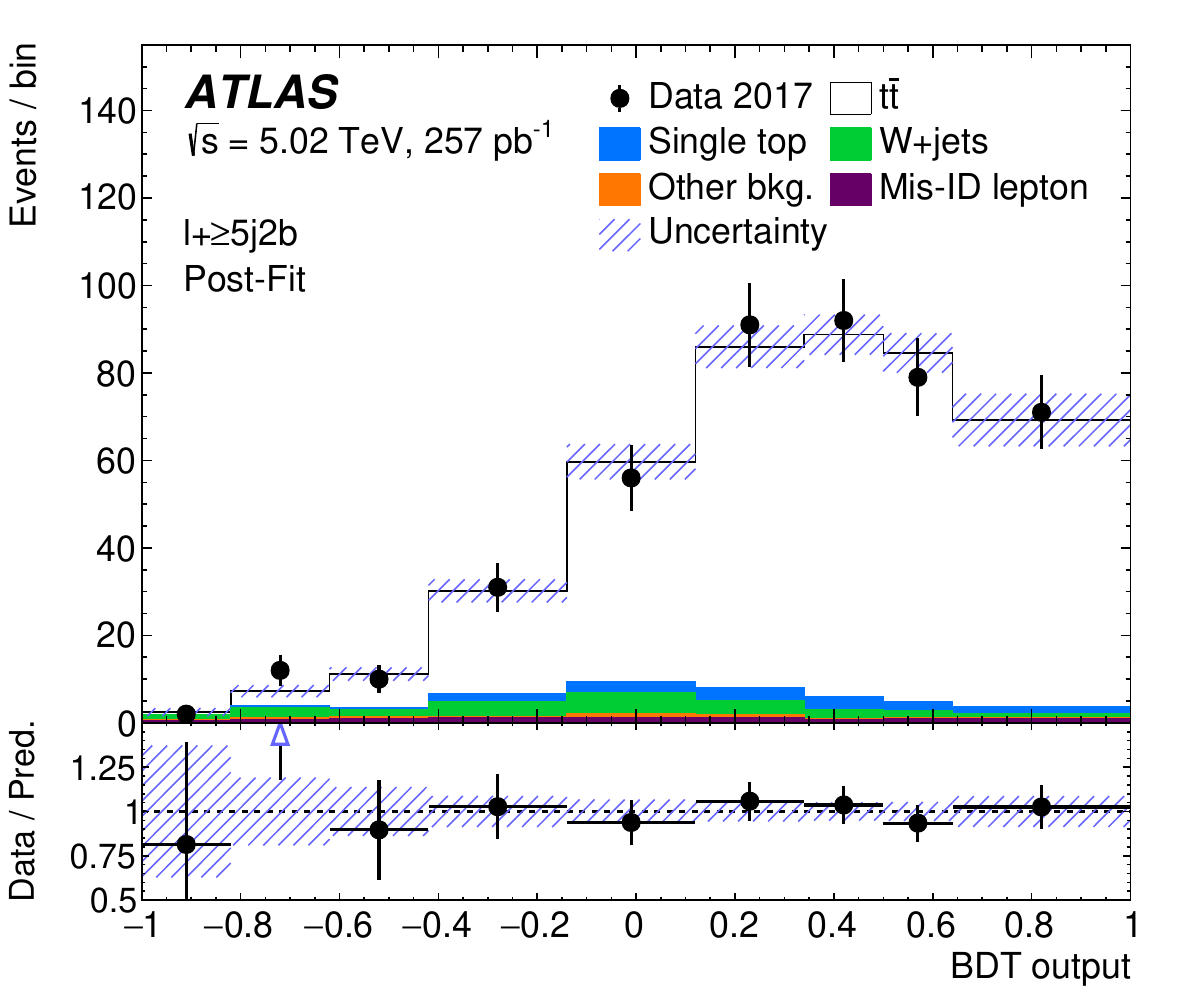}}
\caption{Comparison between data and the results of the fit to the BDT output distributions
in the
(a) {\CRzero} region,
(b) {\CRone} region,
(c) {\CRtwo} region,
(d) {\SRone} region,
(e) {\SRtwo} region, and
(f) {\SRthree} region.
The `Other bkg.' includes \Zjets and diboson backgrounds.
The bottom panel displays the ratio of data to the total prediction.
The hashed area represents the total uncertainty of the background.}
\label{fig:postfit_BDT}
\end{center}
\end{figure}

The \ttbar~cross-section in the single-lepton channel was determined by fitting a model
consisting of the estimated background and \ttbar\ signal distributions
to the observed BDT distributions in the six regions simultaneously.
The likelihood fit incorporated the systematic uncertainties as Gaussian nuisance parameters that modified the shape and/or normalisation of the predicted distributions as defined in Section~\ref{sec:systematics}.
Additional bin-by-bin uncertainties were included to account for the statistical uncertainties
in the predicted backgrounds.
The fitted background parameters in the model were found to be consistent with their central values, taking into account
uncertainties.
The results of the fit to the BDT distributions are shown in Figure~\ref{fig:postfit_BDT}\ and confirm that the fit
gives a good description of the data distributions. The pulls and constraints
of the nuisance parameters were found to be reasonable, with none of the
parameters being pulled by more than half a standard deviation from their input
value, or constrained to less than 50\% of their input uncertainty.
 
The single-lepton channel fit results in
\begin{equation}\nonumber
\xtt = \ttxljval \pm \ttxljstat\,\mathrm{(stat.)}\ \pm \ttxljsyst\,\mathrm{(syst.)}\ \pm \ttxljlumi\,\mathrm{(lumi.)}\ \pm \ttxljebeam\,\mathrm{(beam)}\,\mathrm{pb}.
\end{equation}
The \ttbar~cross-sections at \sxut~measured in the single-lepton and
dilepton channels are consistent.
 
The `single lepton' column of Table~\ref{tab:sys_impact_data} shows a breakdown
of the individual uncertainties.
The largest contributions to the total uncertainty of \ttxljerrtot\,pb come
from modelling of the \Wjets background, the integrated luminosity
uncertainty, and the data statistical uncertainty.
The result changes by $\mp \dmtopvallj$\% for a $\pm 1$\,\GeV\ change in \mtop,
a larger sensitivity than in the dilepton channel because the simulation is
also used to predict the efficiencies of the jet requirements.
This result is slightly more precise than the \ttbar\ cross-section measurement
in the single-lepton channel at \sxyt~\cite{TOPQ-2020-02}, despite the much
smaller data sample.
This arises from the use of six regions and the fit of the shapes of the BDT distributions in each region to constrain the background contributions.
It also reflects the fact that both analyses are dominated by systematic
uncertainties, and although the non-\ttbar\ backgrounds are larger than at
\sxyt, the uncertainties due to QCD radiation in \ttbar\ events are less
important.
 
\FloatBarrier


\section{Combined cross-section and comparison with predictions}
 
\label{sec:combination}

The cross-section measurements in the dilepton and single-lepton channels were combined to provide a more precise result.
The dilepton measurement used a fit of the \ttbar\ signal, background contributions, and $b$-tagging efficiency, while
the single-lepton measurement used a binned profile-likelihood fit, resulting
in post-fit uncertainty correlations that must be accounted for in a combination.
To combine these two measurements, taking into account their different techniques,
the Convino~\cite{Convino} algorithm was used.
It minimises a \chiS~with
three terms: one encoding the results and the statistical uncertainty of each measurement,
the second encoding the correlations between uncertainties in a given measurement, and the third term accounting for the
prior knowledge of correlations between the uncertainties of the different measurements.

The prior probabilities (priors) for the correlations between uncertainties in the
two measurements were determined by assigning each systematic uncertainty in
the combination to one of three categories: unique uncertainties,
one-to-one uncertainties, and one-to-many uncertainties.
Unique uncertainties are ones that exist only in one of the two measurements,
or which are uncorrelated between the two measurements.
One-to-one uncertainties are those that affect both measurements in a similar
way.
Finally, one-to-many uncertainties are the set of uncertainties represented by single nuisance parameters in one
measurement and multiple nuisance parameters in the other.
The latter category covers most of the \ttbar\ modelling uncertainties, which
were represented by a single nuisance parameter in the dilepton analysis and
separate nuisance parameters for each of the fit regions in the single-lepton
analysis, due to the need to account for uncertainties in the shapes of the BDT
input distributions.
 
This categorisation was used to determine the correlations between the various systematic uncertainties in
the Convino minimisation procedure.
By definition, unique uncertainties in one measurement are uncorrelated with all other uncertainties.
One-to-one uncertainties are fully correlated between the two measurements.
The one-to-many uncertainties in this combination were investigated by considering numerous correlation
priors  between the one nuisance parameter in one channel and the set of nuisance parameters in the other.
The results of the combination were found to be insensitive to the prior employed for these one-to-many uncertainties
and so the assumption of fully correlated uncertainties was employed in the fit.
In the combination, the $b$-tagging uncertainties were treated as one-to-one uncertainties, even though different tagger
working points were used in the two measurements.
 
The \ttbar cross-section value at \sxut\ obtained from the combination of the dilepton and single-lepton measurements is
\begin{equation*}
\xtt = \ttxval \pm \ttxstat (\text{stat.}) \pm \ttxsyst (\text{syst.}) \pm \ttxlumi (\text{lumi.}) \pm \ttxebeam (\text{beam})\,\text{pb}.
\end{equation*}
Table~\ref{tab:sys_impact_data} shows a breakdown of uncertainties in the individual measurements and the combined result.
The combination gives a lower overall uncertainty due to reductions in the signal modelling, electron reconstruction,
flavour-tagging, and jet-energy-scale uncertainties.
The uncertainties arising from theory and MC modelling are larger than the experimental uncertainties.
The combined result changes by $\mp\dmtopval$\% for a $\pm 1$\,\GeV\
change in \mtop.
 
\begin{table}[tp]
\centering
\caption{Breakdown of uncertainties in the dilepton, single-lepton, and
combined measurements of the inclusive \ttbar\ cross-section at \sxut.
For each category, the dilepton uncertainties are calculated by summing all
the contributing uncertainties in quadrature.
The single-lepton and combination uncertainties are calculated by fixing the set of nuisance parameters corresponding
to a category, repeating the fit, and subtracting in quadrature the resulting uncertainty from the total uncertainty of the nominal fit. Categories
that include unique uncertainties, uncorrelated between dilepton and
single-lepton measurements, are denoted by *, and those with one-to-many
uncertainties by $^\dagger$ (see text). Other categories include only one-to-one
uncertainty sources, correlated between channels.
The total uncertainty is the sum in quadrature of the total systematic uncertainty, the statistical uncertainty, and the effects
of the uncertainties in the integrated luminosity and beam energy.
The systematic uncertainties for the single-lepton and combination measurements do not add up in quadrature to the total
systematic uncertainty because of their correlations in the fit.
}
\begin{tabular}{ l  c  c  c }
\toprule
Category & \multicolumn{3}{c}{$\delta \xtt$ [\%] } \\
& Dilepton  &  Single lepton & Combination  \\
\toprule
$t\bar{t}$ generator$^\dagger$            & 1.2   & 1.0 & 0.8 \\
$t\bar{t}$ parton-shower/hadronisation*$^{,\dagger}$                  & 0.3 & 0.9 & 0.7 \\
$t\bar{t}$ $h_{\text{damp}}$ and scale variations$^\dagger$ 	& 1.0   & 1.1 & 0.8  \\
$t\bar{t}$ parton distribution functions$^\dagger$            & 0.2   & 0.2 & 0.2  \\
\midrule
Single-top background & 1.1 & 0.8 & 0.6 \\
$W$/$Z$\,+\,jets background* & 0.8 & 2.4 & 1.8  \\
Diboson background & 0.3   & 0.1 & $< 0.1$~~~~ \\
Misidentified leptons* & 0.7 & 0.3 & 0.3 \\
\midrule
Electron identification/isolation    & 0.8 & 1.2 & 0.8 \\
Electron energy scale/resolution     & 0.1 & 0.1 & $ < 0.1$~~~~ \\
Muon identification/isolation        & 0.6 & 0.2 & 0.3 \\
Muon momentum scale/resolution       & 0.1 & 0.1 & 0.1  \\
Lepton-trigger efficiency            & 0.2 & 0.9 & 0.7 \\
Jet-energy scale/resolution           & 0.1 & 1.1 & 0.8 \\
\sxut\ JES correction    & 0.1 & 0.6 & 0.5 \\
Jet-vertex tagging			& $< 0.1$~~~~ & 0.2 & 0.2 \\
Flavour tagging            & 0.1 & 1.1 & 0.8  \\
$E_{\text{T}}^{\text{miss}}$ & 0.1 & 0.4 &  0.3 \\
\midrule
Simulation statistical uncertainty* 	& 0.2 & 0.6 & 0.5 \\
\midrule
Data statistical uncertainty* & 6.8 & 1.3 & 1.3  \\
Total systematic uncertainty  & 2.5 & 4.2 & 3.4 \\
Integrated luminosity    & 1.8 & 1.6 & 1.6  \\
Beam energy & 0.3 & 0.3 & 0.3 \\
\midrule
Total uncertainty	      & 7.5 & 4.5 & 3.9 \\
\bottomrule
\end{tabular}
\label{tab:sys_impact_data}
\end{table}

\begin{figure}[tp]
\centering
 
\includegraphics[width=150mm]{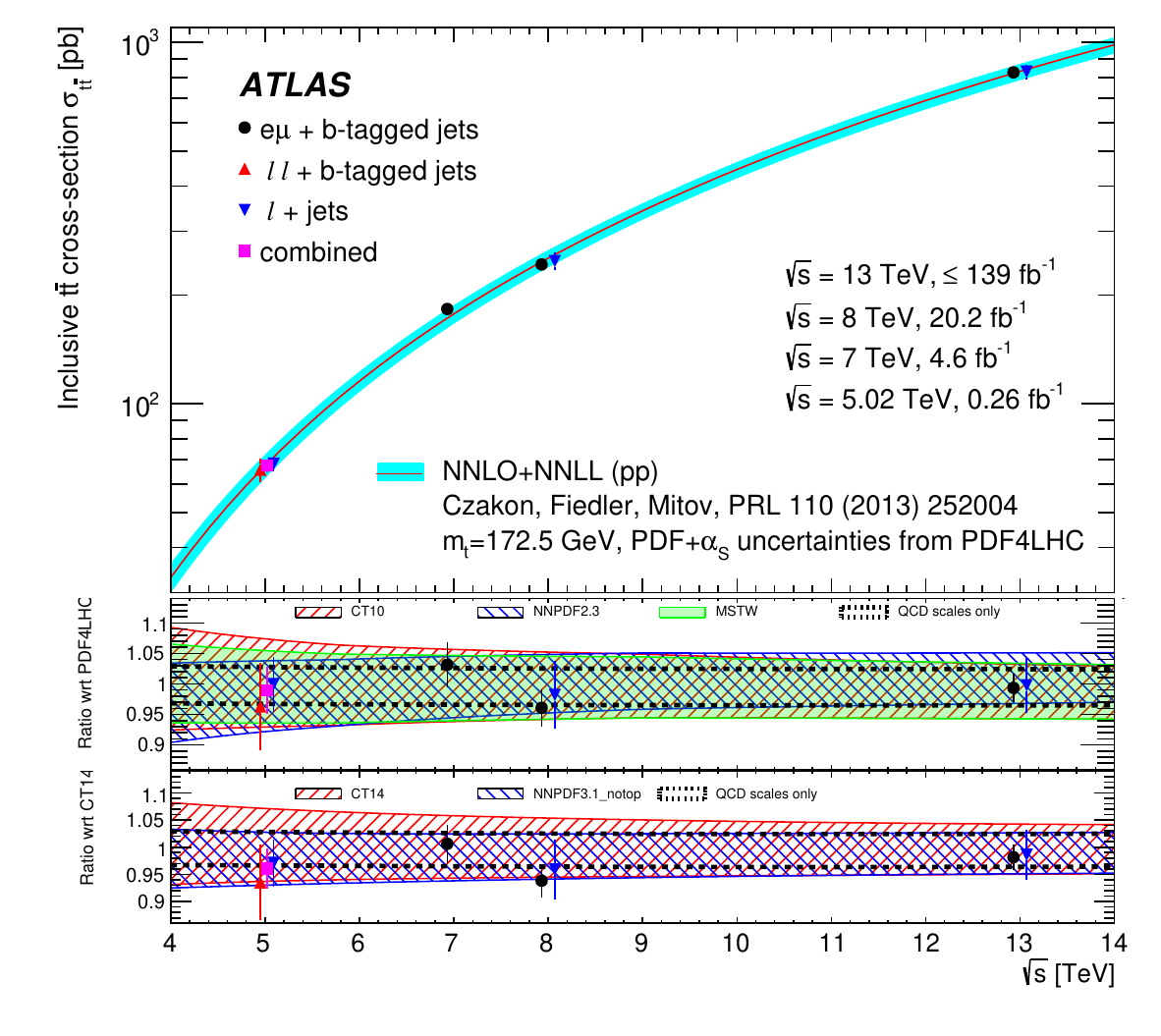}
\caption{\label{f:xsvse}The upper panel shows the inclusive \ttbar\
cross-section \xtt\ as a function
of centre-of-mass energy $\sqrt{s}$, comparing ATLAS Collaboration measurements from this analysis,
from the $e\mu$ plus $b$-tagged-jets final state at
$\sqrt{s}=7$, 8, and 13\,\TeV, and from the single-lepton final state at
$\sqrt{s}=8$ and 13\,\TeV\ with NNLO+NNLL theoretical predictions
\cite{topxtheo5} calculated using \TOPpp~\cite{toppp} with the \PDFforLHC\
prescription for PDF and \alphas\ uncertainties \cite{pdf4lhc},
and $\mtop=172.5$\,\GeV.
The middle panel shows the ratios of the measurements and predictions
to the central value of the prediction using \PDFforLHC\ PDFs.
The total uncertainties when using the  individual \NNPDF[2.3], \MSTW and \CT[10]
PDFs are shown as overlapping hatched or coloured bands, and the dotted lines
show the QCD scale uncertainties alone.
The lower panel shows the ratios of the measurements and predictions from
the CT14 and \NNPDF[3.1\_notop]\ PDFs to the central value from CT14.
Measurements made at the same $\sqrt{s}$ are slightly offset for clarity.
The $\sqrt{s}=7$ and 8\,\TeV\ results are taken from
Refs. \cite{TOPQ-2013-04,TOPQ-2016-08}, with the LHC beam energy uncertainties
reduced according to Ref. \cite{ebeam2}. The $\sqrt{s}=13$\,\TeV\ results
are taken from Refs. \cite{TOPQ-2018-17,TOPQ-2020-02}.}
\end{figure}

The combined result is consistent with the NNLO+NNLL QCD prediction of
$\xtt=68.2^{+5.2}_{-5.3}$\,pb discussed in Section~\ref{sec:intro}, and
with the CMS Collaboration measurement \cite{CMS-TOP-20-004}.
The individual and combined ATLAS Collaboration results are compared in Figure~\ref{f:xsvse}
with measurements in the $e\mu$ and single-lepton channels at other $\sqrt{s}$ values. Predictions
using the \CT[10] \cite{cttenpdf}, \NNPDF[2.3] \cite{nnpdfffn},
\MSTW~\cite{mstwnnlo1,mstwnnlo2},  CT14~\cite{ct14}, and
\NNPDF[3.1\_notop] \cite{nnpdf31} PDF sets are also shown. The first three PDF sets do not
incorporate any constraints from
LHC data, while the last two are PDF sets that use some LHC data
but no \ttbar\ cross-section measurements.
 
\begin{figure}[tp]
\parbox{88mm}{\includegraphics[width=88mm]{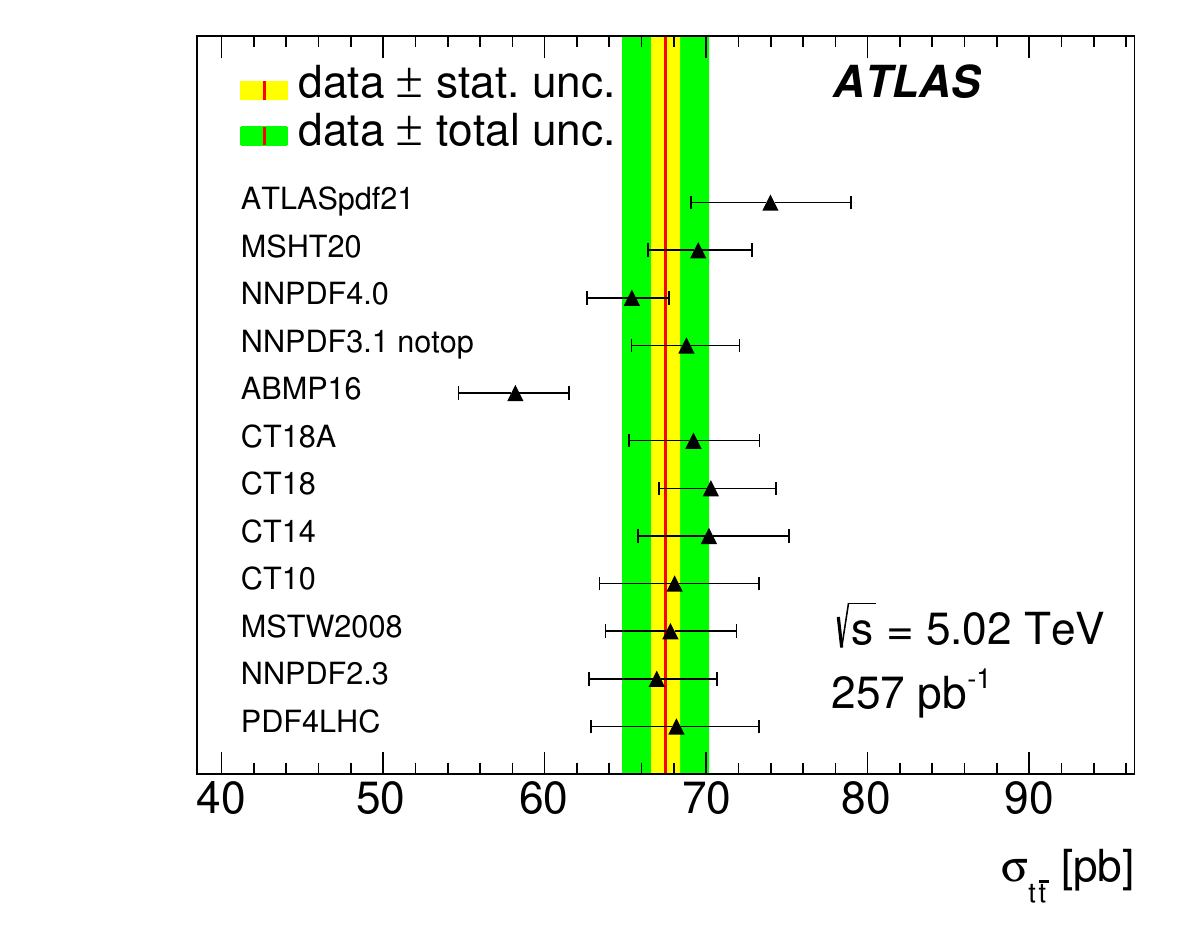}\vspace{-6mm}\center{(a)}}
\parbox{65mm}{\rule[-7mm]{0mm}{50mm}\includegraphics[width=65mm]{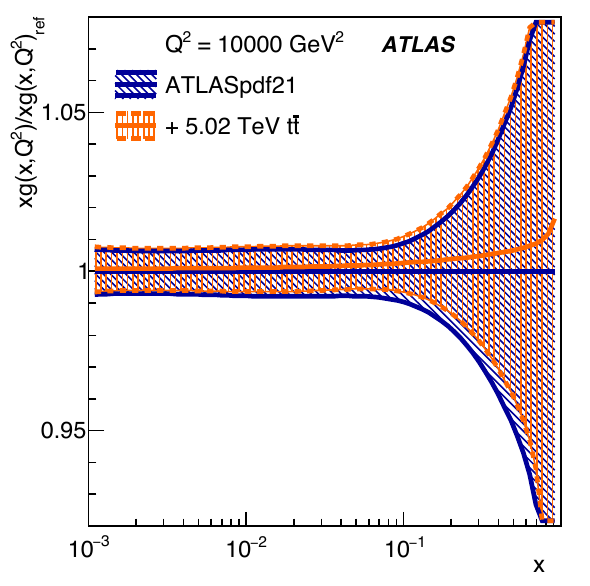}\vspace{-6.5mm}\center{(b)}}
\caption{\label{f:xseccomp}(a) Measurement of the \ttbar\ production cross-section
at \sxut\ in the combined dilepton and single-lepton channels
compared with the predictions from various NNLO PDF sets.
The bands show the experimental measurement, with the statistical
(inner yellow band) and total (outer green band) uncertainties shown
separately.
The last entry shows the prediction using the \PDFforLHC\ recipe,
encompassing the predictions from the \CT[10], \MSTW\ and \NNPDF[2.3] PDF sets.
(b) Ratio of the gluon PDF determined using the data of the ATLASpdf21 PDF fit
plus the constraint from the combined \ttbar\ cross-section measurement
compared with the ATLASpdf21 fit alone, as a function of Bjorken-$x$. The
hatched bands show the uncertainties in the combined fit (orange) and
ATLASpdf21 fit (blue), and the orange line shows the shift in central value
when adding the \sxut\ cross-section measurement.}
\end{figure}

Figure~\ref{f:xseccomp}(a) compares
the measured cross-section at \sxut\ with the predictions from the aforementioned PDF sets, as well as the
MSHT20 \cite{msht20}, NNPDF4.0 \cite{nnpdf40}, \ABMP[16] \cite{abmp16pdf},
CT18 \cite{ct18} and CT18A \cite{ct18} PDFs,
and the combination of the \CT[10], \MSTW, and \NNPDF[2.3] PDFs using the
\PDFforLHC\ recipe. The result is also compared with the recent \ATLASpdf[21]
PDF fit \cite{STDM-2020-32}, which includes a diverse set of ATLAS Collaboration
measurements at $\sqrt{s}=7$, 8 and 13\,\TeV\ (including \ttbar\
differential cross-section measurements at $\sqrt{s}=8$ and 13\,\TeV),
together with deep-inelastic scattering data from $ep$ collisions at the
HERA collider \cite{heradis}.
The measured value is compatible with the predictions from
all the PDF sets considered, except for \ABMP[16], which has
a softer gluon distribution and predicts a \xtt\ value
2.2~standard deviations lower than measured.
 
Figure~\ref{f:xseccomp}(b) shows the effect
of adding the combined \sxut\ \ttbar\ cross-section measurement to the
\ATLASpdf[21] PDF fit.
The combination was performed using the xFitter framework
\cite{xfitter} as described in Ref. \cite{STDM-2020-32}.
At a squared-momentum-transfer value of $Q^2=10^4\,\GeV^2$, the addition
of the new data reduces the gluon PDF uncertainty in the
region of  Bjorken-$x$ (the fraction of the proton momentum carried by the
parton participating in the initial interaction) above $x\approx 0.05$,
and gives e.g.\ a 5\% reduction at $x=0.1$.
The uncertainties in the valance and sea quark PDFs are unaffected.
 
\section{Conclusion}
\label{sec:conclusions}

The inclusive \ttbar\ production cross-section \xtt\ has been measured
in $pp$ collisions at \sxut\ using \intlumi\,\ipb\ of data recorded
by the ATLAS experiment at the LHC in 2017 using events from both the
dilepton and single-lepton channels. The combined result is
\begin{equation}\nonumber
\xtt = \ttxval \pm \ttxstat\,\mathrm{(stat.)}\ \pm \ttxsyst\,\mathrm{(syst.)}\ \pm \ttxlumi\,\mathrm{(lumi.)}\ \pm \ttxebeam\,\mathrm{(beam)}\,\mathrm{pb},
\end{equation}
where the four uncertainties are due to the size of the data sample,
experimental and theoretical systematic effects, and imperfect knowledge of both
the integrated luminosity and the LHC beam energy. The total uncertainty is
\ttxerrtot\,pb, or \ttxrel.
The result is consistent with the NNLO+NNLL QCD prediction of
$68.2^{+5.2}_{-5.3}$\,pb and  with a previous measurement by the
CMS Collaboration, but has a total uncertainty that is almost a factor of two
smaller. This measurement provides additional constraints on the
gluon distribution of the proton PDF for Bjorken-$x>0.05$.


\section*{Acknowledgements}


We thank CERN for the very successful operation of the LHC, as well as the
support staff from our institutions without whom ATLAS could not be
operated efficiently.
 
We acknowledge the support of
ANPCyT, Argentina;
YerPhI, Armenia;
ARC, Australia;
BMWFW and FWF, Austria;
ANAS, Azerbaijan;
CNPq and FAPESP, Brazil;
NSERC, NRC and CFI, Canada;
CERN;
ANID, Chile;
CAS, MOST and NSFC, China;
Minciencias, Colombia;
MEYS CR, Czech Republic;
DNRF and DNSRC, Denmark;
IN2P3-CNRS and CEA-DRF/IRFU, France;
SRNSFG, Georgia;
BMBF, HGF and MPG, Germany;
GSRI, Greece;
RGC and Hong Kong SAR, China;
ISF and Benoziyo Center, Israel;
INFN, Italy;
MEXT and JSPS, Japan;
CNRST, Morocco;
NWO, Netherlands;
RCN, Norway;
MEiN, Poland;
FCT, Portugal;
MNE/IFA, Romania;
MESTD, Serbia;
MSSR, Slovakia;
ARRS and MIZ\v{S}, Slovenia;
DSI/NRF, South Africa;
MICINN, Spain;
SRC and Wallenberg Foundation, Sweden;
SERI, SNSF and Cantons of Bern and Geneva, Switzerland;
MOST, Taiwan;
TENMAK, T\"urkiye;
STFC, United Kingdom;
DOE and NSF, United States of America.
In addition, individual groups and members have received support from
BCKDF, CANARIE, Compute Canada and CRC, Canada;
PRIMUS 21/SCI/017 and UNCE SCI/013, Czech Republic;
COST, ERC, ERDF, Horizon 2020 and Marie Sk{\l}odowska-Curie Actions, European Union;
Investissements d'Avenir Labex, Investissements d'Avenir Idex and ANR, France;
DFG and AvH Foundation, Germany;
Herakleitos, Thales and Aristeia programmes co-financed by EU-ESF and the Greek NSRF, Greece;
BSF-NSF and MINERVA, Israel;
Norwegian Financial Mechanism 2014-2021, Norway;
NCN and NAWA, Poland;
La Caixa Banking Foundation, CERCA Programme Generalitat de Catalunya and PROMETEO and GenT Programmes Generalitat Valenciana, Spain;
G\"{o}ran Gustafssons Stiftelse, Sweden;
The Royal Society and Leverhulme Trust, United Kingdom.
 
The crucial computing support from all WLCG partners is acknowledged gratefully, in particular from CERN, the ATLAS Tier-1 facilities at TRIUMF (Canada), NDGF (Denmark, Norway, Sweden), CC-IN2P3 (France), KIT/GridKA (Germany), INFN-CNAF (Italy), NL-T1 (Netherlands), PIC (Spain), ASGC (Taiwan), RAL (UK) and BNL (USA), the Tier-2 facilities worldwide and large non-WLCG resource providers. Major contributors of computing resources are listed in Ref.~\cite{ATL-SOFT-PUB-2021-003}.


\clearpage
\printbibliography
 
\clearpage
 
\begin{flushleft}
\hypersetup{urlcolor=black}
{\Large The ATLAS Collaboration}

\bigskip

\AtlasOrcid[0000-0002-6665-4934]{G.~Aad}$^\textrm{\scriptsize 101}$,
\AtlasOrcid[0000-0002-5888-2734]{B.~Abbott}$^\textrm{\scriptsize 119}$,
\AtlasOrcid[0000-0002-7248-3203]{D.C.~Abbott}$^\textrm{\scriptsize 102}$,
\AtlasOrcid[0000-0002-1002-1652]{K.~Abeling}$^\textrm{\scriptsize 55}$,
\AtlasOrcid[0000-0002-8496-9294]{S.H.~Abidi}$^\textrm{\scriptsize 29}$,
\AtlasOrcid[0000-0002-9987-2292]{A.~Aboulhorma}$^\textrm{\scriptsize 35e}$,
\AtlasOrcid[0000-0001-5329-6640]{H.~Abramowicz}$^\textrm{\scriptsize 150}$,
\AtlasOrcid[0000-0002-1599-2896]{H.~Abreu}$^\textrm{\scriptsize 149}$,
\AtlasOrcid[0000-0003-0403-3697]{Y.~Abulaiti}$^\textrm{\scriptsize 116}$,
\AtlasOrcid[0000-0003-0762-7204]{A.C.~Abusleme~Hoffman}$^\textrm{\scriptsize 136a}$
% ...
\bigskip
\\

% ...
$^{29}$Physics Department, Brookhaven National Laboratory, Upton NY; United States of America.\\
% ...
$^{35}$$^{(a)}$Facult\'e des Sciences Ain Chock, R\'eseau Universitaire de Physique des Hautes Energies - Universit\'e Hassan II, Casablanca;$^{(b)}$Facult\'{e} des Sciences, Universit\'{e} Ibn-Tofail, K\'{e}nitra;$^{(c)}$Facult\'e des Sciences Semlalia, Universit\'e Cadi Ayyad, LPHEA-Marrakech;$^{(d)}$LPMR, Facult\'e des Sciences, Universit\'e Mohamed Premier, Oujda;$^{(e)}$Facult\'e des sciences, Universit\'e Mohammed V, Rabat;$^{(f)}$Institute of Applied Physics, Mohammed VI Polytechnic University, Ben Guerir; Morocco.\\
% ...
$^{55}$II. Physikalisches Institut, Georg-August-Universit\"{a}t G\"ottingen, G\"ottingen; Germany.\\
% ...
$^{101}$CPPM, Aix-Marseille Universit\'e, CNRS/IN2P3, Marseille; France.\\
$^{102}$Department of Physics, University of Massachusetts, Amherst MA; United States of America.\\
% ...
$^{116}$Department of Physics, New York University, New York NY; United States of America.\\
% ...
$^{119}$Homer L. Dodge Department of Physics and Astronomy, University of Oklahoma, Norman OK; United States of America.\\
% ...
$^{136}$$^{(a)}$Departamento de F\'isica, Pontificia Universidad Cat\'olica de Chile, Santiago;$^{(b)}$Millennium Institute for Subatomic physics at high energy frontier (SAPHIR), Santiago;$^{(c)}$Instituto de Investigaci\'on Multidisciplinario en Ciencia y Tecnolog\'ia, y Departamento de F\'isica, Universidad de La Serena;$^{(d)}$Universidad Andres Bello, Department of Physics, Santiago;$^{(e)}$Instituto de Alta Investigaci\'on, Universidad de Tarapac\'a, Arica;$^{(f)}$Departamento de F\'isica, Universidad T\'ecnica Federico Santa Mar\'ia, Valpara\'iso; Chile.\\
% ...
$^{149}$Department of Physics, Technion, Israel Institute of Technology, Haifa; Israel.\\
$^{150}$Raymond and Beverly Sackler School of Physics and Astronomy, Tel Aviv University, Tel Aviv; Israel.\\
% ...
$^{a}$ Also Affiliated with an institute covered by a cooperation agreement with CERN.\\
$^{b}$ Also at An-Najah National University, Nablus; Palestine.\\
$^{c}$ Also at Borough of Manhattan Community College, City University of New York, New York NY; United States of America.\\
$^{d}$ Also at Bruno Kessler Foundation, Trento; Italy.\\
$^{e}$ Also at Center for High Energy Physics, Peking University; China.\\
$^{f}$ Also at Centro Studi e Ricerche Enrico Fermi; Italy.\\
% ...

\end{flushleft}


\end{document}